\documentclass{article}
\usepackage[centertags]{amsmath}
\usepackage{amssymb}
\usepackage{texdraw}
\usepackage{a4wide}
\usepackage{epsfig}
\numberwithin{equation}{section}

\font\bbigbf=cmbx12 scaled \magstep 1 
\font\bigsc=cmcsc10 scaled \magstep 1 \font\bigrm=cmr12
\font\filt=msbm10 
\def\reals{\hbox{\filt R}}

\def\tr{\mathop{\rm tr}\nolimits}
\def\div{\mathop{\rm div}\nolimits}
\def\curl{\mathop{\rm curl}\nolimits}
\def\dps{\displaystyle}
\def\Ca{C_{\rm a}}
\def\Co{C_\perp}
\def\Cp{C_\parallel}
\def\np{\bv{n}_\perp}
\def\nt{\bv{n}_3}
\def\qc{q_{\rm ch}}
\def\qs{q_{\rm sm}}
\font\bol=cmbx10\font\bolsev=cmbx7\font\bolfiv=cmbx5
\font\mib=cmmib10\font\mibsev=cmmib10 scaled 700
\font\mibfiv=cmmib10 scaled 500 \font\mbsy=cmbsy10
\font\mbsysev=cmbsy10 scaled 700 \font\mbsyfiv=cmbsy10 scaled 500
\def\bold{
   \textfont0=\bol \scriptfont0=\bolsev \scriptscriptfont0=\bolfiv
   \textfont1=\mib \scriptfont1=\mibsev \scriptscriptfont1=\mibfiv
   \textfont2=\mbsy \scriptfont2=\mbsysev \scriptscriptfont2=\mbsyfiv
         }
\def\bv#1{\hbox{$#1\bold$}}
\begin{document}
\begin{titlepage}
\null\vfil
\begin{center}
{\bbigbf Telephone-cord instabilities in thin smectic capillaries}
\end{center}

\vfil

\centerline{\bigrm {\bigsc Paolo Biscari}$^1$ and {\bigsc Maria
Carme Calderer}$^2$}
\vfil\noindent
\small $^1$ Dipartimento di Matematica, Politecnico di Milano, P.\
Leonardo da Vinci 32, 20133 Milano (Italy) and \\
$\phantom{^1}$\ Istituto Nazionale di Fisica della Materia, Via
Ferrata 1, 27100 Pavia (Italy)\\
$\phantom{^1}$\ e-mail: \texttt{paolo.biscari@polimi.it}\\
$^2$ School of Mathematics, University of Minnesota, 206 Church
St.\ SE, 55455 Minneapolis, MN (U.S.A.)\\
$\phantom{^2}$\ e-mail: \texttt{mcc@math.umn.edu}
\normalsize

\vfil

\noindent 2003 PACS: \ \ $\vtop{\hsize 12 cm \noindent 61.30.Dk --
Continuum models and theories of liquid crystal structure
\hfil\break 61.30.Pq -- Microconfined liquid crystals: droplets,
cylinders, \dots \hfil\break 77.84.Nh -- Ferroelectric liquid
crystals, \dots }$

\vfil

\begin{abstract}
Telephone-cord patterns have been recently observed in smectic
liquid crystal capillaries. In this paper we analyse the effects
that may induce them. As long as the capillary keeps its linear
shape, we show that a nonzero chiral cholesteric pitch favors the
SmA*-SmC* transition. However, neither the cholesteric pitch nor
the presence of an intrinsic bending stress are able to give rise
to a curved capillary shape.\\
The key ingredient for the telephone-cord instability is
spontaneous polarization. The free energy minimizer of a
spontaneously polarized SmA* is attained on a planar capillary,
characterized by a nonzero curvature. More interestingly, in the
SmC* phase the combined effect of the molecular tilt and the
spontaneous polarization pushes towards a helicoidal capillary
shape, with nonzero curvature and torsion.

\end{abstract}

\vfil
\noindent Date: \today

\vfil\vfil\vfil\vfil\vfil
\end{titlepage}

\section{Introduction}

Telephone-cord instabilities in carbon films have been identified
as processes which allow the material to relax its residual stress
\cite{02moje}. Recently, similar helicoid fibers have been
observed in bent-shaped liquid crystals (the so-called
\emph{banana liquid crystals\/}) \cite{00jali,03jakr}. In these
observations, the telephone-cord instability occurs in the smectic
phase, and a central role is played by the spontaneous
polarization, which characterizes the banana molecules
\cite{03cofe}. In fact, although some nematics are polar liquids
\cite{95phsa,97teph}, the first well-known liquid crystals
exhibiting significant spontaneous polarizations are smectic-C*.
In this phase, the molecules are tilted with respect to the layer
normal, and thus break the mirror symmetry
\cite{96nise,97lina,01ralu}.

The local polarization vector of the helical phases of a
smectic-C* is perpendicular to both the director and the layer
normal. Consequently, it is free to rotate in the plane of the
layers giving a zero polarization average over one pitch. The
electrooptic effects of the SmC* phase emerge with the unwinding
of the helix by surface stabilization, and result in homogeneous
spontaneous polarization throughout the sample. These homogeneous
director states give rise to the ferroelectric SmC* phases. For a
much more detailed description of the role of polarization in
liquid crystals, we refer the reader to the book by Lagerwall
\cite{99lage}, and more precisely to Sections 4.9-4.10, 5.4-5.6,
6.1-6.2, and 12.2-12.5 therein.

In this paper we analyze how polarization and chirality may
influence the ground state \emph{shapes\/} of thin filaments. Our
main results deal with a smectic liquid crystal endowed with a
nonzero spontaneous polarization. In the SmA* phase we find curved
planar configurations (the capillary axis is bent with nonzero
curvature) that have lower energy than straight ones. More
interestingly, in the SmC* phase the ground state configuration
has also a nonzero torsion. We derive analytical relations linking
the curvature and the torsion of the ground state shapes to the
material parameters.

Another interesting issue, stemming from experimental evidence on
telephone cord instabilities and reflected in our results, is the
fact that the mechanism for the capillary to decrease its energy
is by bending and twisting. For other geometries, the mechanism
for energy minimization may be the formation of domains
\cite{03cofe,99lage}. However, domain formation is often coupled
with the creation of energetically expensive boundary defects. The
defect energy favors the changes in material geometry that we
describe in our analysis.

The plan of the paper is as follows. In Section 2 we present and
discuss the model and the free energy functional. Section 3 is
devoted to linear capillary shapes: in it we show how a nonzero
cholesteric pitch may anticipate the SmA-SmC transition. In
Sections 4 and 5 we analyse the curved domains. In the former we
prove that neither the cholesteric pitch nor the intrinsic bending
stress are able to bend the axis of the smectic capillary. In the
latter we determine the curved shapes induced by the spontaneous
polarization: they are planar or three-dimensional depending on
the SmA*-SmC* phase of the liquid crystal. In the concluding
section we collect and discuss the above results.

\section{Free energy functional}

We consider a liquid crystal occupying a curvilinear cylinder
$\Omega$. The domain is thus the set of points which lie within a
maximum distance $r$ from a smooth curve
$\bv{c}:[0,\ell]\to\reals^3$ (to be determined):
\begin{equation}
\Omega=\big\{P\in\reals^3:P=\bv{c}(s)+\xi\,\bv{e}\,,\text{ for
some } s\in[0,\ell]\,,\,\xi\in[0,r]\,,\,\text{ and
}\bv{e}\cdot\bv{e}=1\big\}\,.
\label{defom}
\end{equation}
Let \bv{N}, \bv{B} be the normal and binormal unit vectors on
\bv{c} (the unit tangent \bv{T} completes an orthogonal basis),
and let $\kappa,\tau$ denote the curvature and torsion along the
same curve. Given a point $P\in\Omega$, the arc-length
$s\in[0,\ell]$ identifies its projection on \bv{c}, while
$\xi\in[0,r]$ yields its distance from \bv{c}. Let $\vartheta\in
[0,2\pi)$ be the angle that the unit vector \bv{e} in
(\ref{defom}) determines with \bv{N}. We show in Appendix A1 that
the coordinate set $(s,\xi,\vartheta)$ is well defined as long as
$\Omega$ is sufficiently thin: $r<\dps\min_{s\in[0,\ell]}
\kappa^{-1} (s)\;$.

According to the experimental conditions in which smectic helices
have been observed \cite{00jali,03jakr}, we consider a freely
suspended capillary, immersed in an isotropic fluid that does not
interact with the surface director. Thus, free-boundary conditions
will be imposed on both the nematic and the smectic variables.
However, an anchoring energy will be necessary in order to take
into account the surface charges induced by spontaneous
polarization. In the absence of nematic anchoring, there is no
energy gain for the system if the smectic and nematic variables
depend on the transverse coordinates $\xi,\vartheta$. Accordingly,
we will assume throughout our calculations that all fields depend
only on the arc-length $s$.

\subsection{Nematic energy}

We introduce the angles $\alpha,\varphi$ to identify the director
orientation as
$$
\bv{n}=\cos\alpha\,\bv{T}+\sin\alpha\cos\varphi\,\bv{N}
+\sin\alpha\sin\varphi\,\bv{B}\,.
$$
Let us also define for future use the unit-vectors
$$
\np:=-\sin\varphi\,\bv{N}+\cos\varphi\,\bv{B}\qquad {\rm and}
\qquad\nt:=-\sin\alpha\,\bv{T}+\cos\alpha\cos\varphi\,\bv{N}
+\cos\alpha\sin\varphi\,\bv{B}\;.
$$
Together with \bv{n}, they complete another orthogonal basis
$\big(\bv{n}\wedge\nt=\np\,$; we arbitrarily define $\varphi=0$
when $\alpha=0$). We have (see again Appendix A1 for technical
details):
$$
\bv{\nabla n}=\left(\frac{\alpha'+\kappa\cos\varphi}
{1-\kappa\xi\cos\vartheta} \,\nt +\frac{
(\varphi'-\tau)\sin\alpha- \kappa\cos\alpha \sin\varphi}
{1-\kappa\xi\cos\vartheta}\,\np\right) \otimes\bv{T}\;,
$$
where a prime denotes differentiation with respect to the
arc-length $s$. The Frank free energy density is given by
\cite{93dgpr}:
\begin{align*}
\sigma_{\rm F}[\alpha,\varphi]&=K_1\, \big(\div\bv{n}\big)^2+
K_2\,\big(\bv{n}\cdot\curl\bv{n}+\qc\big)^2+
K_3\,\big|\bv{n}\wedge\curl\bv{n}+\bv{v}_0\big|^2\\
&+\big(K_2+K_4\big)\,\left(\tr(\bv{\nabla
n})^2-(\div\bv{n})^2\right)\\
&=\frac{(\alpha'+\kappa\cos\varphi)^2}
{(1-\kappa\xi\cos\vartheta)^2}\,\left(K_1\sin^2\alpha+
K_3\cos^2\alpha\right)\\
&+K_2\left(\qc-\frac{\sin\alpha}{1-\kappa\xi\cos\vartheta}\,
\big((\varphi'-\tau)\sin\alpha-
\kappa\cos\alpha\sin\varphi\big)\right)^2\\
&+K_3\left(b_0\sin\alpha-\frac{
(\varphi'-\tau)\sin\alpha-\kappa\cos\alpha\sin\varphi}
{1-\kappa\xi\cos\vartheta}\,\cos\alpha\right)^2\;,
\end{align*}
where $\qc$ is the cholesteric pitch and
$\bv{v}_0=b_0\,\bv{T}\wedge\bv{n}=b_0\sin\alpha\,\np$ is the
intrinsic bending stress.

\subsection{Smectic energy}

Let $\psi(s) =\rho(s)\, {\rm e}^{i\omega(s)}$ be the smectic order
parameter \cite{93dgpr}, so that
$$
\bv{\nabla}\psi=\frac{\left(\rho'+ i\rho\omega'\right){\rm
e}^{i\omega(s)}}{1-\kappa\xi\cos\vartheta} \,\bv{T}\;,
$$
and \bv{T} is also the normal to the smectic layers. The smectic
part of the free energy density is given by:
\begin{align}
\nonumber\sigma_{\rm sm}[\rho,\omega,\alpha]&= \Co\,
\big|\bv{\nabla}\psi-i\qs\psi\bv{n}\big|^2+\Ca\, \big|
\bv{n}\cdot (\bv{\nabla}\psi-i\qs\psi\bv{n})\big|^2+\zeta(\rho)\\
&=\Co \left[ \frac{\rho^{\prime
2}}{(1-\kappa\xi\cos\vartheta)^2}+\rho^2
\left(\frac{\omega'}{1-\kappa\xi\cos\vartheta}-\qs
\cos\alpha\right)^2+\qs^2\rho^2\sin^2\alpha\right]
\label{smener}\\
&\nonumber+\Ca\left[ \frac{\rho^{\prime
2}\cos^2\alpha}{(1-\kappa\xi\cos\vartheta)^2}+\rho^2
\left(\frac{\omega'\cos\alpha}{1-\kappa\xi\cos\vartheta}
-\qs\right)^2\right]+\zeta(\rho)\;,
\end{align}
where $\Ca:=\Cp-\Co$, $\qs$ is the smectic pitch, and $\zeta$ is a
scalar potential depending on the degree of smectic order.

The smectic energy (\ref{smener}) does not rule out the
possibility of the elastic constants $\Cp,\Co$ being different
$(\Ca\neq 0)$. The $\Ca$-term may be neglected when dealing with
SmA materials \cite{00cali,02baca}, but it is necessary to keep it
in the free energy when tilted phases come into play. The SmA
phase may become unstable when $\Co<0$ \cite{76chlu}, and in that
case a higher-order term should be included in the free energy to
ensure the functional to be positive definite. However, we do not
need to insert extra terms in the free energy, since we will prove
that the transition to a tilted phase can be induced by the
cholesteric pitch, even in the presence of a positive $\Co$.
We remark that the free energy density (\ref{smener})
remains positive-defined even if $\Co$ is negative, provided
$\alpha$ is not too large. In fact, $\sigma_{\rm sm}\geq 0$
whenever
$$
\Ca\cos^2\alpha+\Co=\Cp\cos^2\alpha+\Co\sin^2\alpha>0\qquad
i.e.\qquad \Co\geq 0\quad{\rm or}\quad
\tan^2\alpha<-\frac{\Cp}{\Co}\;.
$$

\subsection{Spontaneous polarization}

One important difference between polar smectics and solids is the
freedom of the polarization vector to rotate  in the layer plane
in the former (\bv{P}\/ is a Goldstone variable) as opposed to
taking specific values determined by the solid lattice
\cite{99lage,01jakr,02memm}. Because of this vectorial symmetry
the energy density of the field \bv{P}\/ contains, together with a
term of the form $|\bv{\nabla P}|^2$ which penalizes interfaces in
the material, a term proportional to $(\div\bv{P})^2$:
\begin{equation}
\sigma_{\rm pol}\big[\bv{P}\big]=G_1\,
\big(\div\bv{P}\big)^2+G\,\big|\bv{\nabla
P}\big|^2+\mathcal{G}\big(|\bv{P}|)\;.
\label{polener}
\end{equation}
In (\ref{polener}), $\mathcal{G}$ denotes a scalar potential which
determines the polarization intensity $|\bv{P}|$. When the
permanent molecular polarization is not sufficiently strong to
self-interact, this term avoids the onset of a spontaneous
polarization. This is why we will insert the potential
(\ref{polener}) only in Section 5, when we will be dealing with
spontaneously polarized materials.

A complete description of the polarization energy density can be
found in \cite{87lomo}. We remark that, in materials with strong
permanent polarization, the $G_1$-term can also take the different
form $(\div\bv{P}-c_0)^2$, where $c_0$ can be either positive or
negative. This reflects the preference of the material for a
specific sign of the polarization. However, in the following we
will restrict our attention to the case $c_0=0$. We also neglect
the nonlocal Coulombian interaction of the polarization with the
self-field.

\subsection{Anchoring energy}

The presence of a nonzero polarization induces a surface charge in
the capillary, which in turn requires an opposite charge layer in
the surrounding fluid. This boundary effect can be taken into
account through an effective anchoring energy, which depends on
the polarization \cite{01zado,03syro}:
$$
\sigma_{\rm anch}[\bv{P}]=\omega_P\,P\,\big(1-\bv{p}\cdot
\bv{\nu}\big)\,
$$
where $P$ and \bv{p} respectively denote the intensity and the
direction of the polarization vector \bv{P}, $\bv{\nu}$ is the
outer normal at the external surface, and $\omega_P$ is an
effective anchoring strength. The anchoring potential above may
favor either homeotropic or planar anchoring for the polarization
vector, depending on the sign of $\omega_P$.

\section{Linear shapes}\label{linear}

We first consider a linear smectic capillary, in the absence of
spontaneous polarization. In this section we show that the
presence of a non-zero cholesteric pitch may induce a SmA-SmC
transition in the ground state configuration, even if $\Co>0$.

Let $\kappa=\tau\equiv 0$. The free energy density
$\sigma=\sigma_{\rm F}+\sigma_{\rm sm}$ simplifies into
\begin{align}
\nonumber\sigma[\alpha,\varphi,\rho,\omega]&=
\left(K_1\sin^2\alpha+K_3\cos^2\alpha\right)\alpha^{\prime 2}
+K_2\left(\qc-\varphi'\sin^2\alpha\right)^2\\
\nonumber&+K_3\sin^2\alpha\left(b_0-\varphi'\cos\alpha\right)^2
+\Ca\left[ \rho^{\prime 2}\cos^2\alpha+\rho^2
\left(\omega'\cos\alpha
-\qs\right)^2\right]\\
&+\Co \left[ \rho^{\prime 2}+\rho^2 \left(\omega'-\qs
\cos\alpha\right)^2+\qs^2\rho^2\sin^2\alpha\right]+\zeta(\rho)\;.
\label{lnr}
\end{align}
The Euler-Lagrange equations associated to the the free energy
density (\ref{lnr}), with respect to the variables $\varphi$ and
$\omega$, can be easily integrated once to yield
$$
\frac{\partial\sigma}{\partial\varphi'}=c_1\qquad{\rm and} \qquad
\frac{\partial\sigma}{\partial\omega'}=c_2\;,
$$
with $c_1$ and $c_2$ constants along the capillary. The
free-boundary conditions require $c_1=0$ and $c_2=0$, and thus
$$
\varphi'\equiv\frac{K_2\qc+K_3b_0\cos\alpha}{K_2\sin^2\alpha
+K_3\cos^2\alpha}\qquad
\text{and}\qquad\omega'\equiv\frac{\Cp\qs\cos\alpha}
{\Cp\cos^2\alpha+\Co\sin^2\alpha}\ .
$$
Furthermore, the free energy density (\ref{lnr}) is minimized if
$\alpha'\equiv 0$ and $\rho'\equiv 0$ (which is allowed by the
free-boundary conditions). When these requirements are satisfied,
the free energy density depends only on the constant values of
$\alpha$ and $\rho$:
\begin{equation}
\sigma(\alpha_0,\rho_0)= \frac{K_2K_3\,(\qc
\cos\alpha_0-b_0\sin^2\alpha_0)^2}{K_2\sin^2\alpha_0
+K_3\cos^2\alpha_0}+\frac{\Cp\,\Co\,
\rho_0^2\qs^2\sin^2\alpha_0}{\Cp\cos^2\alpha_0+\Co\sin^2\alpha_0}
+\zeta(\rho_0)\;.
\label{sig}
\end{equation}
The smectic A phase $(\alpha_0=0)$ is always associated to a
stationarity point of (\ref{sig}). However, it becomes unstable
even when $\Co>0$, provided that
\begin{equation}
\Co\qs^2\rho_0^2< \frac{K_2}{K_3}\,\qc\,\big(K_2\qc + 2
b_0K_3\big)\;.
\label{smasmc}
\end{equation}
In fact,
$$\sigma(\alpha_0,\rho_0)=\sigma(0,\rho_0)+\left(\Co\qs^2\rho_0^2-
\frac{K_2\qc\,\big(K_2\qc + 2 b_0K_3\big)}{K_3}\right)\alpha_0^2+
O(\alpha_0^4)\qquad {\rm as}\quad \alpha_0\to 0.
$$
Figure \ref{fig1} shows that, when condition (\ref{smasmc})
applies, the preferred angle moves with continuity from
$\alpha_0=0$. An exceptional situation arises when $b_0=0$ and
$\Co=\Cp$ (bold plot of the right panel). In that case the optimal
value of $\alpha_0$ jumps from $0$ to $\frac{\pi}{2}$ when $\qc$
exceeds $\qs$. In all other cases, the SmA-SmC transition induced
by the cholesteric pitch is second-order.

\begin{figure}
\begin{center}
%
%
\begin{texdraw}
\drawdim truecm \setgray 0
\linewd 0.04
\move (0 0) \lvec (4.8 0)
\lvec (4.8 3.0) \lvec (0 3.0) \lvec (0 0)
\linewd 0.02
\textref h:L v:T \htext (0.00 -.18) {0}
\move (1.200 0) \lvec (1.200 0.120)
\textref h:C v:T \htext (1.20 -.18) {1}
\move (2.400 0) \lvec (2.400 0.120)
\textref h:C v:T \htext (2.40 -.18) {2}
\move (3.600 0) \lvec (3.600 0.120)
\textref h:C v:T \htext (3.60 -.18) {3}
\move (4.800 0) \lvec (4.800 0.120)
\textref h:R v:T \htext (4.80 -.18) {4}
\textref h:R v:C \htext (-.1 0.27) {0}
\move (0 1.500) \lvec (0.15 1.500)
\textref h:R v:C \htext (-.1 1.50) {$\frac{\pi}{4}$}
\move (0 2.727) \lvec (0.15 2.727)
\textref h:R v:C \htext (-.1 2.73) {$\frac{\pi}{2}$}
\textref h:C v:T \htext (2.40 -.6) {$q_{\rm ch}/q_{\rm sm}$}
\textref h:L v:C \htext (.2 2.11) {$\alpha_0^{\rm opt}$}
\linewd 0.04
\move(0.000 0.273) \lvec(0.693 0.273) \lvec(0.696 0.402)
\lvec(0.708 0.554) \lvec(0.720 0.648) \lvec(0.732 0.722)
\lvec(0.744 0.784) \lvec(0.756 0.838) \lvec(0.768 0.887)
\lvec(0.780 0.932) \lvec(0.792 0.973) \lvec(0.804 1.012)
\lvec(0.816 1.048) \lvec(0.828 1.082) \lvec(0.840 1.114)
\lvec(0.852 1.144) \lvec(0.864 1.174) \lvec(0.876 1.202)
\lvec(0.888 1.228) \lvec(0.900 1.254) \lvec(0.912 1.279)
\lvec(0.924 1.303) \lvec(0.936 1.326) \lvec(0.948 1.349)
\lvec(0.960 1.370) \lvec(0.972 1.392) \lvec(0.984 1.412)
\lvec(0.996 1.432) \lvec(1.008 1.452) \lvec(1.020 1.471)
\lvec(1.032 1.489) \lvec(1.044 1.507) \lvec(1.056 1.525)
\lvec(1.068 1.542) \lvec(1.080 1.559) \lvec(1.092 1.576)
\lvec(1.104 1.592) \lvec(1.116 1.608) \lvec(1.128 1.623)
\lvec(1.140 1.639) \lvec(1.152 1.654) \lvec(1.164 1.669)
\lvec(1.176 1.683) \lvec(1.188 1.698) \lvec(1.200 1.712)
\lvec(1.212 1.726) \lvec(1.224 1.740) \lvec(1.236 1.753)
\lvec(1.248 1.767) \lvec(1.260 1.780) \lvec(1.272 1.793)
\lvec(1.284 1.806) \lvec(1.296 1.818) \lvec(1.308 1.831)
\lvec(1.320 1.843) \lvec(1.332 1.856) \lvec(1.344 1.868)
\lvec(1.356 1.880) \lvec(1.368 1.892) \lvec(1.380 1.904)
\lvec(1.392 1.915) \lvec(1.404 1.927) \lvec(1.416 1.938)
\lvec(1.428 1.950) \lvec(1.440 1.961) \lvec(1.452 1.972)
\lvec(1.464 1.984) \lvec(1.476 1.995) \lvec(1.488 2.006)
\lvec(1.500 2.017) \lvec(1.512 2.028) \lvec(1.524 2.039)
\lvec(1.536 2.050) \lvec(1.548 2.060) \lvec(1.560 2.071)
\lvec(1.572 2.082) \lvec(1.584 2.093) \lvec(1.596 2.103)
\lvec(1.608 2.114) \lvec(1.620 2.125) \lvec(1.632 2.135)
\lvec(1.644 2.146) \lvec(1.656 2.157) \lvec(1.668 2.167)
\lvec(1.680 2.178) \lvec(1.692 2.189) \lvec(1.716 2.210)
\lvec(1.728 2.221) \lvec(1.740 2.232) \lvec(1.752 2.243)
\lvec(1.764 2.254) \lvec(1.776 2.265) \lvec(1.788 2.276)
\lvec(1.800 2.287) \lvec(1.812 2.298) \lvec(1.824 2.310)
\lvec(1.836 2.321) \lvec(1.848 2.333) \lvec(1.860 2.345)
\lvec(1.872 2.357) \lvec(1.884 2.369) \lvec(1.896 2.382)
\lvec(1.908 2.395) \lvec(1.920 2.408) \lvec(1.932 2.421)
\lvec(1.944 2.435) \lvec(1.956 2.449) \lvec(1.968 2.464)
\lvec(1.980 2.480) \lvec(1.992 2.496) \lvec(2.004 2.514)
\lvec(2.016 2.532) \lvec(2.028 2.553) \lvec(2.040 2.575)
\lvec(2.052 2.602) \lvec(2.064 2.635) \lvec(2.076 2.689)
\lvec(2.079 2.727) \lvec(4.800 2.727)
\linewd 0.02
\move(0.000 0.273) \lvec(0.455 0.273) \lvec(0.456 0.331)
\lvec(0.468 0.482) \lvec(0.480 0.562) \lvec(0.492 0.624)
\lvec(0.504 0.676) \lvec(0.516 0.721) \lvec(0.528 0.763)
\lvec(0.540 0.800) \lvec(0.552 0.835) \lvec(0.564 0.868)
\lvec(0.576 0.898) \lvec(0.588 0.927) \lvec(0.600 0.955)
\lvec(0.612 0.981) \lvec(0.624 1.006) \lvec(0.636 1.030)
\lvec(0.648 1.053) \lvec(0.660 1.076) \lvec(0.672 1.097)
\lvec(0.684 1.118) \lvec(0.696 1.138) \lvec(0.708 1.158)
\lvec(0.720 1.177) \lvec(0.732 1.196) \lvec(0.744 1.214)
\lvec(0.756 1.231) \lvec(0.768 1.248) \lvec(0.780 1.265)
\lvec(0.792 1.282) \lvec(0.804 1.298) \lvec(0.816 1.313)
\lvec(0.828 1.328) \lvec(0.840 1.343) \lvec(0.852 1.358)
\lvec(0.864 1.373) \lvec(0.876 1.387) \lvec(0.888 1.401)
\lvec(0.900 1.414) \lvec(0.912 1.428) \lvec(0.924 1.441)
\lvec(0.936 1.454) \lvec(0.948 1.466) \lvec(0.960 1.479)
\lvec(0.972 1.491) \lvec(0.984 1.503) \lvec(0.996 1.515)
\lvec(1.008 1.527) \lvec(1.020 1.538) \lvec(1.032 1.550)
\lvec(1.044 1.561) \lvec(1.056 1.572) \lvec(1.068 1.583)
\lvec(1.080 1.594) \lvec(1.092 1.604) \lvec(1.104 1.615)
\lvec(1.116 1.625) \lvec(1.128 1.636) \lvec(1.140 1.646)
\lvec(1.152 1.656) \lvec(1.164 1.665) \lvec(1.176 1.675)
\lvec(1.188 1.685) \lvec(1.200 1.694) \lvec(1.212 1.704)
\lvec(1.224 1.713) \lvec(1.236 1.722) \lvec(1.248 1.731)
\lvec(1.260 1.740) \lvec(1.272 1.749) \lvec(1.284 1.758)
\lvec(1.296 1.767) \lvec(1.308 1.775) \lvec(1.320 1.784)
\lvec(1.332 1.792) \lvec(1.344 1.801) \lvec(1.356 1.809)
\lvec(1.368 1.817) \lvec(1.380 1.825) \lvec(1.392 1.833)
\lvec(1.404 1.841) \lvec(1.416 1.849) \lvec(1.428 1.857)
\lvec(1.440 1.865) \lvec(1.452 1.873) \lvec(1.464 1.880)
\lvec(1.476 1.888) \lvec(1.488 1.895) \lvec(1.500 1.903)
\lvec(1.512 1.910) \lvec(1.524 1.917) \lvec(1.536 1.925)
\lvec(1.548 1.932) \lvec(1.560 1.939) \lvec(1.572 1.946)
\lvec(1.584 1.953) \lvec(1.596 1.960) \lvec(1.608 1.967)
\lvec(1.620 1.974) \lvec(1.632 1.980) \lvec(1.644 1.987)
\lvec(1.656 1.994) \lvec(1.668 2.000) \lvec(1.680 2.007)
\lvec(1.692 2.013) \lvec(1.704 2.020) \lvec(1.716 2.026)
\lvec(1.728 2.033) \lvec(1.740 2.039) \lvec(1.752 2.045)
\lvec(1.764 2.052) \lvec(1.776 2.058) \lvec(1.788 2.064)
\lvec(1.800 2.070) \lvec(1.812 2.076) \lvec(1.824 2.082)
\lvec(1.836 2.088) \lvec(1.848 2.094) \lvec(1.860 2.100)
\lvec(1.872 2.105) \lvec(1.884 2.111) \lvec(1.896 2.117)
\lvec(1.908 2.123) \lvec(1.920 2.128) \lvec(1.932 2.134)
\lvec(1.944 2.139) \lvec(1.956 2.145) \lvec(1.968 2.150)
\lvec(1.980 2.156) \lvec(1.992 2.161) \lvec(2.004 2.167)
\lvec(2.016 2.172) \lvec(2.028 2.177) \lvec(2.040 2.182)
\lvec(2.052 2.187) \lvec(2.064 2.193) \lvec(2.076 2.198)
\lvec(2.088 2.203) \lvec(2.100 2.208) \lvec(2.112 2.213)
\lvec(2.124 2.218) \lvec(2.136 2.222) \lvec(2.148 2.227)
\lvec(2.160 2.232) \lvec(2.172 2.237) \lvec(2.184 2.242)
\lvec(2.196 2.246) \lvec(2.208 2.251) \lvec(2.220 2.255)
\lvec(2.232 2.260) \lvec(2.244 2.264) \lvec(2.256 2.269)
\lvec(2.268 2.273) \lvec(2.280 2.278) \lvec(2.292 2.282)
\lvec(2.304 2.286) \lvec(2.316 2.291) \lvec(2.328 2.295)
\lvec(2.340 2.299) \lvec(2.352 2.303) \lvec(2.364 2.307)
\lvec(2.376 2.311) \lvec(2.388 2.315) \lvec(2.400 2.319)
\lvec(2.412 2.323) \lvec(2.424 2.327) \lvec(2.436 2.331)
\lvec(2.448 2.334) \lvec(2.460 2.338) \lvec(2.472 2.342)
\lvec(2.484 2.345) \lvec(2.496 2.349) \lvec(2.508 2.353)
\lvec(2.520 2.356) \lvec(2.532 2.360) \lvec(2.544 2.363)
\lvec(2.556 2.366) \lvec(2.568 2.370) \lvec(2.580 2.373)
\lvec(2.592 2.376) \lvec(2.604 2.380) \lvec(2.616 2.383)
\lvec(2.628 2.386) \lvec(2.640 2.389) \lvec(2.652 2.392)
\lvec(2.664 2.395) \lvec(2.676 2.398) \lvec(2.688 2.401)
\lvec(2.700 2.404) \lvec(2.712 2.407) \lvec(2.724 2.410)
\lvec(2.736 2.412) \lvec(2.748 2.415) \lvec(2.760 2.418)
\lvec(2.772 2.421) \lvec(2.784 2.423) \lvec(2.796 2.426)
\lvec(2.808 2.429) \lvec(2.820 2.431) \lvec(2.832 2.434)
\lvec(2.844 2.436) \lvec(2.856 2.439) \lvec(2.868 2.441)
\lvec(2.880 2.443) \lvec(2.892 2.446) \lvec(2.904 2.448)
\lvec(2.916 2.450) \lvec(2.928 2.453) \lvec(2.940 2.455)
\lvec(2.952 2.457) \lvec(2.964 2.459) \lvec(2.976 2.461)
\lvec(2.988 2.463) \lvec(3.000 2.466) \lvec(3.012 2.468)
\lvec(3.024 2.470) \lvec(3.036 2.472) \lvec(3.048 2.474)
\lvec(3.060 2.476) \lvec(3.072 2.477) \lvec(3.084 2.479)
\lvec(3.096 2.481) \lvec(3.108 2.483) \lvec(3.120 2.485)
\lvec(3.132 2.487) \lvec(3.144 2.488) \lvec(3.156 2.490)
\lvec(3.168 2.492) \lvec(3.180 2.494) \lvec(3.192 2.495)
\lvec(3.204 2.497) \lvec(3.216 2.499) \lvec(3.228 2.500)
\lvec(3.240 2.502) \lvec(3.252 2.503) \lvec(3.264 2.505)
\lvec(3.276 2.506) \lvec(3.288 2.508) \lvec(3.300 2.509)
\lvec(3.312 2.511) \lvec(3.324 2.512) \lvec(3.336 2.514)
\lvec(3.348 2.515) \lvec(3.360 2.517) \lvec(3.372 2.518)
\lvec(3.384 2.519) \lvec(3.396 2.521) \lvec(3.408 2.522)
\lvec(3.420 2.523) \lvec(3.432 2.525) \lvec(3.444 2.526)
\lvec(3.456 2.527) \lvec(3.468 2.528) \lvec(3.480 2.530)
\lvec(3.492 2.531) \lvec(3.504 2.532) \lvec(3.516 2.533)
\lvec(3.528 2.534) \lvec(3.540 2.535) \lvec(3.552 2.537)
\lvec(3.564 2.538) \lvec(3.576 2.539) \lvec(3.588 2.540)
\lvec(3.600 2.541) \lvec(3.612 2.542) \lvec(3.624 2.543)
\lvec(3.636 2.544) \lvec(3.648 2.545) \lvec(3.660 2.546)
\lvec(3.672 2.547) \lvec(3.684 2.548) \lvec(3.696 2.549)
\lvec(3.708 2.550) \lvec(3.720 2.551) \lvec(3.732 2.552)
\lvec(3.744 2.553) \lvec(3.756 2.554) \lvec(3.768 2.555)
\lvec(3.780 2.556) \lvec(3.792 2.557) \lvec(3.804 2.558)
\lvec(3.816 2.559) \lvec(3.828 2.560) \lvec(3.840 2.560)
\lvec(3.852 2.561) \lvec(3.864 2.562) \lvec(3.876 2.563)
\lvec(3.888 2.564) \lvec(3.900 2.565) \lvec(3.912 2.565)
\lvec(3.924 2.566) \lvec(3.936 2.567) \lvec(3.948 2.568)
\lvec(3.960 2.569) \lvec(3.972 2.569) \lvec(3.984 2.570)
\lvec(3.996 2.571) \lvec(4.008 2.572) \lvec(4.020 2.572)
\lvec(4.032 2.573) \lvec(4.044 2.574) \lvec(4.056 2.575)
\lvec(4.068 2.575) \lvec(4.080 2.576) \lvec(4.092 2.577)
\lvec(4.104 2.577) \lvec(4.116 2.578) \lvec(4.128 2.579)
\lvec(4.140 2.579) \lvec(4.152 2.580) \lvec(4.164 2.581)
\lvec(4.176 2.581) \lvec(4.188 2.582) \lvec(4.200 2.583)
\lvec(4.212 2.583) \lvec(4.224 2.584) \lvec(4.236 2.585)
\lvec(4.248 2.585) \lvec(4.260 2.586) \lvec(4.272 2.586)
\lvec(4.284 2.587) \lvec(4.296 2.588) \lvec(4.308 2.588)
\lvec(4.320 2.589) \lvec(4.332 2.589) \lvec(4.344 2.590)
\lvec(4.356 2.590) \lvec(4.368 2.591) \lvec(4.380 2.592)
\lvec(4.392 2.592) \lvec(4.404 2.593) \lvec(4.416 2.593)
\lvec(4.428 2.594) \lvec(4.440 2.594) \lvec(4.452 2.595)
\lvec(4.464 2.595) \lvec(4.476 2.596) \lvec(4.488 2.596)
\lvec(4.500 2.597) \lvec(4.512 2.597) \lvec(4.524 2.598)
\lvec(4.536 2.598) \lvec(4.548 2.599) \lvec(4.560 2.599)
\lvec(4.572 2.600) \lvec(4.584 2.600) \lvec(4.596 2.601)
\lvec(4.608 2.601) \lvec(4.620 2.602) \lvec(4.632 2.602)
\lvec(4.644 2.603) \lvec(4.656 2.603) \lvec(4.668 2.604)
\lvec(4.680 2.604) \lvec(4.692 2.605) \lvec(4.704 2.605)
\lvec(4.716 2.606) \lvec(4.728 2.606) \lvec(4.740 2.606)
\lvec(4.752 2.607) \lvec(4.764 2.607) \lvec(4.776 2.608)
\lvec(4.788 2.608) \lvec(4.800 2.609)
\move(0.000 0.273) \lvec(0.317 0.273) \lvec(0.324 0.411)
\lvec(0.336 0.496) \lvec(0.348 0.556) \lvec(0.360 0.605)
\lvec(0.372 0.648) \lvec(0.384 0.686) \lvec(0.396 0.720)
\lvec(0.408 0.752) \lvec(0.420 0.782) \lvec(0.432 0.810)
\lvec(0.444 0.836) \lvec(0.456 0.861) \lvec(0.468 0.885)
\lvec(0.480 0.908) \lvec(0.492 0.930) \lvec(0.504 0.951)
\lvec(0.516 0.971) \lvec(0.528 0.991) \lvec(0.540 1.009)
\lvec(0.552 1.028) \lvec(0.564 1.046) \lvec(0.576 1.063)
\lvec(0.588 1.080) \lvec(0.600 1.096) \lvec(0.612 1.112)
\lvec(0.624 1.128) \lvec(0.636 1.143) \lvec(0.648 1.158)
\lvec(0.660 1.173) \lvec(0.672 1.187) \lvec(0.684 1.201)
\lvec(0.696 1.215) \lvec(0.708 1.228) \lvec(0.720 1.242)
\lvec(0.732 1.254) \lvec(0.744 1.267) \lvec(0.756 1.280)
\lvec(0.768 1.292) \lvec(0.780 1.304) \lvec(0.792 1.316)
\lvec(0.804 1.327) \lvec(0.816 1.339) \lvec(0.828 1.350)
\lvec(0.840 1.361) \lvec(0.852 1.372) \lvec(0.864 1.383)
\lvec(0.876 1.394) \lvec(0.888 1.404) \lvec(0.900 1.414)
\lvec(0.912 1.425) \lvec(0.924 1.435) \lvec(0.936 1.444)
\lvec(0.948 1.454) \lvec(0.960 1.464) \lvec(0.972 1.473)
\lvec(0.984 1.483) \lvec(0.996 1.492) \lvec(1.008 1.501)
\lvec(1.020 1.510) \lvec(1.032 1.519) \lvec(1.044 1.528)
\lvec(1.056 1.537) \lvec(1.068 1.545) \lvec(1.080 1.554)
\lvec(1.092 1.562) \lvec(1.104 1.571) \lvec(1.116 1.579)
\lvec(1.128 1.587) \lvec(1.140 1.595) \lvec(1.152 1.603)
\lvec(1.164 1.611) \lvec(1.176 1.619) \lvec(1.188 1.626)
\lvec(1.200 1.634) \lvec(1.212 1.642) \lvec(1.224 1.649)
\lvec(1.236 1.657) \lvec(1.248 1.664) \lvec(1.260 1.671)
\lvec(1.272 1.678) \lvec(1.284 1.686) \lvec(1.296 1.693)
\lvec(1.308 1.700) \lvec(1.320 1.707) \lvec(1.332 1.713)
\lvec(1.344 1.720) \lvec(1.356 1.727) \lvec(1.368 1.734)
\lvec(1.380 1.740) \lvec(1.392 1.747) \lvec(1.404 1.753)
\lvec(1.416 1.760) \lvec(1.428 1.766) \lvec(1.440 1.773)
\lvec(1.452 1.779) \lvec(1.464 1.785) \lvec(1.476 1.791)
\lvec(1.488 1.797) \lvec(1.500 1.804) \lvec(1.512 1.810)
\lvec(1.524 1.816) \lvec(1.536 1.821) \lvec(1.548 1.827)
\lvec(1.560 1.833) \lvec(1.572 1.839) \lvec(1.584 1.845)
\lvec(1.596 1.850) \lvec(1.608 1.856) \lvec(1.620 1.862)
\lvec(1.632 1.867) \lvec(1.644 1.873) \lvec(1.656 1.878)
\lvec(1.668 1.884) \lvec(1.680 1.889) \lvec(1.692 1.894)
\lvec(1.704 1.900) \lvec(1.716 1.905) \lvec(1.728 1.910)
\lvec(1.740 1.915) \lvec(1.752 1.920) \lvec(1.764 1.926)
\lvec(1.776 1.931) \lvec(1.788 1.936) \lvec(1.800 1.941)
\lvec(1.812 1.946) \lvec(1.824 1.951) \lvec(1.836 1.955)
\lvec(1.848 1.960) \lvec(1.860 1.965) \lvec(1.872 1.970)
\lvec(1.884 1.975) \lvec(1.896 1.979) \lvec(1.908 1.984)
\lvec(1.920 1.989) \lvec(1.932 1.993) \lvec(1.944 1.998)
\lvec(1.956 2.002) \lvec(1.968 2.007) \lvec(1.980 2.011)
\lvec(1.992 2.016) \lvec(2.004 2.020) \lvec(2.016 2.025)
\lvec(2.028 2.029) \lvec(2.040 2.033) \lvec(2.052 2.037)
\lvec(2.064 2.042) \lvec(2.076 2.046) \lvec(2.088 2.050)
\lvec(2.100 2.054) \lvec(2.112 2.058) \lvec(2.124 2.063)
\lvec(2.136 2.067) \lvec(2.148 2.071) \lvec(2.160 2.075)
\lvec(2.172 2.079) \lvec(2.184 2.083) \lvec(2.196 2.087)
\lvec(2.208 2.090) \lvec(2.220 2.094) \lvec(2.232 2.098)
\lvec(2.244 2.102) \lvec(2.256 2.106) \lvec(2.268 2.110)
\lvec(2.280 2.113) \lvec(2.292 2.117) \lvec(2.304 2.121)
\lvec(2.316 2.124) \lvec(2.328 2.128) \lvec(2.340 2.131)
\lvec(2.352 2.135) \lvec(2.364 2.139) \lvec(2.376 2.142)
\lvec(2.388 2.146) \lvec(2.400 2.149) \lvec(2.412 2.153)
\lvec(2.424 2.156) \lvec(2.436 2.159) \lvec(2.448 2.163)
\lvec(2.460 2.166) \lvec(2.472 2.169) \lvec(2.484 2.173)
\lvec(2.496 2.176) \lvec(2.508 2.179) \lvec(2.520 2.182)
\lvec(2.532 2.186) \lvec(2.544 2.189) \lvec(2.556 2.192)
\lvec(2.568 2.195) \lvec(2.580 2.198) \lvec(2.592 2.201)
\lvec(2.604 2.204) \lvec(2.616 2.207) \lvec(2.628 2.210)
\lvec(2.640 2.213) \lvec(2.652 2.216) \lvec(2.664 2.219)
\lvec(2.676 2.222) \lvec(2.688 2.225) \lvec(2.700 2.228)
\lvec(2.712 2.231) \lvec(2.724 2.233) \lvec(2.736 2.236)
\lvec(2.748 2.239) \lvec(2.760 2.242) \lvec(2.772 2.245)
\lvec(2.784 2.247) \lvec(2.796 2.250) \lvec(2.808 2.253)
\lvec(2.820 2.255) \lvec(2.832 2.258) \lvec(2.844 2.261)
\lvec(2.856 2.263) \lvec(2.868 2.266) \lvec(2.880 2.268)
\lvec(2.892 2.271) \lvec(2.904 2.273) \lvec(2.916 2.276)
\lvec(2.928 2.278) \lvec(2.940 2.281) \lvec(2.952 2.283)
\lvec(2.964 2.285) \lvec(2.976 2.288) \lvec(2.988 2.290)
\lvec(3.000 2.293) \lvec(3.012 2.295) \lvec(3.024 2.297)
\lvec(3.036 2.300) \lvec(3.048 2.302) \lvec(3.060 2.304)
\lvec(3.072 2.306) \lvec(3.084 2.309) \lvec(3.096 2.311)
\lvec(3.108 2.313) \lvec(3.120 2.315) \lvec(3.132 2.317)
\lvec(3.144 2.319) \lvec(3.156 2.321) \lvec(3.168 2.324)
\lvec(3.180 2.326) \lvec(3.192 2.328) \lvec(3.204 2.330)
\lvec(3.216 2.332) \lvec(3.228 2.334) \lvec(3.240 2.336)
\lvec(3.252 2.338) \lvec(3.264 2.340) \lvec(3.276 2.342)
\lvec(3.288 2.344) \lvec(3.300 2.346) \lvec(3.312 2.347)
\lvec(3.324 2.349) \lvec(3.336 2.351) \lvec(3.348 2.353)
\lvec(3.360 2.355) \lvec(3.372 2.357) \lvec(3.384 2.359)
\lvec(3.396 2.360) \lvec(3.408 2.362) \lvec(3.420 2.364)
\lvec(3.432 2.366) \lvec(3.444 2.367) \lvec(3.456 2.369)
\lvec(3.468 2.371) \lvec(3.480 2.372) \lvec(3.492 2.374)
\lvec(3.504 2.376) \lvec(3.516 2.377) \lvec(3.528 2.379)
\lvec(3.540 2.381) \lvec(3.552 2.382) \lvec(3.564 2.384)
\lvec(3.576 2.386) \lvec(3.588 2.387) \lvec(3.600 2.389)
\lvec(3.612 2.390) \lvec(3.624 2.392) \lvec(3.636 2.393)
\lvec(3.648 2.395) \lvec(3.660 2.396) \lvec(3.672 2.398)
\lvec(3.684 2.399) \lvec(3.696 2.401) \lvec(3.708 2.402)
\lvec(3.720 2.404) \lvec(3.732 2.405) \lvec(3.744 2.406)
\lvec(3.756 2.408) \lvec(3.768 2.409) \lvec(3.780 2.411)
\lvec(3.792 2.412) \lvec(3.804 2.413) \lvec(3.816 2.415)
\lvec(3.828 2.416) \lvec(3.840 2.417) \lvec(3.852 2.419)
\lvec(3.864 2.420) \lvec(3.876 2.421) \lvec(3.888 2.423)
\lvec(3.900 2.424) \lvec(3.912 2.425) \lvec(3.924 2.426)
\lvec(3.936 2.428) \lvec(3.948 2.429) \lvec(3.960 2.430)
\lvec(3.972 2.431) \lvec(3.984 2.433) \lvec(3.996 2.434)
\lvec(4.008 2.435) \lvec(4.020 2.436) \lvec(4.032 2.437)
\lvec(4.044 2.439) \lvec(4.056 2.440) \lvec(4.068 2.441)
\lvec(4.080 2.442) \lvec(4.092 2.443) \lvec(4.104 2.444)
\lvec(4.116 2.445) \lvec(4.128 2.446) \lvec(4.140 2.448)
\lvec(4.152 2.449) \lvec(4.164 2.450) \lvec(4.176 2.451)
\lvec(4.188 2.452) \lvec(4.200 2.453) \lvec(4.212 2.454)
\lvec(4.224 2.455) \lvec(4.236 2.456) \lvec(4.248 2.457)
\lvec(4.260 2.458) \lvec(4.272 2.459) \lvec(4.284 2.460)
\lvec(4.296 2.461) \lvec(4.308 2.462) \lvec(4.320 2.463)
\lvec(4.332 2.464) \lvec(4.344 2.465) \lvec(4.356 2.466)
\lvec(4.368 2.467) \lvec(4.380 2.468) \lvec(4.392 2.469)
\lvec(4.404 2.470) \lvec(4.416 2.471) \lvec(4.428 2.472)
\lvec(4.440 2.472) \lvec(4.452 2.473) \lvec(4.464 2.474)
\lvec(4.476 2.475) \lvec(4.488 2.476) \lvec(4.500 2.477)
\lvec(4.512 2.478) \lvec(4.524 2.479) \lvec(4.536 2.480)
\lvec(4.548 2.480) \lvec(4.560 2.481) \lvec(4.572 2.482)
\lvec(4.584 2.483) \lvec(4.596 2.484) \lvec(4.608 2.485)
\lvec(4.620 2.485) \lvec(4.632 2.486) \lvec(4.644 2.487)
\lvec(4.656 2.488) \lvec(4.668 2.489) \lvec(4.680 2.490)
\lvec(4.692 2.490) \lvec(4.704 2.491) \lvec(4.716 2.492)
\lvec(4.728 2.493) \lvec(4.740 2.493) \lvec(4.752 2.494)
\lvec(4.764 2.495) \lvec(4.776 2.496) \lvec(4.788 2.496)
\lvec(4.800 2.497)
\move(0.012 0.273) \lvec(0.235 0.273) \lvec(0.240 0.366)
\lvec(0.252 0.455) \lvec(0.264 0.514) \lvec(0.276 0.560)
\lvec(0.288 0.599) \lvec(0.300 0.635) \lvec(0.312 0.667)
\lvec(0.324 0.696) \lvec(0.336 0.723) \lvec(0.348 0.749)
\lvec(0.360 0.773) \lvec(0.372 0.796) \lvec(0.384 0.818)
\lvec(0.396 0.839) \lvec(0.408 0.859) \lvec(0.420 0.879)
\lvec(0.432 0.897) \lvec(0.444 0.915) \lvec(0.456 0.933)
\lvec(0.468 0.950) \lvec(0.480 0.966) \lvec(0.492 0.982)
\lvec(0.504 0.998) \lvec(0.516 1.013) \lvec(0.528 1.028)
\lvec(0.540 1.042) \lvec(0.552 1.056) \lvec(0.564 1.070)
\lvec(0.576 1.084) \lvec(0.588 1.097) \lvec(0.600 1.110)
\lvec(0.612 1.123) \lvec(0.624 1.135) \lvec(0.636 1.147)
\lvec(0.648 1.159) \lvec(0.660 1.171) \lvec(0.672 1.183)
\lvec(0.684 1.194) \lvec(0.696 1.205) \lvec(0.708 1.216)
\lvec(0.720 1.227) \lvec(0.732 1.238) \lvec(0.744 1.248)
\lvec(0.756 1.259) \lvec(0.768 1.269) \lvec(0.780 1.279)
\lvec(0.792 1.289) \lvec(0.804 1.298) \lvec(0.816 1.308)
\lvec(0.828 1.318) \lvec(0.840 1.327) \lvec(0.852 1.336)
\lvec(0.864 1.345) \lvec(0.876 1.354) \lvec(0.888 1.363)
\lvec(0.900 1.372) \lvec(0.912 1.381) \lvec(0.924 1.389)
\lvec(0.936 1.398) \lvec(0.948 1.406) \lvec(0.960 1.414)
\lvec(0.972 1.422) \lvec(0.984 1.431) \lvec(0.996 1.439)
\lvec(1.008 1.446) \lvec(1.020 1.454) \lvec(1.032 1.462)
\lvec(1.044 1.470) \lvec(1.056 1.477) \lvec(1.068 1.485)
\lvec(1.080 1.492) \lvec(1.092 1.499) \lvec(1.104 1.507)
\lvec(1.116 1.514) \lvec(1.128 1.521) \lvec(1.140 1.528)
\lvec(1.152 1.535) \lvec(1.164 1.542) \lvec(1.176 1.549)
\lvec(1.188 1.556) \lvec(1.200 1.562) \lvec(1.212 1.569)
\lvec(1.224 1.576) \lvec(1.236 1.582) \lvec(1.248 1.589)
\lvec(1.260 1.595) \lvec(1.272 1.601) \lvec(1.284 1.608)
\lvec(1.296 1.614) \lvec(1.308 1.620) \lvec(1.320 1.626)
\lvec(1.332 1.632) \lvec(1.344 1.638) \lvec(1.356 1.644)
\lvec(1.368 1.650) \lvec(1.380 1.656) \lvec(1.392 1.662)
\lvec(1.404 1.668) \lvec(1.416 1.673) \lvec(1.428 1.679)
\lvec(1.440 1.685) \lvec(1.452 1.690) \lvec(1.464 1.696)
\lvec(1.476 1.701) \lvec(1.488 1.707) \lvec(1.500 1.712)
\lvec(1.512 1.718) \lvec(1.524 1.723) \lvec(1.536 1.728)
\lvec(1.548 1.733) \lvec(1.560 1.739) \lvec(1.572 1.744)
\lvec(1.584 1.749) \lvec(1.596 1.754) \lvec(1.608 1.759)
\lvec(1.620 1.764) \lvec(1.632 1.769) \lvec(1.644 1.774)
\lvec(1.656 1.779) \lvec(1.668 1.784) \lvec(1.680 1.789)
\lvec(1.692 1.793) \lvec(1.704 1.798) \lvec(1.716 1.803)
\lvec(1.728 1.808) \lvec(1.740 1.812) \lvec(1.752 1.817)
\lvec(1.764 1.822) \lvec(1.776 1.826) \lvec(1.788 1.831)
\lvec(1.800 1.835) \lvec(1.812 1.840) \lvec(1.824 1.844)
\lvec(1.836 1.848) \lvec(1.848 1.853) \lvec(1.860 1.857)
\lvec(1.872 1.862) \lvec(1.884 1.866) \lvec(1.896 1.870)
\lvec(1.908 1.874) \lvec(1.920 1.879) \lvec(1.932 1.883)
\lvec(1.944 1.887) \lvec(1.956 1.891) \lvec(1.968 1.895)
\lvec(1.980 1.899) \lvec(1.992 1.903) \lvec(2.004 1.907)
\lvec(2.016 1.911) \lvec(2.028 1.915) \lvec(2.040 1.919)
\lvec(2.052 1.923) \lvec(2.064 1.927) \lvec(2.076 1.931)
\lvec(2.088 1.934) \lvec(2.100 1.938) \lvec(2.112 1.942)
\lvec(2.124 1.946) \lvec(2.136 1.950) \lvec(2.148 1.953)
\lvec(2.160 1.957) \lvec(2.172 1.961) \lvec(2.184 1.964)
\lvec(2.196 1.968) \lvec(2.208 1.971) \lvec(2.220 1.975)
\lvec(2.232 1.979) \lvec(2.244 1.982) \lvec(2.256 1.986)
\lvec(2.268 1.989) \lvec(2.280 1.992) \lvec(2.292 1.996)
\lvec(2.304 1.999) \lvec(2.316 2.003) \lvec(2.328 2.006)
\lvec(2.340 2.009) \lvec(2.352 2.013) \lvec(2.364 2.016)
\lvec(2.376 2.019) \lvec(2.388 2.023) \lvec(2.400 2.026)
\lvec(2.412 2.029) \lvec(2.424 2.032) \lvec(2.436 2.035)
\lvec(2.448 2.039) \lvec(2.460 2.042) \lvec(2.472 2.045)
\lvec(2.484 2.048) \lvec(2.496 2.051) \lvec(2.508 2.054)
\lvec(2.520 2.057) \lvec(2.532 2.060) \lvec(2.544 2.063)
\lvec(2.556 2.066) \lvec(2.568 2.069) \lvec(2.580 2.072)
\lvec(2.592 2.075) \lvec(2.604 2.078) \lvec(2.616 2.081)
\lvec(2.628 2.084) \lvec(2.640 2.087) \lvec(2.652 2.089)
\lvec(2.664 2.092) \lvec(2.676 2.095) \lvec(2.688 2.098)
\lvec(2.700 2.101) \lvec(2.712 2.103) \lvec(2.724 2.106)
\lvec(2.736 2.109) \lvec(2.748 2.112) \lvec(2.760 2.114)
\lvec(2.772 2.117) \lvec(2.784 2.120) \lvec(2.796 2.122)
\lvec(2.808 2.125) \lvec(2.820 2.127) \lvec(2.832 2.130)
\lvec(2.844 2.133) \lvec(2.856 2.135) \lvec(2.868 2.138)
\lvec(2.880 2.140) \lvec(2.892 2.143) \lvec(2.904 2.145)
\lvec(2.916 2.148) \lvec(2.928 2.150) \lvec(2.940 2.153)
\lvec(2.952 2.155) \lvec(2.964 2.157) \lvec(2.976 2.160)
\lvec(2.988 2.162) \lvec(3.000 2.165) \lvec(3.012 2.167)
\lvec(3.024 2.169) \lvec(3.036 2.172) \lvec(3.048 2.174)
\lvec(3.060 2.176) \lvec(3.072 2.179) \lvec(3.084 2.181)
\lvec(3.096 2.183) \lvec(3.108 2.185) \lvec(3.120 2.188)
\lvec(3.132 2.190) \lvec(3.144 2.192) \lvec(3.156 2.194)
\lvec(3.168 2.196) \lvec(3.180 2.198) \lvec(3.192 2.201)
\lvec(3.204 2.203) \lvec(3.216 2.205) \lvec(3.228 2.207)
\lvec(3.240 2.209) \lvec(3.252 2.211) \lvec(3.264 2.213)
\lvec(3.276 2.215) \lvec(3.288 2.217) \lvec(3.300 2.219)
\lvec(3.312 2.221) \lvec(3.324 2.223) \lvec(3.336 2.225)
\lvec(3.348 2.227) \lvec(3.360 2.229) \lvec(3.372 2.231)
\lvec(3.384 2.233) \lvec(3.396 2.235) \lvec(3.408 2.237)
\lvec(3.420 2.239) \lvec(3.432 2.241) \lvec(3.444 2.243)
\lvec(3.456 2.245) \lvec(3.468 2.247) \lvec(3.480 2.248)
\lvec(3.492 2.250) \lvec(3.504 2.252) \lvec(3.516 2.254)
\lvec(3.528 2.256) \lvec(3.540 2.257) \lvec(3.552 2.259)
\lvec(3.564 2.261) \lvec(3.576 2.263) \lvec(3.588 2.265)
\lvec(3.600 2.266) \lvec(3.612 2.268) \lvec(3.624 2.270)
\lvec(3.636 2.271) \lvec(3.648 2.273) \lvec(3.660 2.275)
\lvec(3.672 2.276) \lvec(3.684 2.278) \lvec(3.696 2.280)
\lvec(3.708 2.281) \lvec(3.720 2.283) \lvec(3.732 2.285)
\lvec(3.744 2.286) \lvec(3.756 2.288) \lvec(3.768 2.290)
\lvec(3.780 2.291) \lvec(3.792 2.293) \lvec(3.804 2.294)
\lvec(3.816 2.296) \lvec(3.828 2.297) \lvec(3.840 2.299)
\lvec(3.852 2.300) \lvec(3.864 2.302) \lvec(3.876 2.303)
\lvec(3.888 2.305) \lvec(3.900 2.306) \lvec(3.912 2.308)
\lvec(3.924 2.309) \lvec(3.936 2.311) \lvec(3.948 2.312)
\lvec(3.960 2.314) \lvec(3.972 2.315) \lvec(3.984 2.317)
\lvec(3.996 2.318) \lvec(4.008 2.319) \lvec(4.020 2.321)
\lvec(4.032 2.322) \lvec(4.044 2.324) \lvec(4.056 2.325)
\lvec(4.068 2.326) \lvec(4.080 2.328) \lvec(4.092 2.329)
\lvec(4.104 2.330) \lvec(4.116 2.332) \lvec(4.128 2.333)
\lvec(4.140 2.334) \lvec(4.152 2.336) \lvec(4.164 2.337)
\lvec(4.176 2.338) \lvec(4.188 2.340) \lvec(4.200 2.341)
\lvec(4.212 2.342) \lvec(4.224 2.344) \lvec(4.236 2.345)
\lvec(4.248 2.346) \lvec(4.260 2.347) \lvec(4.272 2.349)
\lvec(4.284 2.350) \lvec(4.296 2.351) \lvec(4.308 2.352)
\lvec(4.320 2.353) \lvec(4.332 2.355) \lvec(4.344 2.356)
\lvec(4.356 2.357) \lvec(4.368 2.358) \lvec(4.380 2.359)
\lvec(4.392 2.361) \lvec(4.404 2.362) \lvec(4.416 2.363)
\lvec(4.428 2.364) \lvec(4.440 2.365) \lvec(4.452 2.366)
\lvec(4.464 2.367) \lvec(4.476 2.369) \lvec(4.488 2.370)
\lvec(4.500 2.371) \lvec(4.512 2.372) \lvec(4.524 2.373)
\lvec(4.536 2.374) \lvec(4.548 2.375) \lvec(4.560 2.376)
\lvec(4.572 2.377) \lvec(4.584 2.378) \lvec(4.596 2.379)
\lvec(4.800 2.392)
\lpatt (.03 .1)
\move(0.000 0.273) \lvec(0.185 0.273) \lvec(0.192 0.377)
\lvec(0.204 0.449) \lvec(0.216 0.500) \lvec(0.228 0.541)
\lvec(0.240 0.576) \lvec(0.252 0.608) \lvec(0.264 0.636)
\lvec(0.276 0.663) \lvec(0.288 0.687) \lvec(0.300 0.711)
\lvec(0.312 0.733) \lvec(0.324 0.753) \lvec(0.336 0.773)
\lvec(0.348 0.793) \lvec(0.360 0.811) \lvec(0.372 0.829)
\lvec(0.384 0.846) \lvec(0.396 0.862) \lvec(0.408 0.878)
\lvec(0.420 0.894) \lvec(0.432 0.909) \lvec(0.444 0.924)
\lvec(0.456 0.938) \lvec(0.468 0.952) \lvec(0.480 0.966)
\lvec(0.492 0.979) \lvec(0.504 0.992) \lvec(0.516 1.005)
\lvec(0.528 1.017) \lvec(0.540 1.029) \lvec(0.552 1.041)
\lvec(0.564 1.053) \lvec(0.576 1.065) \lvec(0.588 1.076)
\lvec(0.600 1.087) \lvec(0.612 1.098) \lvec(0.624 1.109)
\lvec(0.636 1.119) \lvec(0.648 1.130) \lvec(0.660 1.140)
\lvec(0.672 1.150) \lvec(0.684 1.160) \lvec(0.696 1.170)
\lvec(0.708 1.180) \lvec(0.720 1.189) \lvec(0.732 1.199)
\lvec(0.744 1.208) \lvec(0.756 1.217) \lvec(0.768 1.226)
\lvec(0.780 1.235) \lvec(0.792 1.244) \lvec(0.804 1.252)
\lvec(0.816 1.261) \lvec(0.828 1.269) \lvec(0.840 1.278)
\lvec(0.852 1.286) \lvec(0.864 1.294) \lvec(0.876 1.302)
\lvec(0.888 1.310) \lvec(0.900 1.318) \lvec(0.912 1.326)
\lvec(0.924 1.333) \lvec(0.936 1.341) \lvec(0.948 1.349)
\lvec(0.960 1.356) \lvec(0.972 1.363) \lvec(0.984 1.371)
\lvec(0.996 1.378) \lvec(1.008 1.385) \lvec(1.020 1.392)
\lvec(1.032 1.399) \lvec(1.044 1.406) \lvec(1.056 1.413)
\lvec(1.068 1.420) \lvec(1.080 1.426) \lvec(1.092 1.433)
\lvec(1.104 1.440) \lvec(1.116 1.446) \lvec(1.128 1.453)
\lvec(1.140 1.459) \lvec(1.152 1.466) \lvec(1.164 1.472)
\lvec(1.176 1.478) \lvec(1.188 1.484) \lvec(1.200 1.490)
\lvec(1.212 1.497) \lvec(1.224 1.503) \lvec(1.236 1.509)
\lvec(1.248 1.514) \lvec(1.260 1.520) \lvec(1.272 1.526)
\lvec(1.284 1.532) \lvec(1.296 1.538) \lvec(1.308 1.543)
\lvec(1.320 1.549) \lvec(1.332 1.555) \lvec(1.344 1.560)
\lvec(1.356 1.566) \lvec(1.368 1.571) \lvec(1.380 1.577)
\lvec(1.392 1.582) \lvec(1.404 1.587) \lvec(1.416 1.593)
\lvec(1.428 1.598) \lvec(1.440 1.603) \lvec(1.452 1.608)
\lvec(1.464 1.613) \lvec(1.476 1.619) \lvec(1.488 1.624)
\lvec(1.500 1.629) \lvec(1.512 1.634) \lvec(1.524 1.639)
\lvec(1.536 1.644) \lvec(1.548 1.648) \lvec(1.560 1.653)
\lvec(1.572 1.658) \lvec(1.584 1.663) \lvec(1.596 1.668)
\lvec(1.608 1.672) \lvec(1.620 1.677) \lvec(1.632 1.682)
\lvec(1.644 1.686) \lvec(1.656 1.691) \lvec(1.668 1.695)
\lvec(1.680 1.700) \lvec(1.692 1.704) \lvec(1.704 1.709)
\lvec(1.716 1.713) \lvec(1.728 1.718) \lvec(1.740 1.722)
\lvec(1.752 1.726) \lvec(1.764 1.731) \lvec(1.776 1.735)
\lvec(1.788 1.739) \lvec(1.800 1.743) \lvec(1.812 1.748)
\lvec(1.824 1.752) \lvec(1.836 1.756) \lvec(1.848 1.760)
\lvec(1.860 1.764) \lvec(1.872 1.768) \lvec(1.884 1.772)
\lvec(1.896 1.776) \lvec(1.908 1.780) \lvec(1.920 1.784)
\lvec(1.932 1.788) \lvec(1.944 1.792) \lvec(1.956 1.796)
\lvec(1.968 1.800) \lvec(1.980 1.804) \lvec(1.992 1.807)
\lvec(2.004 1.811) \lvec(2.016 1.815) \lvec(2.028 1.819)
\lvec(2.040 1.822) \lvec(2.052 1.826) \lvec(2.064 1.830)
\lvec(2.076 1.833) \lvec(2.088 1.837) \lvec(2.100 1.841)
\lvec(2.112 1.844) \lvec(2.124 1.848) \lvec(2.136 1.851)
\lvec(2.148 1.855) \lvec(2.160 1.858) \lvec(2.172 1.862)
\lvec(2.184 1.865) \lvec(2.196 1.869) \lvec(2.208 1.872)
\lvec(2.220 1.876) \lvec(2.232 1.879) \lvec(2.244 1.882)
\lvec(2.256 1.886) \lvec(2.268 1.889) \lvec(2.280 1.892)
\lvec(2.292 1.896) \lvec(2.304 1.899) \lvec(2.316 1.902)
\lvec(2.328 1.905) \lvec(2.340 1.909) \lvec(2.352 1.912)
\lvec(2.364 1.915) \lvec(2.376 1.918) \lvec(2.388 1.921)
\lvec(2.400 1.924) \lvec(2.412 1.927) \lvec(2.424 1.930)
\lvec(2.436 1.934) \lvec(2.448 1.937) \lvec(2.460 1.940)
\lvec(2.472 1.943) \lvec(2.484 1.946) \lvec(2.496 1.949)
\lvec(2.508 1.952) \lvec(2.520 1.955) \lvec(2.532 1.957)
\lvec(2.544 1.960) \lvec(2.556 1.963) \lvec(2.568 1.966)
\lvec(2.580 1.969) \lvec(2.592 1.972) \lvec(2.604 1.975)
\lvec(2.616 1.978) \lvec(2.628 1.980) \lvec(2.640 1.983)
\lvec(2.652 1.986) \lvec(2.664 1.989) \lvec(2.676 1.991)
\lvec(2.688 1.994) \lvec(2.700 1.997) \lvec(2.712 1.999)
\lvec(2.724 2.002) \lvec(2.736 2.005) \lvec(2.748 2.007)
\lvec(2.760 2.010) \lvec(2.772 2.013) \lvec(2.784 2.015)
\lvec(2.796 2.018) \lvec(2.808 2.021) \lvec(2.820 2.023)
\lvec(2.832 2.026) \lvec(2.844 2.028) \lvec(2.856 2.031)
\lvec(2.868 2.033) \lvec(2.880 2.036) \lvec(2.892 2.038)
\lvec(2.904 2.041) \lvec(2.916 2.043) \lvec(2.928 2.046)
\lvec(2.940 2.048) \lvec(2.952 2.050) \lvec(2.964 2.053)
\lvec(2.976 2.055) \lvec(2.988 2.058) \lvec(3.000 2.060)
\lvec(3.012 2.062) \lvec(3.024 2.065) \lvec(3.036 2.067)
\lvec(3.048 2.069) \lvec(3.060 2.072) \lvec(3.072 2.074)
\lvec(3.084 2.076) \lvec(3.096 2.078) \lvec(3.108 2.081)
\lvec(3.120 2.083) \lvec(3.132 2.085) \lvec(3.144 2.087)
\lvec(3.156 2.090) \lvec(3.168 2.092) \lvec(3.180 2.094)
\lvec(3.192 2.096) \lvec(3.204 2.098) \lvec(3.216 2.101)
\lvec(3.228 2.103) \lvec(3.240 2.105) \lvec(3.252 2.107)
\lvec(3.264 2.109) \lvec(3.276 2.111) \lvec(3.288 2.113)
\lvec(3.300 2.115) \lvec(3.312 2.117) \lvec(3.324 2.119)
\lvec(3.336 2.121) \lvec(3.348 2.123) \lvec(3.360 2.125)
\lvec(3.372 2.127) \lvec(3.384 2.129) \lvec(3.396 2.131)
\lvec(3.408 2.133) \lvec(3.420 2.135) \lvec(3.432 2.137)
\lvec(3.444 2.139) \lvec(3.456 2.141) \lvec(3.468 2.143)
\lvec(3.480 2.145) \lvec(3.492 2.147) \lvec(3.504 2.149)
\lvec(3.516 2.151) \lvec(3.528 2.153) \lvec(3.540 2.155)
\lvec(3.552 2.156) \lvec(3.564 2.158) \lvec(3.576 2.160)
\lvec(3.588 2.162) \lvec(3.600 2.164) \lvec(3.612 2.166)
\lvec(3.624 2.167) \lvec(3.636 2.169) \lvec(3.648 2.171)
\lvec(3.660 2.173) \lvec(3.672 2.174) \lvec(3.684 2.176)
\lvec(3.696 2.178) \lvec(3.708 2.180) \lvec(3.720 2.181)
\lvec(3.732 2.183) \lvec(3.744 2.185) \lvec(3.756 2.187)
\lvec(3.768 2.188) \lvec(3.780 2.190) \lvec(3.792 2.192)
\lvec(3.804 2.193) \lvec(3.816 2.195) \lvec(3.828 2.197)
\lvec(3.840 2.198) \lvec(3.852 2.200) \lvec(3.864 2.202)
\lvec(3.876 2.203) \lvec(3.888 2.205) \lvec(3.900 2.206)
\lvec(3.912 2.208) \lvec(3.924 2.210) \lvec(3.936 2.211)
\lvec(3.948 2.213) \lvec(3.960 2.214) \lvec(3.972 2.216)
\lvec(3.984 2.217) \lvec(3.996 2.219) \lvec(4.008 2.220)
\lvec(4.020 2.222) \lvec(4.032 2.223) \lvec(4.044 2.225)
\lvec(4.056 2.226) \lvec(4.068 2.228) \lvec(4.080 2.229)
\lvec(4.092 2.231) \lvec(4.104 2.232) \lvec(4.116 2.234)
\lvec(4.128 2.235) \lvec(4.140 2.237) \lvec(4.152 2.238)
\lvec(4.164 2.240) \lvec(4.176 2.241) \lvec(4.188 2.242)
\lvec(4.200 2.244) \lvec(4.212 2.245) \lvec(4.224 2.247)
\lvec(4.236 2.248) \lvec(4.248 2.249) \lvec(4.260 2.251)
\lvec(4.272 2.252) \lvec(4.284 2.254) \lvec(4.296 2.255)
\lvec(4.308 2.256) \lvec(4.320 2.258) \lvec(4.332 2.259)
\lvec(4.344 2.260) \lvec(4.356 2.262) \lvec(4.368 2.263)
\lvec(4.380 2.264) \lvec(4.392 2.266) \lvec(4.404 2.267)
\lvec(4.416 2.268) \lvec(4.428 2.269) \lvec(4.440 2.271)
\lvec(4.452 2.272) \lvec(4.464 2.273) \lvec(4.476 2.275)
\lvec(4.488 2.276) \lvec(4.500 2.277) \lvec(4.512 2.278)
\lvec(4.524 2.280) \lvec(4.536 2.281) \lvec(4.548 2.282)
\lvec(4.560 2.283) \lvec(4.572 2.284) \lvec(4.584 2.286)
\lvec(4.596 2.287) \lvec(4.608 2.288) \lvec(4.620 2.289)
\lvec(4.632 2.290) \lvec(4.644 2.292) \lvec(4.656 2.293)
\lvec(4.668 2.294) \lvec(4.680 2.295) \lvec(4.692 2.296)
\lvec(4.704 2.297) \lvec(4.716 2.299) \lvec(4.728 2.300)
\lvec(4.740 2.301) \lvec(4.752 2.302) \lvec(4.764 2.303)
\lvec(4.776 2.304) \lvec(4.788 2.305) \lvec(4.800 2.306) \lpatt ()
\end{texdraw}
\hfill
%
%
\begin{texdraw}
\drawdim truecm \setgray 0
\linewd 0.04
\move (0 0) \lvec (4.8 0)
\lvec (4.8 3.0) \lvec (0 3.0) \lvec (0 0)
\linewd 0.02
\textref h:L v:T \htext (0.00 -.18) {0}
\move (1.200 0) \lvec (1.200 0.120)
\textref h:C v:T \htext (1.20 -.18) {1}
\move (2.400 0) \lvec (2.400 0.120)
\textref h:C v:T \htext (2.40 -.18) {2}
\move (3.600 0) \lvec (3.600 0.120)
\textref h:C v:T \htext (3.60 -.18) {3}
\move (4.800 0) \lvec (4.800 0.120)
\textref h:R v:T \htext (4.80 -.18) {4}
\move (0 1.500) \lvec (0.15 1.500)
\move (0 2.727) \lvec (0.15 2.727)
\move (4.8 0.273) \lvec (4.65 0.273)
\move (4.8 1.500) \lvec (4.65 1.500)
\textref h:C v:T \htext (2.40 -.6) {$q_{\rm ch}/q_{\rm sm}$}
\textref h:L v:C \htext (.2 2.11) {$\alpha_0^{\rm opt}$}
\linewd 0.04
\move(0.000 0.273) \lvec(0.980 0.273) \lvec(0.984 0.450)
\lvec(0.996 0.621) \lvec(1.008 0.732) \lvec(1.020 0.821)
\lvec(1.032 0.898) \lvec(1.044 0.966) \lvec(1.056 1.029)
\lvec(1.068 1.087) \lvec(1.080 1.141) \lvec(1.092 1.192)
\lvec(1.104 1.241) \lvec(1.128 1.333) \lvec(1.140 1.377)
\lvec(1.152 1.419) \lvec(1.164 1.461) \lvec(1.176 1.501)
\lvec(1.188 1.540) \lvec(1.200 1.579) \lvec(1.212 1.617)
\lvec(1.224 1.655) \lvec(1.236 1.692) \lvec(1.248 1.729)
\lvec(1.260 1.766) \lvec(1.272 1.803) \lvec(1.284 1.840)
\lvec(1.296 1.877) \lvec(1.308 1.914) \lvec(1.320 1.952)
\lvec(1.332 1.990) \lvec(1.344 2.028) \lvec(1.356 2.068)
\lvec(1.368 2.109) \lvec(1.380 2.151) \lvec(1.392 2.195)
\lvec(1.404 2.241) \lvec(1.416 2.291) \lvec(1.428 2.346)
\lvec(1.440 2.408) \lvec(1.452 2.482) \lvec(1.464 2.589)
\lvec(1.470 2.727) \lvec(4.800 2.727)
\linewd 0.02
\move(0.000 0.273) \lvec(0.725 0.273) \lvec(0.732 0.443)
\lvec(0.744 0.550) \lvec(0.756 0.626) \lvec(0.768 0.688)
\lvec(0.780 0.741) \lvec(0.792 0.790) \lvec(0.804 0.834)
\lvec(0.816 0.874) \lvec(0.828 0.912) \lvec(0.840 0.948)
\lvec(0.852 0.982) \lvec(0.864 1.014) \lvec(0.876 1.045)
\lvec(0.888 1.075) \lvec(0.900 1.104) \lvec(0.912 1.132)
\lvec(0.924 1.158) \lvec(0.936 1.184) \lvec(0.948 1.209)
\lvec(0.960 1.234) \lvec(0.972 1.258) \lvec(0.984 1.281)
\lvec(0.996 1.304) \lvec(1.008 1.326) \lvec(1.020 1.348)
\lvec(1.032 1.369) \lvec(1.044 1.390) \lvec(1.056 1.410)
\lvec(1.068 1.430) \lvec(1.080 1.449) \lvec(1.092 1.469)
\lvec(1.104 1.488) \lvec(1.116 1.506) \lvec(1.128 1.524)
\lvec(1.140 1.542) \lvec(1.152 1.560) \lvec(1.164 1.578)
\lvec(1.176 1.595) \lvec(1.188 1.612) \lvec(1.200 1.628)
\lvec(1.212 1.645) \lvec(1.224 1.661) \lvec(1.236 1.677)
\lvec(1.248 1.692) \lvec(1.260 1.708) \lvec(1.272 1.723)
\lvec(1.284 1.738) \lvec(1.296 1.753) \lvec(1.308 1.767)
\lvec(1.320 1.782) \lvec(1.332 1.796) \lvec(1.344 1.810)
\lvec(1.356 1.824) \lvec(1.368 1.837) \lvec(1.380 1.851)
\lvec(1.392 1.864) \lvec(1.404 1.877) \lvec(1.416 1.890)
\lvec(1.428 1.903) \lvec(1.440 1.915) \lvec(1.452 1.927)
\lvec(1.464 1.940) \lvec(1.476 1.952) \lvec(1.488 1.963)
\lvec(1.500 1.975) \lvec(1.512 1.986) \lvec(1.524 1.998)
\lvec(1.536 2.009) \lvec(1.548 2.020) \lvec(1.560 2.030)
\lvec(1.572 2.041) \lvec(1.584 2.051) \lvec(1.596 2.061)
\lvec(1.608 2.071) \lvec(1.620 2.081) \lvec(1.632 2.091)
\lvec(1.644 2.100) \lvec(1.656 2.110) \lvec(1.668 2.119)
\lvec(1.680 2.128) \lvec(1.692 2.137) \lvec(1.704 2.145)
\lvec(1.716 2.154) \lvec(1.728 2.162) \lvec(1.740 2.170)
\lvec(1.752 2.179) \lvec(1.764 2.186) \lvec(1.776 2.194)
\lvec(1.788 2.202) \lvec(1.800 2.209) \lvec(1.812 2.216)
\lvec(1.824 2.223) \lvec(1.836 2.230) \lvec(1.848 2.237)
\lvec(1.860 2.244) \lvec(1.872 2.250) \lvec(1.884 2.257)
\lvec(1.896 2.263) \lvec(1.908 2.269) \lvec(1.920 2.275)
\lvec(1.932 2.281) \lvec(1.944 2.287) \lvec(1.956 2.293)
\lvec(1.968 2.298) \lvec(1.980 2.304) \lvec(1.992 2.309)
\lvec(2.004 2.314) \lvec(2.016 2.319) \lvec(2.028 2.324)
\lvec(2.040 2.329) \lvec(2.052 2.334) \lvec(2.064 2.338)
\lvec(2.076 2.343) \lvec(2.088 2.347) \lvec(2.100 2.352)
\lvec(2.112 2.356) \lvec(2.124 2.360) \lvec(2.136 2.364)
\lvec(2.148 2.368) \lvec(2.160 2.372) \lvec(2.172 2.376)
\lvec(2.184 2.380) \lvec(2.196 2.384) \lvec(2.208 2.387)
\lvec(2.220 2.391) \lvec(2.232 2.394) \lvec(2.244 2.398)
\lvec(2.256 2.401) \lvec(2.268 2.405) \lvec(2.280 2.408)
\lvec(2.292 2.411) \lvec(2.304 2.414) \lvec(2.316 2.417)
\lvec(2.328 2.420) \lvec(2.340 2.423) \lvec(2.352 2.426)
\lvec(2.364 2.429) \lvec(2.376 2.432) \lvec(2.388 2.434)
\lvec(2.400 2.437) \lvec(2.412 2.440) \lvec(2.424 2.442)
\lvec(2.436 2.445) \lvec(2.448 2.447) \lvec(2.460 2.450)
\lvec(2.472 2.452) \lvec(2.484 2.454) \lvec(2.496 2.457)
\lvec(2.508 2.459) \lvec(2.520 2.461) \lvec(2.532 2.463)
\lvec(2.544 2.466) \lvec(2.556 2.468) \lvec(2.568 2.470)
\lvec(2.580 2.472) \lvec(2.592 2.474) \lvec(2.604 2.476)
\lvec(2.616 2.478) \lvec(2.628 2.480) \lvec(2.640 2.482)
\lvec(2.652 2.483) \lvec(2.664 2.485) \lvec(2.676 2.487)
\lvec(2.688 2.489) \lvec(2.700 2.491) \lvec(2.712 2.492)
\lvec(2.724 2.494) \lvec(2.736 2.496) \lvec(2.748 2.497)
\lvec(2.760 2.499) \lvec(2.772 2.501) \lvec(2.784 2.502)
\lvec(2.796 2.504) \lvec(2.808 2.505) \lvec(2.820 2.507)
\lvec(2.832 2.508) \lvec(2.844 2.510) \lvec(2.856 2.511)
\lvec(2.868 2.513) \lvec(2.880 2.514) \lvec(2.892 2.515)
\lvec(2.904 2.517) \lvec(2.916 2.518) \lvec(2.928 2.519)
\lvec(2.940 2.521) \lvec(2.952 2.522) \lvec(2.964 2.523)
\lvec(2.976 2.525) \lvec(2.988 2.526) \lvec(3.000 2.527)
\lvec(3.012 2.528) \lvec(3.024 2.529) \lvec(3.036 2.531)
\lvec(3.048 2.532) \lvec(3.060 2.533) \lvec(3.072 2.534)
\lvec(3.084 2.535) \lvec(3.096 2.536) \lvec(3.108 2.537)
\lvec(3.120 2.538) \lvec(3.132 2.540) \lvec(3.144 2.541)
\lvec(3.156 2.542) \lvec(3.168 2.543) \lvec(3.180 2.544)
\lvec(3.192 2.545) \lvec(3.204 2.546) \lvec(3.216 2.547)
\lvec(3.228 2.548) \lvec(3.240 2.549) \lvec(3.252 2.550)
\lvec(3.264 2.551) \lvec(3.276 2.551) \lvec(3.288 2.552)
\lvec(3.300 2.553) \lvec(3.312 2.554) \lvec(3.324 2.555)
\lvec(3.336 2.556) \lvec(3.348 2.557) \lvec(3.360 2.558)
\lvec(3.372 2.559) \lvec(3.384 2.559) \lvec(3.396 2.560)
\lvec(3.408 2.561) \lvec(3.420 2.562) \lvec(3.432 2.563)
\lvec(3.444 2.563) \lvec(3.456 2.564) \lvec(3.468 2.565)
\lvec(3.480 2.566) \lvec(3.492 2.567) \lvec(3.504 2.567)
\lvec(3.516 2.568) \lvec(3.528 2.569) \lvec(3.540 2.570)
\lvec(3.552 2.570) \lvec(3.564 2.571) \lvec(3.576 2.572)
\lvec(3.588 2.572) \lvec(3.600 2.573) \lvec(3.612 2.574)
\lvec(3.624 2.574) \lvec(3.636 2.575) \lvec(3.648 2.576)
\lvec(3.660 2.576) \lvec(3.672 2.577) \lvec(3.684 2.578)
\lvec(3.696 2.578) \lvec(3.708 2.579) \lvec(3.720 2.580)
\lvec(3.732 2.580) \lvec(3.744 2.581) \lvec(3.756 2.582)
\lvec(3.768 2.582) \lvec(3.780 2.583) \lvec(3.792 2.583)
\lvec(3.804 2.584) \lvec(3.816 2.585) \lvec(3.828 2.585)
\lvec(3.840 2.586) \lvec(3.852 2.586) \lvec(3.864 2.587)
\lvec(3.876 2.587) \lvec(3.888 2.588) \lvec(3.900 2.589)
\lvec(3.912 2.589) \lvec(3.924 2.590) \lvec(3.936 2.590)
\lvec(3.948 2.591) \lvec(3.960 2.591) \lvec(3.972 2.592)
\lvec(3.984 2.592) \lvec(3.996 2.593) \lvec(4.008 2.593)
\lvec(4.020 2.594) \lvec(4.032 2.594) \lvec(4.044 2.595)
\lvec(4.056 2.595) \lvec(4.068 2.596) \lvec(4.080 2.596)
\lvec(4.092 2.597) \lvec(4.104 2.597) \lvec(4.116 2.598)
\lvec(4.128 2.598) \lvec(4.140 2.599) \lvec(4.152 2.599)
\lvec(4.164 2.600) \lvec(4.176 2.600) \lvec(4.188 2.601)
\lvec(4.200 2.601) \lvec(4.212 2.602) \lvec(4.224 2.602)
\lvec(4.236 2.602) \lvec(4.248 2.603) \lvec(4.260 2.603)
\lvec(4.272 2.604) \lvec(4.284 2.604) \lvec(4.296 2.605)
\lvec(4.308 2.605) \lvec(4.320 2.605) \lvec(4.332 2.606)
\lvec(4.344 2.606) \lvec(4.356 2.607) \lvec(4.368 2.607)
\lvec(4.380 2.608) \lvec(4.392 2.608) \lvec(4.404 2.608)
\lvec(4.416 2.609) \lvec(4.428 2.609) \lvec(4.440 2.609)
\lvec(4.452 2.610) \lvec(4.464 2.610) \lvec(4.476 2.611)
\lvec(4.488 2.611) \lvec(4.500 2.611) \lvec(4.512 2.612)
\lvec(4.524 2.612) \lvec(4.536 2.613) \lvec(4.548 2.613)
\lvec(4.560 2.613) \lvec(4.572 2.614) \lvec(4.584 2.614)
\lvec(4.596 2.614) \lvec(4.608 2.615) \lvec(4.620 2.615)
\lvec(4.632 2.615) \lvec(4.644 2.616) \lvec(4.656 2.616)
\lvec(4.668 2.616) \lvec(4.680 2.617) \lvec(4.692 2.617)
\lvec(4.704 2.617) \lvec(4.716 2.618) \lvec(4.728 2.618)
\lvec(4.740 2.618) \lvec(4.752 2.619) \lvec(4.764 2.619)
\lvec(4.776 2.619) \lvec(4.788 2.620) \lvec(4.800 2.620)
\move(0.000 0.273) \lvec(0.549 0.273) \lvec(0.552 0.365)
\lvec(0.564 0.477) \lvec(0.576 0.547) \lvec(0.588 0.602)
\lvec(0.600 0.649) \lvec(0.612 0.690) \lvec(0.624 0.728)
\lvec(0.636 0.763) \lvec(0.648 0.795) \lvec(0.660 0.825)
\lvec(0.672 0.854) \lvec(0.684 0.881) \lvec(0.696 0.907)
\lvec(0.708 0.932) \lvec(0.720 0.956) \lvec(0.732 0.980)
\lvec(0.744 1.002) \lvec(0.756 1.024) \lvec(0.768 1.045)
\lvec(0.780 1.065) \lvec(0.792 1.085) \lvec(0.804 1.104)
\lvec(0.816 1.123) \lvec(0.828 1.141) \lvec(0.840 1.159)
\lvec(0.852 1.177) \lvec(0.864 1.194) \lvec(0.876 1.211)
\lvec(0.888 1.227) \lvec(0.900 1.243) \lvec(0.912 1.259)
\lvec(0.924 1.274) \lvec(0.936 1.290) \lvec(0.948 1.305)
\lvec(0.960 1.319) \lvec(0.972 1.334) \lvec(0.984 1.348)
\lvec(0.996 1.362) \lvec(1.008 1.376) \lvec(1.020 1.389)
\lvec(1.032 1.403) \lvec(1.044 1.416) \lvec(1.056 1.429)
\lvec(1.068 1.442) \lvec(1.080 1.454) \lvec(1.092 1.467)
\lvec(1.104 1.479) \lvec(1.116 1.491) \lvec(1.128 1.503)
\lvec(1.140 1.515) \lvec(1.152 1.527) \lvec(1.164 1.538)
\lvec(1.176 1.549) \lvec(1.188 1.561) \lvec(1.200 1.572)
\lvec(1.212 1.583) \lvec(1.224 1.593) \lvec(1.236 1.604)
\lvec(1.248 1.615) \lvec(1.260 1.625) \lvec(1.272 1.635)
\lvec(1.284 1.645) \lvec(1.296 1.655) \lvec(1.308 1.665)
\lvec(1.320 1.675) \lvec(1.332 1.685) \lvec(1.344 1.694)
\lvec(1.356 1.704) \lvec(1.368 1.713) \lvec(1.380 1.722)
\lvec(1.392 1.732) \lvec(1.404 1.741) \lvec(1.416 1.750)
\lvec(1.428 1.758) \lvec(1.440 1.767) \lvec(1.452 1.776)
\lvec(1.464 1.784) \lvec(1.476 1.793) \lvec(1.488 1.801)
\lvec(1.500 1.809) \lvec(1.512 1.818) \lvec(1.524 1.826)
\lvec(1.536 1.834) \lvec(1.548 1.842) \lvec(1.560 1.849)
\lvec(1.572 1.857) \lvec(1.584 1.865) \lvec(1.596 1.872)
\lvec(1.608 1.880) \lvec(1.620 1.887) \lvec(1.632 1.895)
\lvec(1.644 1.902) \lvec(1.656 1.909) \lvec(1.668 1.916)
\lvec(1.680 1.923) \lvec(1.692 1.930) \lvec(1.704 1.937)
\lvec(1.716 1.943) \lvec(1.728 1.950) \lvec(1.740 1.957)
\lvec(1.752 1.963) \lvec(1.764 1.970) \lvec(1.776 1.976)
\lvec(1.788 1.982) \lvec(1.800 1.988) \lvec(1.812 1.995)
\lvec(1.824 2.001) \lvec(1.836 2.007) \lvec(1.848 2.013)
\lvec(1.860 2.018) \lvec(1.872 2.024) \lvec(1.884 2.030)
\lvec(1.896 2.036) \lvec(1.908 2.041) \lvec(1.920 2.047)
\lvec(1.932 2.052) \lvec(1.944 2.058) \lvec(1.956 2.063)
\lvec(1.968 2.068) \lvec(1.980 2.073) \lvec(1.992 2.079)
\lvec(2.004 2.084) \lvec(2.016 2.089) \lvec(2.028 2.094)
\lvec(2.040 2.099) \lvec(2.052 2.104) \lvec(2.064 2.108)
\lvec(2.076 2.113) \lvec(2.088 2.118) \lvec(2.100 2.122)
\lvec(2.112 2.127) \lvec(2.124 2.131) \lvec(2.136 2.136)
\lvec(2.148 2.140) \lvec(2.160 2.145) \lvec(2.172 2.149)
\lvec(2.184 2.153) \lvec(2.196 2.157) \lvec(2.208 2.162)
\lvec(2.220 2.166) \lvec(2.232 2.170) \lvec(2.244 2.174)
\lvec(2.256 2.178) \lvec(2.268 2.182) \lvec(2.280 2.185)
\lvec(2.292 2.189) \lvec(2.304 2.193) \lvec(2.316 2.197)
\lvec(2.328 2.200) \lvec(2.340 2.204) \lvec(2.352 2.208)
\lvec(2.364 2.211) \lvec(2.376 2.215) \lvec(2.388 2.218)
\lvec(2.400 2.222) \lvec(2.412 2.225) \lvec(2.424 2.228)
\lvec(2.436 2.232) \lvec(2.448 2.235) \lvec(2.460 2.238)
\lvec(2.472 2.241) \lvec(2.484 2.245) \lvec(2.496 2.248)
\lvec(2.508 2.251) \lvec(2.520 2.254) \lvec(2.532 2.257)
\lvec(2.544 2.260) \lvec(2.556 2.263) \lvec(2.568 2.266)
\lvec(2.580 2.269) \lvec(2.592 2.271) \lvec(2.604 2.274)
\lvec(2.616 2.277) \lvec(2.628 2.280) \lvec(2.640 2.283)
\lvec(2.652 2.285) \lvec(2.664 2.288) \lvec(2.676 2.291)
\lvec(2.688 2.293) \lvec(2.700 2.296) \lvec(2.712 2.298)
\lvec(2.724 2.301) \lvec(2.736 2.303) \lvec(2.748 2.306)
\lvec(2.760 2.308) \lvec(2.772 2.311) \lvec(2.784 2.313)
\lvec(2.796 2.315) \lvec(2.808 2.318) \lvec(2.820 2.320)
\lvec(2.832 2.322) \lvec(2.844 2.324) \lvec(2.856 2.327)
\lvec(2.868 2.329) \lvec(2.880 2.331) \lvec(2.892 2.333)
\lvec(2.904 2.335) \lvec(2.916 2.338) \lvec(2.928 2.340)
\lvec(2.940 2.342) \lvec(2.952 2.344) \lvec(2.964 2.346)
\lvec(2.976 2.348) \lvec(2.988 2.350) \lvec(3.000 2.352)
\lvec(3.012 2.354) \lvec(3.024 2.356) \lvec(3.036 2.358)
\lvec(3.048 2.359) \lvec(3.060 2.361) \lvec(3.072 2.363)
\lvec(3.084 2.365) \lvec(3.096 2.367) \lvec(3.108 2.369)
\lvec(3.120 2.370) \lvec(3.132 2.372) \lvec(3.144 2.374)
\lvec(3.156 2.376) \lvec(3.168 2.377) \lvec(3.180 2.379)
\lvec(3.192 2.381) \lvec(3.204 2.382) \lvec(3.216 2.384)
\lvec(3.228 2.386) \lvec(3.240 2.387) \lvec(3.252 2.389)
\lvec(3.264 2.391) \lvec(3.276 2.392) \lvec(3.288 2.394)
\lvec(3.300 2.395) \lvec(3.312 2.397) \lvec(3.324 2.398)
\lvec(3.336 2.400) \lvec(3.348 2.401) \lvec(3.360 2.403)
\lvec(3.372 2.404) \lvec(3.384 2.406) \lvec(3.396 2.407)
\lvec(3.408 2.408) \lvec(3.420 2.410) \lvec(3.432 2.411)
\lvec(3.444 2.413) \lvec(3.456 2.414) \lvec(3.468 2.415)
\lvec(3.480 2.417) \lvec(3.492 2.418) \lvec(3.504 2.419)
\lvec(3.516 2.421) \lvec(3.528 2.422) \lvec(3.540 2.423)
\lvec(3.552 2.425) \lvec(3.564 2.426) \lvec(3.576 2.427)
\lvec(3.588 2.428) \lvec(3.600 2.430) \lvec(3.612 2.431)
\lvec(3.624 2.432) \lvec(3.636 2.433) \lvec(3.648 2.434)
\lvec(3.660 2.436) \lvec(3.672 2.437) \lvec(3.684 2.438)
\lvec(3.696 2.439) \lvec(3.708 2.440) \lvec(3.720 2.441)
\lvec(3.732 2.443) \lvec(3.744 2.444) \lvec(3.756 2.445)
\lvec(3.768 2.446) \lvec(3.780 2.447) \lvec(3.792 2.448)
\lvec(3.804 2.449) \lvec(3.816 2.450) \lvec(3.828 2.451)
\lvec(3.840 2.452) \lvec(3.852 2.453) \lvec(3.864 2.454)
\lvec(3.876 2.455) \lvec(3.888 2.456) \lvec(3.900 2.457)
\lvec(3.912 2.458) \lvec(3.924 2.459) \lvec(3.936 2.460)
\lvec(3.948 2.461) \lvec(3.960 2.462) \lvec(3.972 2.463)
\lvec(3.984 2.464) \lvec(3.996 2.465) \lvec(4.008 2.466)
\lvec(4.020 2.467) \lvec(4.032 2.468) \lvec(4.044 2.469)
\lvec(4.056 2.470) \lvec(4.068 2.471) \lvec(4.080 2.472)
\lvec(4.092 2.473) \lvec(4.104 2.474) \lvec(4.116 2.474)
\lvec(4.128 2.475) \lvec(4.140 2.476) \lvec(4.152 2.477)
\lvec(4.164 2.478) \lvec(4.176 2.479) \lvec(4.188 2.480)
\lvec(4.200 2.480) \lvec(4.212 2.481) \lvec(4.224 2.482)
\lvec(4.236 2.483) \lvec(4.248 2.484) \lvec(4.260 2.485)
\lvec(4.272 2.485) \lvec(4.284 2.486) \lvec(4.296 2.487)
\lvec(4.308 2.488) \lvec(4.320 2.489) \lvec(4.332 2.489)
\lvec(4.344 2.490) \lvec(4.356 2.491) \lvec(4.368 2.492)
\lvec(4.380 2.492) \lvec(4.392 2.493) \lvec(4.404 2.494)
\lvec(4.416 2.495) \lvec(4.428 2.495) \lvec(4.440 2.496)
\lvec(4.452 2.497) \lvec(4.464 2.498) \lvec(4.476 2.498)
\lvec(4.488 2.499) \lvec(4.500 2.500) \lvec(4.512 2.500)
\lvec(4.524 2.501) \lvec(4.536 2.502) \lvec(4.548 2.503)
\lvec(4.560 2.503) \lvec(4.572 2.504) \lvec(4.584 2.505)
\lvec(4.596 2.505) \lvec(4.608 2.506) \lvec(4.620 2.507)
\lvec(4.632 2.507) \lvec(4.644 2.508) \lvec(4.656 2.509)
\lvec(4.668 2.509) \lvec(4.680 2.510) \lvec(4.692 2.510)
\lvec(4.704 2.511) \lvec(4.716 2.512) \lvec(4.728 2.512)
\lvec(4.740 2.513) \lvec(4.752 2.514) \lvec(4.764 2.514)
\lvec(4.776 2.515) \lvec(4.788 2.515) \lvec(4.800 2.516)
\move(0.000 0.273) \lvec(0.430 0.273) \lvec(0.432 0.331)
\lvec(0.444 0.443) \lvec(0.456 0.506) \lvec(0.468 0.556)
\lvec(0.480 0.598) \lvec(0.492 0.635) \lvec(0.504 0.668)
\lvec(0.516 0.699) \lvec(0.528 0.728) \lvec(0.540 0.754)
\lvec(0.552 0.780) \lvec(0.564 0.804) \lvec(0.576 0.827)
\lvec(0.588 0.849) \lvec(0.600 0.870) \lvec(0.612 0.890)
\lvec(0.624 0.910) \lvec(0.636 0.929) \lvec(0.648 0.948)
\lvec(0.660 0.966) \lvec(0.672 0.983) \lvec(0.684 1.000)
\lvec(0.696 1.016) \lvec(0.708 1.033) \lvec(0.720 1.048)
\lvec(0.732 1.064) \lvec(0.744 1.079) \lvec(0.756 1.094)
\lvec(0.768 1.108) \lvec(0.780 1.122) \lvec(0.792 1.136)
\lvec(0.804 1.150) \lvec(0.816 1.163) \lvec(0.828 1.176)
\lvec(0.840 1.189) \lvec(0.852 1.202) \lvec(0.864 1.214)
\lvec(0.876 1.227) \lvec(0.888 1.239) \lvec(0.900 1.251)
\lvec(0.912 1.262) \lvec(0.924 1.274) \lvec(0.936 1.285)
\lvec(0.948 1.297) \lvec(0.960 1.308) \lvec(0.972 1.319)
\lvec(0.984 1.330) \lvec(0.996 1.340) \lvec(1.008 1.351)
\lvec(1.020 1.361) \lvec(1.032 1.371) \lvec(1.044 1.381)
\lvec(1.056 1.391) \lvec(1.068 1.401) \lvec(1.080 1.411)
\lvec(1.092 1.421) \lvec(1.104 1.430) \lvec(1.116 1.440)
\lvec(1.128 1.449) \lvec(1.140 1.458) \lvec(1.152 1.467)
\lvec(1.164 1.476) \lvec(1.176 1.485) \lvec(1.188 1.494)
\lvec(1.200 1.503) \lvec(1.212 1.511) \lvec(1.224 1.520)
\lvec(1.236 1.528) \lvec(1.248 1.537) \lvec(1.260 1.545)
\lvec(1.272 1.553) \lvec(1.284 1.561) \lvec(1.296 1.569)
\lvec(1.308 1.577) \lvec(1.320 1.585) \lvec(1.332 1.593)
\lvec(1.344 1.600) \lvec(1.356 1.608) \lvec(1.368 1.616)
\lvec(1.380 1.623) \lvec(1.392 1.631) \lvec(1.404 1.638)
\lvec(1.416 1.645) \lvec(1.428 1.652) \lvec(1.440 1.659)
\lvec(1.452 1.667) \lvec(1.464 1.674) \lvec(1.476 1.680)
\lvec(1.488 1.687) \lvec(1.500 1.694) \lvec(1.512 1.701)
\lvec(1.524 1.708) \lvec(1.536 1.714) \lvec(1.548 1.721)
\lvec(1.560 1.727) \lvec(1.572 1.734) \lvec(1.584 1.740)
\lvec(1.596 1.746) \lvec(1.608 1.753) \lvec(1.620 1.759)
\lvec(1.632 1.765) \lvec(1.644 1.771) \lvec(1.656 1.777)
\lvec(1.668 1.783) \lvec(1.680 1.789) \lvec(1.692 1.795)
\lvec(1.704 1.801) \lvec(1.716 1.807) \lvec(1.728 1.812)
\lvec(1.740 1.818) \lvec(1.752 1.824) \lvec(1.764 1.829)
\lvec(1.776 1.835) \lvec(1.788 1.840) \lvec(1.800 1.846)
\lvec(1.812 1.851) \lvec(1.824 1.856) \lvec(1.836 1.862)
\lvec(1.848 1.867) \lvec(1.860 1.872) \lvec(1.872 1.877)
\lvec(1.884 1.882) \lvec(1.896 1.887) \lvec(1.908 1.892)
\lvec(1.920 1.897) \lvec(1.932 1.902) \lvec(1.944 1.907)
\lvec(1.956 1.912) \lvec(1.968 1.917) \lvec(1.980 1.921)
\lvec(1.992 1.926) \lvec(2.004 1.931) \lvec(2.016 1.935)
\lvec(2.028 1.940) \lvec(2.040 1.945) \lvec(2.052 1.949)
\lvec(2.064 1.954) \lvec(2.076 1.958) \lvec(2.088 1.962)
\lvec(2.100 1.967) \lvec(2.112 1.971) \lvec(2.124 1.975)
\lvec(2.136 1.980) \lvec(2.148 1.984) \lvec(2.160 1.988)
\lvec(2.172 1.992) \lvec(2.184 1.996) \lvec(2.196 2.000)
\lvec(2.208 2.004) \lvec(2.220 2.008) \lvec(2.232 2.012)
\lvec(2.244 2.016) \lvec(2.256 2.020) \lvec(2.268 2.024)
\lvec(2.280 2.028) \lvec(2.292 2.032) \lvec(2.304 2.035)
\lvec(2.316 2.039) \lvec(2.328 2.043) \lvec(2.340 2.046)
\lvec(2.352 2.050) \lvec(2.364 2.054) \lvec(2.376 2.057)
\lvec(2.388 2.061) \lvec(2.400 2.064) \lvec(2.412 2.068)
\lvec(2.424 2.071) \lvec(2.436 2.075) \lvec(2.448 2.078)
\lvec(2.460 2.081) \lvec(2.472 2.085) \lvec(2.484 2.088)
\lvec(2.496 2.091) \lvec(2.508 2.094) \lvec(2.520 2.098)
\lvec(2.532 2.101) \lvec(2.544 2.104) \lvec(2.556 2.107)
\lvec(2.568 2.110) \lvec(2.580 2.113) \lvec(2.592 2.116)
\lvec(2.604 2.119) \lvec(2.616 2.122) \lvec(2.628 2.125)
\lvec(2.640 2.128) \lvec(2.652 2.131) \lvec(2.664 2.134)
\lvec(2.676 2.137) \lvec(2.688 2.140) \lvec(2.700 2.143)
\lvec(2.712 2.146) \lvec(2.724 2.148) \lvec(2.736 2.151)
\lvec(2.748 2.154) \lvec(2.760 2.157) \lvec(2.772 2.159)
\lvec(2.784 2.162) \lvec(2.796 2.165) \lvec(2.808 2.167)
\lvec(2.820 2.170) \lvec(2.832 2.173) \lvec(2.844 2.175)
\lvec(2.856 2.178) \lvec(2.868 2.180) \lvec(2.880 2.183)
\lvec(2.892 2.185) \lvec(2.904 2.188) \lvec(2.916 2.190)
\lvec(2.928 2.193) \lvec(2.940 2.195) \lvec(2.952 2.197)
\lvec(2.964 2.200) \lvec(2.976 2.202) \lvec(2.988 2.204)
\lvec(3.000 2.207) \lvec(3.012 2.209) \lvec(3.024 2.211)
\lvec(3.036 2.213) \lvec(3.048 2.216) \lvec(3.060 2.218)
\lvec(3.072 2.220) \lvec(3.084 2.222) \lvec(3.096 2.225)
\lvec(3.108 2.227) \lvec(3.120 2.229) \lvec(3.132 2.231)
\lvec(3.144 2.233) \lvec(3.156 2.235) \lvec(3.168 2.237)
\lvec(3.180 2.239) \lvec(3.192 2.241) \lvec(3.204 2.243)
\lvec(3.216 2.245) \lvec(3.228 2.247) \lvec(3.240 2.249)
\lvec(3.252 2.251) \lvec(3.264 2.253) \lvec(3.276 2.255)
\lvec(3.288 2.257) \lvec(3.300 2.259) \lvec(3.312 2.261)
\lvec(3.324 2.263) \lvec(3.336 2.265) \lvec(3.348 2.266)
\lvec(3.360 2.268) \lvec(3.372 2.270) \lvec(3.384 2.272)
\lvec(3.396 2.274) \lvec(3.408 2.275) \lvec(3.420 2.277)
\lvec(3.432 2.279) \lvec(3.444 2.281) \lvec(3.456 2.282)
\lvec(3.468 2.284) \lvec(3.480 2.286) \lvec(3.492 2.287)
\lvec(3.504 2.289) \lvec(3.516 2.291) \lvec(3.528 2.292)
\lvec(3.540 2.294) \lvec(3.552 2.296) \lvec(3.564 2.297)
\lvec(3.576 2.299) \lvec(3.588 2.301) \lvec(3.600 2.302)
\lvec(3.612 2.304) \lvec(3.624 2.305) \lvec(3.636 2.307)
\lvec(3.648 2.308) \lvec(3.660 2.310) \lvec(3.672 2.311)
\lvec(3.684 2.313) \lvec(3.696 2.314) \lvec(3.708 2.316)
\lvec(3.720 2.317) \lvec(3.732 2.319) \lvec(3.744 2.320)
\lvec(3.756 2.322) \lvec(3.768 2.323) \lvec(3.780 2.325)
\lvec(3.792 2.326) \lvec(3.804 2.327) \lvec(3.816 2.329)
\lvec(3.828 2.330) \lvec(3.840 2.332) \lvec(3.852 2.333)
\lvec(3.864 2.334) \lvec(3.876 2.336) \lvec(3.888 2.337)
\lvec(3.900 2.338) \lvec(3.912 2.340) \lvec(3.924 2.341)
\lvec(3.936 2.342) \lvec(3.948 2.344) \lvec(3.960 2.345)
\lvec(3.972 2.346) \lvec(3.984 2.347) \lvec(3.996 2.349)
\lvec(4.008 2.350) \lvec(4.020 2.351) \lvec(4.032 2.352)
\lvec(4.044 2.354) \lvec(4.056 2.355) \lvec(4.068 2.356)
\lvec(4.080 2.357) \lvec(4.092 2.358) \lvec(4.104 2.360)
\lvec(4.116 2.361) \lvec(4.128 2.362) \lvec(4.140 2.363)
\lvec(4.152 2.364) \lvec(4.164 2.366) \lvec(4.176 2.367)
\lvec(4.188 2.368) \lvec(4.200 2.369) \lvec(4.212 2.370)
\lvec(4.224 2.371) \lvec(4.236 2.372) \lvec(4.248 2.373)
\lvec(4.260 2.374) \lvec(4.272 2.376) \lvec(4.284 2.377)
\lvec(4.296 2.378) \lvec(4.308 2.379) \lvec(4.320 2.380)
\lvec(4.332 2.381) \lvec(4.344 2.382) \lvec(4.356 2.383)
\lvec(4.368 2.384) \lvec(4.380 2.385) \lvec(4.392 2.386)
\lvec(4.404 2.387) \lvec(4.416 2.388) \lvec(4.428 2.389)
\lvec(4.440 2.390) \lvec(4.452 2.391) \lvec(4.464 2.392)
\lvec(4.476 2.393) \lvec(4.488 2.394) \lvec(4.500 2.395)
\lvec(4.512 2.396) \lvec(4.524 2.397) \lvec(4.536 2.398)
\lvec(4.548 2.399) \lvec(4.560 2.400) \lvec(4.572 2.401)
\lvec(4.584 2.402) \lvec(4.596 2.403) \lvec(4.608 2.404)
\lvec(4.620 2.405) \lvec(4.632 2.405) \lvec(4.644 2.406)
\lvec(4.656 2.407) \lvec(4.668 2.408) \lvec(4.680 2.409)
\lvec(4.692 2.410) \lvec(4.704 2.411) \lvec(4.716 2.412)
\lvec(4.728 2.413) \lvec(4.740 2.413) \lvec(4.752 2.414)
\lvec(4.764 2.415) \lvec(4.776 2.416) \lvec(4.788 2.417)
\lvec(4.800 2.418)
\lpatt (.03 .1)
\move(0.000 0.273) \lvec(0.348 0.273) \lvec(0.360 0.410)
\lvec(0.372 0.471) \lvec(0.384 0.518) \lvec(0.396 0.557)
\lvec(0.408 0.591) \lvec(0.420 0.622) \lvec(0.432 0.650)
\lvec(0.444 0.677) \lvec(0.456 0.701) \lvec(0.468 0.724)
\lvec(0.480 0.746) \lvec(0.492 0.767) \lvec(0.504 0.787)
\lvec(0.516 0.807) \lvec(0.528 0.825) \lvec(0.540 0.843)
\lvec(0.552 0.860) \lvec(0.564 0.877) \lvec(0.576 0.893)
\lvec(0.588 0.909) \lvec(0.600 0.925) \lvec(0.612 0.940)
\lvec(0.624 0.954) \lvec(0.636 0.969) \lvec(0.648 0.983)
\lvec(0.660 0.996) \lvec(0.672 1.010) \lvec(0.684 1.023)
\lvec(0.696 1.036) \lvec(0.708 1.048) \lvec(0.720 1.061)
\lvec(0.732 1.073) \lvec(0.744 1.085) \lvec(0.756 1.096)
\lvec(0.768 1.108) \lvec(0.780 1.119) \lvec(0.792 1.131)
\lvec(0.804 1.142) \lvec(0.816 1.152) \lvec(0.828 1.163)
\lvec(0.840 1.174) \lvec(0.852 1.184) \lvec(0.864 1.194)
\lvec(0.876 1.204) \lvec(0.888 1.214) \lvec(0.900 1.224)
\lvec(0.912 1.234) \lvec(0.924 1.243) \lvec(0.936 1.253)
\lvec(0.948 1.262) \lvec(0.960 1.271) \lvec(0.972 1.280)
\lvec(0.984 1.289) \lvec(0.996 1.298) \lvec(1.008 1.307)
\lvec(1.020 1.316) \lvec(1.032 1.324) \lvec(1.044 1.333)
\lvec(1.056 1.341) \lvec(1.068 1.350) \lvec(1.080 1.358)
\lvec(1.092 1.366) \lvec(1.104 1.374) \lvec(1.116 1.382)
\lvec(1.128 1.390) \lvec(1.140 1.398) \lvec(1.152 1.406)
\lvec(1.164 1.413) \lvec(1.176 1.421) \lvec(1.188 1.428)
\lvec(1.200 1.436) \lvec(1.212 1.443) \lvec(1.224 1.450)
\lvec(1.236 1.458) \lvec(1.248 1.465) \lvec(1.260 1.472)
\lvec(1.272 1.479) \lvec(1.284 1.486) \lvec(1.296 1.493)
\lvec(1.308 1.500) \lvec(1.320 1.507) \lvec(1.332 1.513)
\lvec(1.344 1.520) \lvec(1.356 1.527) \lvec(1.368 1.533)
\lvec(1.380 1.540) \lvec(1.392 1.546) \lvec(1.404 1.552)
\lvec(1.416 1.559) \lvec(1.428 1.565) \lvec(1.440 1.571)
\lvec(1.452 1.577) \lvec(1.464 1.584) \lvec(1.476 1.590)
\lvec(1.488 1.596) \lvec(1.500 1.602) \lvec(1.512 1.608)
\lvec(1.524 1.613) \lvec(1.536 1.619) \lvec(1.548 1.625)
\lvec(1.560 1.631) \lvec(1.572 1.636) \lvec(1.584 1.642)
\lvec(1.596 1.648) \lvec(1.608 1.653) \lvec(1.620 1.659)
\lvec(1.632 1.664) \lvec(1.644 1.670) \lvec(1.656 1.675)
\lvec(1.668 1.680) \lvec(1.680 1.686) \lvec(1.692 1.691)
\lvec(1.704 1.696) \lvec(1.716 1.701) \lvec(1.728 1.706)
\lvec(1.740 1.712) \lvec(1.752 1.717) \lvec(1.764 1.722)
\lvec(1.776 1.727) \lvec(1.788 1.732) \lvec(1.800 1.737)
\lvec(1.812 1.741) \lvec(1.824 1.746) \lvec(1.836 1.751)
\lvec(1.848 1.756) \lvec(1.860 1.761) \lvec(1.872 1.765)
\lvec(1.884 1.770) \lvec(1.896 1.775) \lvec(1.908 1.779)
\lvec(1.920 1.784) \lvec(1.932 1.788) \lvec(1.944 1.793)
\lvec(1.956 1.797) \lvec(1.968 1.802) \lvec(1.980 1.806)
\lvec(1.992 1.810) \lvec(2.004 1.815) \lvec(2.016 1.819)
\lvec(2.028 1.823) \lvec(2.040 1.828) \lvec(2.052 1.832)
\lvec(2.064 1.836) \lvec(2.076 1.840) \lvec(2.088 1.844)
\lvec(2.100 1.848) \lvec(2.112 1.852) \lvec(2.124 1.856)
\lvec(2.136 1.860) \lvec(2.148 1.864) \lvec(2.160 1.868)
\lvec(2.172 1.872) \lvec(2.184 1.876) \lvec(2.196 1.880)
\lvec(2.208 1.884) \lvec(2.220 1.888) \lvec(2.232 1.891)
\lvec(2.244 1.895) \lvec(2.256 1.899) \lvec(2.268 1.903)
\lvec(2.280 1.906) \lvec(2.292 1.910) \lvec(2.304 1.914)
\lvec(2.316 1.917) \lvec(2.328 1.921) \lvec(2.340 1.924)
\lvec(2.352 1.928) \lvec(2.364 1.931) \lvec(2.376 1.935)
\lvec(2.388 1.938) \lvec(2.400 1.942) \lvec(2.412 1.945)
\lvec(2.424 1.948) \lvec(2.436 1.952) \lvec(2.448 1.955)
\lvec(2.460 1.959) \lvec(2.472 1.962) \lvec(2.484 1.965)
\lvec(2.496 1.968) \lvec(2.508 1.972) \lvec(2.520 1.975)
\lvec(2.532 1.978) \lvec(2.544 1.981) \lvec(2.556 1.984)
\lvec(2.568 1.987) \lvec(2.580 1.990) \lvec(2.592 1.993)
\lvec(2.604 1.997) \lvec(2.616 2.000) \lvec(2.628 2.003)
\lvec(2.640 2.006) \lvec(2.652 2.009) \lvec(2.664 2.011)
\lvec(2.676 2.014) \lvec(2.688 2.017) \lvec(2.700 2.020)
\lvec(2.712 2.023) \lvec(2.724 2.026) \lvec(2.736 2.029)
\lvec(2.748 2.032) \lvec(2.760 2.034) \lvec(2.772 2.037)
\lvec(2.784 2.040) \lvec(2.796 2.043) \lvec(2.808 2.045)
\lvec(2.820 2.048) \lvec(2.832 2.051) \lvec(2.844 2.053)
\lvec(2.856 2.056) \lvec(2.868 2.059) \lvec(2.880 2.061)
\lvec(2.892 2.064) \lvec(2.904 2.066) \lvec(2.916 2.069)
\lvec(2.928 2.072) \lvec(2.940 2.074) \lvec(2.952 2.077)
\lvec(2.964 2.079) \lvec(2.976 2.082) \lvec(2.988 2.084)
\lvec(3.000 2.086) \lvec(3.012 2.089) \lvec(3.024 2.091)
\lvec(3.036 2.094) \lvec(3.048 2.096) \lvec(3.060 2.098)
\lvec(3.072 2.101) \lvec(3.084 2.103) \lvec(3.096 2.105)
\lvec(3.108 2.108) \lvec(3.120 2.110) \lvec(3.132 2.112)
\lvec(3.144 2.115) \lvec(3.156 2.117) \lvec(3.168 2.119)
\lvec(3.180 2.121) \lvec(3.192 2.124) \lvec(3.204 2.126)
\lvec(3.216 2.128) \lvec(3.228 2.130) \lvec(3.240 2.132)
\lvec(3.252 2.134) \lvec(3.264 2.136) \lvec(3.276 2.139)
\lvec(3.288 2.141) \lvec(3.300 2.143) \lvec(3.312 2.145)
\lvec(3.324 2.147) \lvec(3.336 2.149) \lvec(3.348 2.151)
\lvec(3.360 2.153) \lvec(3.372 2.155) \lvec(3.384 2.157)
\lvec(3.396 2.159) \lvec(3.408 2.161) \lvec(3.420 2.163)
\lvec(3.432 2.165) \lvec(3.444 2.167) \lvec(3.456 2.169)
\lvec(3.468 2.171) \lvec(3.480 2.172) \lvec(3.492 2.174)
\lvec(3.504 2.176) \lvec(3.516 2.178) \lvec(3.528 2.180)
\lvec(3.540 2.182) \lvec(3.552 2.184) \lvec(3.564 2.185)
\lvec(3.576 2.187) \lvec(3.588 2.189) \lvec(3.600 2.191)
\lvec(3.612 2.193) \lvec(3.624 2.194) \lvec(3.636 2.196)
\lvec(3.648 2.198) \lvec(3.660 2.200) \lvec(3.672 2.201)
\lvec(3.684 2.203) \lvec(3.696 2.205) \lvec(3.708 2.206)
\lvec(3.720 2.208) \lvec(3.732 2.210) \lvec(3.744 2.211)
\lvec(3.756 2.213) \lvec(3.768 2.215) \lvec(3.780 2.216)
\lvec(3.792 2.218) \lvec(3.804 2.220) \lvec(3.816 2.221)
\lvec(3.828 2.223) \lvec(3.840 2.224) \lvec(3.852 2.226)
\lvec(3.864 2.227) \lvec(3.876 2.229) \lvec(3.888 2.231)
\lvec(3.900 2.232) \lvec(3.912 2.234) \lvec(3.924 2.235)
\lvec(3.936 2.237) \lvec(3.948 2.238) \lvec(3.960 2.240)
\lvec(3.972 2.241) \lvec(3.984 2.243) \lvec(3.996 2.244)
\lvec(4.008 2.246) \lvec(4.020 2.247) \lvec(4.032 2.248)
\lvec(4.044 2.250) \lvec(4.056 2.251) \lvec(4.068 2.253)
\lvec(4.080 2.254) \lvec(4.092 2.256) \lvec(4.104 2.257)
\lvec(4.116 2.258) \lvec(4.128 2.260) \lvec(4.140 2.261)
\lvec(4.152 2.263) \lvec(4.164 2.264) \lvec(4.176 2.265)
\lvec(4.188 2.267) \lvec(4.200 2.268) \lvec(4.212 2.269)
\lvec(4.224 2.271) \lvec(4.236 2.272) \lvec(4.248 2.273)
\lvec(4.260 2.274) \lvec(4.272 2.276) \lvec(4.284 2.277)
\lvec(4.296 2.278) \lvec(4.308 2.280) \lvec(4.320 2.281)
\lvec(4.332 2.282) \lvec(4.344 2.283) \lvec(4.356 2.285)
\lvec(4.368 2.286) \lvec(4.380 2.287) \lvec(4.392 2.288)
\lvec(4.404 2.290) \lvec(4.416 2.291) \lvec(4.428 2.292)
\lvec(4.440 2.293) \lvec(4.452 2.294) \lvec(4.464 2.296)
\lvec(4.476 2.297) \lvec(4.488 2.298) \lvec(4.500 2.299)
\lvec(4.512 2.300) \lvec(4.524 2.301) \lvec(4.536 2.303)
\lvec(4.548 2.304) \lvec(4.560 2.305) \lvec(4.572 2.306)
\lvec(4.584 2.307) \lvec(4.596 2.308) \lvec(4.608 2.309)
\lvec(4.620 2.310) \lvec(4.632 2.312) \lvec(4.644 2.313)
\lvec(4.656 2.314) \lvec(4.668 2.315) \lvec(4.680 2.316)
\lvec(4.692 2.317) \lvec(4.704 2.318) \lvec(4.716 2.319)
\lvec(4.728 2.320) \lvec(4.740 2.321) \lvec(4.752 2.322)
\lvec(4.764 2.323) \lvec(4.776 2.324) \lvec(4.788 2.325)
\lvec(4.800 2.326) \lpatt()
\end{texdraw}
\hfill
%
%
\begin{texdraw}
\drawdim truecm \setgray 0
\linewd 0.04
\move (0 0) \lvec (4.8 0)
\lvec (4.8 3.0) \lvec (0 3.0) \lvec (0 0)
\linewd 0.02
\textref h:L v:T \htext (0.00 -.18) {0}
\move (1.200 0) \lvec (1.200 0.120)
\textref h:C v:T \htext (1.20 -.18) {1}
\move (2.400 0) \lvec (2.400 0.120)
\textref h:C v:T \htext (2.40 -.18) {2}
\move (3.600 0) \lvec (3.600 0.120)
\textref h:C v:T \htext (3.60 -.18) {3}
\move (4.800 0) \lvec (4.800 0.120)
\textref h:R v:T \htext (4.80 -.18) {4}
\move (4.8 0.273) \lvec (4.65 0.273)
\textref h:L v:C \htext (4.9 0.27) {0}
\move (4.8 1.500) \lvec (4.65 1.500)
\textref h:L v:C \htext (4.9 1.50) {$\frac{\pi}{4}$}
\textref h:L v:C \htext (4.9 2.73) {$\frac{\pi}{2}$}
\textref h:C v:T \htext (2.40 -.6) {$q_{\rm ch}/q_{\rm sm}$}
\textref h:L v:C \htext (.2 2.11) {$\alpha_0^{\rm opt}$}
\linewd 0.04
\move(0.000 0.273) \lvec(1.200 0.273) \lvec(1.200 2.727)
\lvec(4.800 2.727)
\linewd 0.02
\move(0.000 0.273) \lvec(0.938 0.273) \lvec(0.948 0.572)
\lvec(0.960 0.705) \lvec(0.972 0.805) \lvec(0.984 0.888)
\lvec(0.996 0.961) \lvec(1.008 1.026) \lvec(1.020 1.085)
\lvec(1.032 1.140) \lvec(1.044 1.191) \lvec(1.056 1.239)
\lvec(1.068 1.285) \lvec(1.080 1.327) \lvec(1.092 1.368)
\lvec(1.104 1.407) \lvec(1.116 1.444) \lvec(1.128 1.479)
\lvec(1.140 1.513) \lvec(1.152 1.545) \lvec(1.164 1.576)
\lvec(1.176 1.606) \lvec(1.188 1.635) \lvec(1.200 1.663)
\lvec(1.212 1.689) \lvec(1.224 1.715) \lvec(1.236 1.740)
\lvec(1.248 1.764) \lvec(1.260 1.787) \lvec(1.272 1.809)
\lvec(1.284 1.830) \lvec(1.296 1.851) \lvec(1.308 1.871)
\lvec(1.320 1.890) \lvec(1.332 1.909) \lvec(1.344 1.927)
\lvec(1.356 1.944) \lvec(1.368 1.961) \lvec(1.380 1.977)
\lvec(1.392 1.993) \lvec(1.404 2.008) \lvec(1.416 2.023)
\lvec(1.428 2.037) \lvec(1.440 2.050) \lvec(1.452 2.064)
\lvec(1.464 2.077) \lvec(1.476 2.089) \lvec(1.488 2.101)
\lvec(1.500 2.113) \lvec(1.512 2.124) \lvec(1.524 2.135)
\lvec(1.536 2.145) \lvec(1.548 2.155) \lvec(1.560 2.165)
\lvec(1.572 2.175) \lvec(1.584 2.184) \lvec(1.596 2.193)
\lvec(1.608 2.202) \lvec(1.620 2.211) \lvec(1.632 2.219)
\lvec(1.644 2.227) \lvec(1.656 2.235) \lvec(1.668 2.242)
\lvec(1.680 2.249) \lvec(1.692 2.256) \lvec(1.704 2.263)
\lvec(1.716 2.270) \lvec(1.728 2.277) \lvec(1.740 2.283)
\lvec(1.752 2.289) \lvec(1.764 2.295) \lvec(1.776 2.301)
\lvec(1.788 2.307) \lvec(1.800 2.312) \lvec(1.812 2.318)
\lvec(1.824 2.323) \lvec(1.836 2.328) \lvec(1.848 2.333)
\lvec(1.860 2.338) \lvec(1.872 2.343) \lvec(1.884 2.347)
\lvec(1.896 2.352) \lvec(1.908 2.356) \lvec(1.920 2.361)
\lvec(1.932 2.365) \lvec(1.944 2.369) \lvec(1.956 2.373)
\lvec(1.968 2.377) \lvec(1.980 2.381) \lvec(1.992 2.385)
\lvec(2.004 2.388) \lvec(2.016 2.392) \lvec(2.028 2.395)
\lvec(2.040 2.399) \lvec(2.052 2.402) \lvec(2.064 2.405)
\lvec(2.076 2.409) \lvec(2.088 2.412) \lvec(2.100 2.415)
\lvec(2.112 2.418) \lvec(2.124 2.421) \lvec(2.136 2.424)
\lvec(2.148 2.427) \lvec(2.160 2.429) \lvec(2.172 2.432)
\lvec(2.184 2.435) \lvec(2.196 2.437) \lvec(2.208 2.440)
\lvec(2.220 2.443) \lvec(2.232 2.445) \lvec(2.244 2.448)
\lvec(2.256 2.450) \lvec(2.268 2.452) \lvec(2.280 2.455)
\lvec(2.292 2.457) \lvec(2.304 2.459) \lvec(2.316 2.461)
\lvec(2.328 2.463) \lvec(2.340 2.466) \lvec(2.352 2.468)
\lvec(2.364 2.470) \lvec(2.376 2.472) \lvec(2.388 2.474)
\lvec(2.400 2.476) \lvec(2.412 2.478) \lvec(2.424 2.479)
\lvec(2.436 2.481) \lvec(2.448 2.483) \lvec(2.460 2.485)
\lvec(2.472 2.487) \lvec(2.484 2.488) \lvec(2.496 2.490)
\lvec(2.508 2.492) \lvec(2.520 2.494) \lvec(2.532 2.495)
\lvec(2.544 2.497) \lvec(2.556 2.498) \lvec(2.568 2.500)
\lvec(2.580 2.501) \lvec(2.592 2.503) \lvec(2.604 2.504)
\lvec(2.616 2.506) \lvec(2.628 2.507) \lvec(2.640 2.509)
\lvec(2.652 2.510) \lvec(2.664 2.512) \lvec(2.676 2.513)
\lvec(2.688 2.514) \lvec(2.700 2.516) \lvec(2.712 2.517)
\lvec(2.724 2.518) \lvec(2.736 2.520) \lvec(2.748 2.521)
\lvec(2.760 2.522) \lvec(2.772 2.523) \lvec(2.784 2.525)
\lvec(2.796 2.526) \lvec(2.808 2.527) \lvec(2.820 2.528)
\lvec(2.832 2.529) \lvec(2.844 2.530) \lvec(2.856 2.532)
\lvec(2.868 2.533) \lvec(2.880 2.534) \lvec(2.892 2.535)
\lvec(2.904 2.536) \lvec(2.916 2.537) \lvec(2.928 2.538)
\lvec(2.940 2.539) \lvec(2.952 2.540) \lvec(2.964 2.541)
\lvec(2.976 2.542) \lvec(2.988 2.543) \lvec(3.000 2.544)
\lvec(3.012 2.545) \lvec(3.024 2.546) \lvec(3.036 2.547)
\lvec(3.048 2.548) \lvec(3.060 2.549) \lvec(3.072 2.550)
\lvec(3.084 2.551) \lvec(3.096 2.552) \lvec(3.108 2.553)
\lvec(3.120 2.553) \lvec(3.132 2.554) \lvec(3.144 2.555)
\lvec(3.156 2.556) \lvec(3.168 2.557) \lvec(3.180 2.558)
\lvec(3.192 2.558) \lvec(3.204 2.559) \lvec(3.216 2.560)
\lvec(3.228 2.561) \lvec(3.240 2.562) \lvec(3.252 2.562)
\lvec(3.264 2.563) \lvec(3.276 2.564) \lvec(3.288 2.565)
\lvec(3.300 2.565) \lvec(3.312 2.566) \lvec(3.324 2.567)
\lvec(3.336 2.568) \lvec(3.348 2.568) \lvec(3.360 2.569)
\lvec(3.372 2.570) \lvec(3.384 2.571) \lvec(3.396 2.571)
\lvec(3.408 2.572) \lvec(3.420 2.573) \lvec(3.432 2.573)
\lvec(3.444 2.574) \lvec(3.456 2.575) \lvec(3.468 2.575)
\lvec(3.480 2.576) \lvec(3.492 2.577) \lvec(3.504 2.577)
\lvec(3.516 2.578) \lvec(3.528 2.578) \lvec(3.540 2.579)
\lvec(3.552 2.580) \lvec(3.564 2.580) \lvec(3.576 2.581)
\lvec(3.588 2.582) \lvec(3.600 2.582) \lvec(3.612 2.583)
\lvec(3.624 2.583) \lvec(3.636 2.584) \lvec(3.648 2.584)
\lvec(3.660 2.585) \lvec(3.672 2.586) \lvec(3.684 2.586)
\lvec(3.696 2.587) \lvec(3.708 2.587) \lvec(3.720 2.588)
\lvec(3.732 2.588) \lvec(3.744 2.589) \lvec(3.756 2.589)
\lvec(3.768 2.590) \lvec(3.780 2.590) \lvec(3.792 2.591)
\lvec(3.804 2.591) \lvec(3.816 2.592) \lvec(3.828 2.593)
\lvec(3.840 2.593) \lvec(3.852 2.594) \lvec(3.864 2.594)
\lvec(3.876 2.594) \lvec(3.888 2.595) \lvec(3.900 2.595)
\lvec(3.912 2.596) \lvec(3.924 2.596) \lvec(3.936 2.597)
\lvec(3.948 2.597) \lvec(3.960 2.598) \lvec(3.972 2.598)
\lvec(3.984 2.599) \lvec(3.996 2.599) \lvec(4.008 2.600)
\lvec(4.020 2.600) \lvec(4.032 2.601) \lvec(4.044 2.601)
\lvec(4.056 2.601) \lvec(4.068 2.602) \lvec(4.080 2.602)
\lvec(4.092 2.603) \lvec(4.104 2.603) \lvec(4.116 2.604)
\lvec(4.128 2.604) \lvec(4.140 2.604) \lvec(4.152 2.605)
\lvec(4.164 2.605) \lvec(4.176 2.606) \lvec(4.188 2.606)
\lvec(4.200 2.606) \lvec(4.212 2.607) \lvec(4.224 2.607)
\lvec(4.236 2.608) \lvec(4.248 2.608) \lvec(4.260 2.608)
\lvec(4.272 2.609) \lvec(4.284 2.609) \lvec(4.296 2.610)
\lvec(4.308 2.610) \lvec(4.320 2.610) \lvec(4.332 2.611)
\lvec(4.344 2.611) \lvec(4.356 2.611) \lvec(4.368 2.612)
\lvec(4.380 2.612) \lvec(4.392 2.613) \lvec(4.404 2.613)
\lvec(4.416 2.613) \lvec(4.428 2.614) \lvec(4.440 2.614)
\lvec(4.452 2.614) \lvec(4.464 2.615) \lvec(4.476 2.615)
\lvec(4.488 2.615) \lvec(4.500 2.616) \lvec(4.512 2.616)
\lvec(4.524 2.616) \lvec(4.536 2.617) \lvec(4.548 2.617)
\lvec(4.560 2.617) \lvec(4.572 2.618) \lvec(4.584 2.618)
\lvec(4.596 2.618) \lvec(4.608 2.619) \lvec(4.620 2.619)
\lvec(4.632 2.619) \lvec(4.644 2.620) \lvec(4.656 2.620)
\lvec(4.668 2.620) \lvec(4.680 2.621) \lvec(4.692 2.621)
\lvec(4.704 2.621) \lvec(4.716 2.621) \lvec(4.728 2.622)
\lvec(4.740 2.622) \lvec(4.752 2.622) \lvec(4.764 2.623)
\lvec(4.776 2.623) \lvec(4.788 2.623) \lvec(4.800 2.624)
\move(0.000 0.273) \lvec(0.742 0.273) \lvec(0.744 0.371)
\lvec(0.756 0.514) \lvec(0.768 0.600) \lvec(0.780 0.667)
\lvec(0.792 0.724) \lvec(0.804 0.774) \lvec(0.816 0.819)
\lvec(0.828 0.861) \lvec(0.840 0.900) \lvec(0.852 0.936)
\lvec(0.864 0.971) \lvec(0.876 1.003) \lvec(0.888 1.034)
\lvec(0.900 1.064) \lvec(0.912 1.092) \lvec(0.924 1.120)
\lvec(0.936 1.146) \lvec(0.948 1.171) \lvec(0.960 1.196)
\lvec(0.972 1.219) \lvec(0.984 1.242) \lvec(0.996 1.265)
\lvec(1.008 1.286) \lvec(1.020 1.308) \lvec(1.032 1.328)
\lvec(1.044 1.348) \lvec(1.056 1.368) \lvec(1.068 1.387)
\lvec(1.080 1.405) \lvec(1.092 1.423) \lvec(1.104 1.441)
\lvec(1.116 1.458) \lvec(1.128 1.475) \lvec(1.140 1.492)
\lvec(1.152 1.508) \lvec(1.164 1.524) \lvec(1.176 1.539)
\lvec(1.188 1.554) \lvec(1.200 1.569) \lvec(1.212 1.584)
\lvec(1.224 1.598) \lvec(1.236 1.612) \lvec(1.248 1.626)
\lvec(1.260 1.639) \lvec(1.272 1.652) \lvec(1.284 1.665)
\lvec(1.296 1.678) \lvec(1.308 1.690) \lvec(1.320 1.703)
\lvec(1.332 1.715) \lvec(1.344 1.726) \lvec(1.356 1.738)
\lvec(1.368 1.749) \lvec(1.380 1.760) \lvec(1.392 1.771)
\lvec(1.404 1.782) \lvec(1.416 1.793) \lvec(1.428 1.803)
\lvec(1.440 1.813) \lvec(1.452 1.823) \lvec(1.464 1.833)
\lvec(1.476 1.843) \lvec(1.488 1.852) \lvec(1.500 1.861)
\lvec(1.512 1.870) \lvec(1.524 1.879) \lvec(1.536 1.888)
\lvec(1.548 1.897) \lvec(1.560 1.906) \lvec(1.572 1.914)
\lvec(1.584 1.922) \lvec(1.596 1.930) \lvec(1.608 1.938)
\lvec(1.620 1.946) \lvec(1.632 1.954) \lvec(1.644 1.961)
\lvec(1.656 1.969) \lvec(1.668 1.976) \lvec(1.680 1.983)
\lvec(1.692 1.990) \lvec(1.704 1.997) \lvec(1.716 2.004)
\lvec(1.728 2.011) \lvec(1.740 2.018) \lvec(1.752 2.024)
\lvec(1.764 2.031) \lvec(1.776 2.037) \lvec(1.788 2.043)
\lvec(1.800 2.049) \lvec(1.812 2.055) \lvec(1.824 2.061)
\lvec(1.836 2.067) \lvec(1.848 2.073) \lvec(1.860 2.079)
\lvec(1.872 2.084) \lvec(1.884 2.090) \lvec(1.896 2.095)
\lvec(1.908 2.100) \lvec(1.920 2.106) \lvec(1.932 2.111)
\lvec(1.944 2.116) \lvec(1.956 2.121) \lvec(1.968 2.126)
\lvec(1.980 2.131) \lvec(1.992 2.135) \lvec(2.004 2.140)
\lvec(2.016 2.145) \lvec(2.028 2.149) \lvec(2.040 2.154)
\lvec(2.052 2.158) \lvec(2.064 2.163) \lvec(2.076 2.167)
\lvec(2.088 2.171) \lvec(2.100 2.175) \lvec(2.112 2.180)
\lvec(2.124 2.184) \lvec(2.136 2.188) \lvec(2.148 2.192)
\lvec(2.160 2.196) \lvec(2.172 2.199) \lvec(2.184 2.203)
\lvec(2.196 2.207) \lvec(2.208 2.211) \lvec(2.220 2.214)
\lvec(2.232 2.218) \lvec(2.244 2.221) \lvec(2.256 2.225)
\lvec(2.268 2.228) \lvec(2.280 2.232) \lvec(2.292 2.235)
\lvec(2.304 2.239) \lvec(2.316 2.242) \lvec(2.328 2.245)
\lvec(2.340 2.248) \lvec(2.352 2.251) \lvec(2.364 2.254)
\lvec(2.376 2.258) \lvec(2.388 2.261) \lvec(2.400 2.264)
\lvec(2.412 2.267) \lvec(2.424 2.269) \lvec(2.436 2.272)
\lvec(2.448 2.275) \lvec(2.460 2.278) \lvec(2.472 2.281)
\lvec(2.484 2.284) \lvec(2.496 2.286) \lvec(2.508 2.289)
\lvec(2.520 2.292) \lvec(2.532 2.294) \lvec(2.544 2.297)
\lvec(2.556 2.299) \lvec(2.568 2.302) \lvec(2.580 2.304)
\lvec(2.592 2.307) \lvec(2.604 2.309) \lvec(2.616 2.312)
\lvec(2.628 2.314) \lvec(2.640 2.316) \lvec(2.652 2.319)
\lvec(2.664 2.321) \lvec(2.676 2.323) \lvec(2.688 2.325)
\lvec(2.700 2.328) \lvec(2.712 2.330) \lvec(2.724 2.332)
\lvec(2.736 2.334) \lvec(2.748 2.336) \lvec(2.760 2.338)
\lvec(2.772 2.341) \lvec(2.784 2.343) \lvec(2.796 2.345)
\lvec(2.808 2.347) \lvec(2.820 2.349) \lvec(2.832 2.351)
\lvec(2.844 2.353) \lvec(2.856 2.354) \lvec(2.868 2.356)
\lvec(2.880 2.358) \lvec(2.892 2.360) \lvec(2.904 2.362)
\lvec(2.916 2.364) \lvec(2.928 2.366) \lvec(2.940 2.367)
\lvec(2.952 2.369) \lvec(2.964 2.371) \lvec(2.976 2.373)
\lvec(2.988 2.374) \lvec(3.000 2.376) \lvec(3.012 2.378)
\lvec(3.024 2.380) \lvec(3.036 2.381) \lvec(3.048 2.383)
\lvec(3.060 2.384) \lvec(3.072 2.386) \lvec(3.084 2.388)
\lvec(3.096 2.389) \lvec(3.108 2.391) \lvec(3.120 2.392)
\lvec(3.132 2.394) \lvec(3.144 2.395) \lvec(3.156 2.397)
\lvec(3.168 2.398) \lvec(3.180 2.400) \lvec(3.192 2.401)
\lvec(3.204 2.403) \lvec(3.216 2.404) \lvec(3.228 2.406)
\lvec(3.240 2.407) \lvec(3.252 2.408) \lvec(3.264 2.410)
\lvec(3.276 2.411) \lvec(3.288 2.413) \lvec(3.300 2.414)
\lvec(3.312 2.415) \lvec(3.324 2.417) \lvec(3.336 2.418)
\lvec(3.348 2.419) \lvec(3.360 2.421) \lvec(3.372 2.422)
\lvec(3.384 2.423) \lvec(3.396 2.424) \lvec(3.408 2.426)
\lvec(3.420 2.427) \lvec(3.432 2.428) \lvec(3.444 2.429)
\lvec(3.456 2.430) \lvec(3.468 2.432) \lvec(3.480 2.433)
\lvec(3.492 2.434) \lvec(3.504 2.435) \lvec(3.516 2.436)
\lvec(3.528 2.438) \lvec(3.540 2.439) \lvec(3.552 2.440)
\lvec(3.564 2.441) \lvec(3.576 2.442) \lvec(3.588 2.443)
\lvec(3.600 2.444) \lvec(3.612 2.445) \lvec(3.624 2.446)
\lvec(3.636 2.447) \lvec(3.648 2.449) \lvec(3.660 2.450)
\lvec(3.672 2.451) \lvec(3.684 2.452) \lvec(3.696 2.453)
\lvec(3.708 2.454) \lvec(3.720 2.455) \lvec(3.732 2.456)
\lvec(3.744 2.457) \lvec(3.756 2.458) \lvec(3.768 2.459)
\lvec(3.780 2.460) \lvec(3.792 2.461) \lvec(3.804 2.462)
\lvec(3.816 2.463) \lvec(3.828 2.464) \lvec(3.840 2.464)
\lvec(3.852 2.465) \lvec(3.864 2.466) \lvec(3.876 2.467)
\lvec(3.888 2.468) \lvec(3.900 2.469) \lvec(3.912 2.470)
\lvec(3.924 2.471) \lvec(3.936 2.472) \lvec(3.948 2.473)
\lvec(3.960 2.473) \lvec(3.972 2.474) \lvec(3.984 2.475)
\lvec(3.996 2.476) \lvec(4.008 2.477) \lvec(4.020 2.478)
\lvec(4.032 2.479) \lvec(4.044 2.479) \lvec(4.056 2.480)
\lvec(4.068 2.481) \lvec(4.080 2.482) \lvec(4.092 2.483)
\lvec(4.104 2.483) \lvec(4.116 2.484) \lvec(4.128 2.485)
\lvec(4.140 2.486) \lvec(4.152 2.487) \lvec(4.164 2.487)
\lvec(4.176 2.488) \lvec(4.188 2.489) \lvec(4.200 2.490)
\lvec(4.212 2.490) \lvec(4.224 2.491) \lvec(4.236 2.492)
\lvec(4.248 2.493) \lvec(4.260 2.493) \lvec(4.272 2.494)
\lvec(4.284 2.495) \lvec(4.296 2.496) \lvec(4.308 2.496)
\lvec(4.320 2.497) \lvec(4.332 2.498) \lvec(4.344 2.498)
\lvec(4.356 2.499) \lvec(4.368 2.500) \lvec(4.380 2.501)
\lvec(4.392 2.501) \lvec(4.404 2.502) \lvec(4.416 2.503)
\lvec(4.428 2.503) \lvec(4.440 2.504) \lvec(4.452 2.505)
\lvec(4.464 2.505) \lvec(4.476 2.506) \lvec(4.488 2.507)
\lvec(4.500 2.507) \lvec(4.512 2.508) \lvec(4.524 2.509)
\lvec(4.536 2.509) \lvec(4.548 2.510) \lvec(4.560 2.510)
\lvec(4.572 2.511) \lvec(4.584 2.512) \lvec(4.596 2.512)
\lvec(4.608 2.513) \lvec(4.620 2.514) \lvec(4.632 2.514)
\lvec(4.644 2.515) \lvec(4.656 2.515) \lvec(4.668 2.516)
\lvec(4.680 2.517) \lvec(4.692 2.517) \lvec(4.704 2.518)
\lvec(4.716 2.518) \lvec(4.728 2.519) \lvec(4.740 2.519)
\lvec(4.752 2.520) \lvec(4.764 2.521) \lvec(4.776 2.521)
\lvec(4.788 2.522) \lvec(4.800 2.522)
\move(0.000 0.273) \lvec(0.600 0.273) \lvec(0.612 0.453)
\lvec(0.624 0.527) \lvec(0.636 0.585) \lvec(0.648 0.632)
\lvec(0.660 0.675) \lvec(0.672 0.713) \lvec(0.684 0.747)
\lvec(0.696 0.780) \lvec(0.708 0.810) \lvec(0.720 0.839)
\lvec(0.732 0.866) \lvec(0.744 0.892) \lvec(0.756 0.916)
\lvec(0.768 0.940) \lvec(0.780 0.963) \lvec(0.792 0.985)
\lvec(0.804 1.006) \lvec(0.816 1.027) \lvec(0.828 1.046)
\lvec(0.840 1.066) \lvec(0.852 1.085) \lvec(0.864 1.103)
\lvec(0.876 1.121) \lvec(0.888 1.138) \lvec(0.900 1.155)
\lvec(0.912 1.172) \lvec(0.924 1.188) \lvec(0.936 1.204)
\lvec(0.948 1.219) \lvec(0.960 1.234) \lvec(0.972 1.249)
\lvec(0.984 1.264) \lvec(0.996 1.278) \lvec(1.008 1.292)
\lvec(1.020 1.306) \lvec(1.032 1.319) \lvec(1.044 1.333)
\lvec(1.056 1.346) \lvec(1.068 1.359) \lvec(1.080 1.371)
\lvec(1.092 1.384) \lvec(1.104 1.396) \lvec(1.116 1.408)
\lvec(1.128 1.420) \lvec(1.140 1.431) \lvec(1.152 1.443)
\lvec(1.164 1.454) \lvec(1.176 1.465) \lvec(1.188 1.476)
\lvec(1.200 1.487) \lvec(1.212 1.497) \lvec(1.224 1.508)
\lvec(1.236 1.518) \lvec(1.248 1.528) \lvec(1.260 1.538)
\lvec(1.272 1.548) \lvec(1.284 1.558) \lvec(1.296 1.567)
\lvec(1.308 1.577) \lvec(1.320 1.586) \lvec(1.332 1.595)
\lvec(1.344 1.604) \lvec(1.356 1.613) \lvec(1.368 1.622)
\lvec(1.380 1.631) \lvec(1.392 1.640) \lvec(1.404 1.648)
\lvec(1.416 1.656) \lvec(1.428 1.665) \lvec(1.440 1.673)
\lvec(1.452 1.681) \lvec(1.464 1.689) \lvec(1.476 1.697)
\lvec(1.488 1.705) \lvec(1.500 1.712) \lvec(1.512 1.720)
\lvec(1.524 1.727) \lvec(1.536 1.735) \lvec(1.548 1.742)
\lvec(1.560 1.749) \lvec(1.572 1.756) \lvec(1.584 1.764)
\lvec(1.596 1.770) \lvec(1.608 1.777) \lvec(1.620 1.784)
\lvec(1.632 1.791) \lvec(1.644 1.798) \lvec(1.656 1.804)
\lvec(1.668 1.811) \lvec(1.680 1.817) \lvec(1.692 1.823)
\lvec(1.704 1.830) \lvec(1.716 1.836) \lvec(1.728 1.842)
\lvec(1.740 1.848) \lvec(1.752 1.854) \lvec(1.764 1.860)
\lvec(1.776 1.866) \lvec(1.788 1.871) \lvec(1.800 1.877)
\lvec(1.812 1.883) \lvec(1.824 1.888) \lvec(1.836 1.894)
\lvec(1.848 1.899) \lvec(1.860 1.905) \lvec(1.872 1.910)
\lvec(1.884 1.915) \lvec(1.896 1.921) \lvec(1.908 1.926)
\lvec(1.920 1.931) \lvec(1.932 1.936) \lvec(1.944 1.941)
\lvec(1.956 1.946) \lvec(1.968 1.951) \lvec(1.980 1.955)
\lvec(1.992 1.960) \lvec(2.004 1.965) \lvec(2.016 1.970)
\lvec(2.028 1.974) \lvec(2.040 1.979) \lvec(2.052 1.983)
\lvec(2.064 1.988) \lvec(2.076 1.992) \lvec(2.088 1.997)
\lvec(2.100 2.001) \lvec(2.112 2.005) \lvec(2.124 2.010)
\lvec(2.136 2.014) \lvec(2.148 2.018) \lvec(2.160 2.022)
\lvec(2.172 2.026) \lvec(2.184 2.030) \lvec(2.196 2.034)
\lvec(2.208 2.038) \lvec(2.220 2.042) \lvec(2.232 2.046)
\lvec(2.244 2.050) \lvec(2.256 2.054) \lvec(2.268 2.057)
\lvec(2.280 2.061) \lvec(2.292 2.065) \lvec(2.304 2.068)
\lvec(2.316 2.072) \lvec(2.328 2.075) \lvec(2.340 2.079)
\lvec(2.352 2.083) \lvec(2.364 2.086) \lvec(2.376 2.089)
\lvec(2.388 2.093) \lvec(2.400 2.096) \lvec(2.412 2.100)
\lvec(2.424 2.103) \lvec(2.436 2.106) \lvec(2.448 2.109)
\lvec(2.460 2.113) \lvec(2.472 2.116) \lvec(2.484 2.119)
\lvec(2.496 2.122) \lvec(2.508 2.125) \lvec(2.520 2.128)
\lvec(2.532 2.131) \lvec(2.544 2.134) \lvec(2.556 2.137)
\lvec(2.568 2.140) \lvec(2.580 2.143) \lvec(2.592 2.146)
\lvec(2.604 2.149) \lvec(2.616 2.152) \lvec(2.628 2.154)
\lvec(2.640 2.157) \lvec(2.652 2.160) \lvec(2.664 2.163)
\lvec(2.676 2.165) \lvec(2.688 2.168) \lvec(2.700 2.171)
\lvec(2.712 2.173) \lvec(2.724 2.176) \lvec(2.736 2.179)
\lvec(2.748 2.181) \lvec(2.760 2.184) \lvec(2.772 2.186)
\lvec(2.784 2.189) \lvec(2.796 2.191) \lvec(2.808 2.194)
\lvec(2.820 2.196) \lvec(2.832 2.198) \lvec(2.844 2.201)
\lvec(2.856 2.203) \lvec(2.868 2.206) \lvec(2.880 2.208)
\lvec(2.892 2.210) \lvec(2.904 2.213) \lvec(2.916 2.215)
\lvec(2.928 2.217) \lvec(2.940 2.219) \lvec(2.952 2.221)
\lvec(2.964 2.224) \lvec(2.976 2.226) \lvec(2.988 2.228)
\lvec(3.000 2.230) \lvec(3.012 2.232) \lvec(3.024 2.234)
\lvec(3.036 2.236) \lvec(3.048 2.239) \lvec(3.060 2.241)
\lvec(3.072 2.243) \lvec(3.084 2.245) \lvec(3.096 2.247)
\lvec(3.108 2.249) \lvec(3.120 2.251) \lvec(3.132 2.253)
\lvec(3.144 2.255) \lvec(3.156 2.256) \lvec(3.168 2.258)
\lvec(3.180 2.260) \lvec(3.192 2.262) \lvec(3.204 2.264)
\lvec(3.216 2.266) \lvec(3.228 2.268) \lvec(3.240 2.270)
\lvec(3.252 2.271) \lvec(3.264 2.273) \lvec(3.276 2.275)
\lvec(3.288 2.277) \lvec(3.300 2.278) \lvec(3.312 2.280)
\lvec(3.324 2.282) \lvec(3.336 2.284) \lvec(3.348 2.285)
\lvec(3.360 2.287) \lvec(3.372 2.289) \lvec(3.384 2.290)
\lvec(3.396 2.292) \lvec(3.408 2.294) \lvec(3.420 2.295)
\lvec(3.432 2.297) \lvec(3.444 2.299) \lvec(3.456 2.300)
\lvec(3.468 2.302) \lvec(3.480 2.303) \lvec(3.492 2.305)
\lvec(3.504 2.306) \lvec(3.516 2.308) \lvec(3.528 2.309)
\lvec(3.540 2.311) \lvec(3.552 2.312) \lvec(3.564 2.314)
\lvec(3.576 2.315) \lvec(3.588 2.317) \lvec(3.600 2.318)
\lvec(3.612 2.320) \lvec(3.624 2.321) \lvec(3.636 2.323)
\lvec(3.648 2.324) \lvec(3.660 2.325) \lvec(3.672 2.327)
\lvec(3.684 2.328) \lvec(3.696 2.330) \lvec(3.708 2.331)
\lvec(3.720 2.332) \lvec(3.732 2.334) \lvec(3.744 2.335)
\lvec(3.756 2.336) \lvec(3.768 2.338) \lvec(3.780 2.339)
\lvec(3.792 2.340) \lvec(3.804 2.342) \lvec(3.816 2.343)
\lvec(3.828 2.344) \lvec(3.840 2.346) \lvec(3.852 2.347)
\lvec(3.864 2.348) \lvec(3.876 2.349) \lvec(3.888 2.351)
\lvec(3.900 2.352) \lvec(3.912 2.353) \lvec(3.924 2.354)
\lvec(3.936 2.355) \lvec(3.948 2.357) \lvec(3.960 2.358)
\lvec(3.972 2.359) \lvec(3.984 2.360) \lvec(3.996 2.361)
\lvec(4.008 2.363) \lvec(4.020 2.364) \lvec(4.032 2.365)
\lvec(4.044 2.366) \lvec(4.056 2.367) \lvec(4.068 2.368)
\lvec(4.080 2.369) \lvec(4.092 2.370) \lvec(4.104 2.372)
\lvec(4.116 2.373) \lvec(4.128 2.374) \lvec(4.140 2.375)
\lvec(4.152 2.376) \lvec(4.164 2.377) \lvec(4.176 2.378)
\lvec(4.188 2.379) \lvec(4.200 2.380) \lvec(4.212 2.381)
\lvec(4.224 2.382) \lvec(4.236 2.383) \lvec(4.248 2.384)
\lvec(4.260 2.385) \lvec(4.272 2.386) \lvec(4.284 2.387)
\lvec(4.296 2.388) \lvec(4.308 2.389) \lvec(4.320 2.390)
\lvec(4.332 2.391) \lvec(4.344 2.392) \lvec(4.356 2.393)
\lvec(4.368 2.394) \lvec(4.380 2.395) \lvec(4.392 2.396)
\lvec(4.404 2.397) \lvec(4.416 2.398) \lvec(4.428 2.399)
\lvec(4.440 2.400) \lvec(4.452 2.401) \lvec(4.464 2.402)
\lvec(4.476 2.403) \lvec(4.488 2.404) \lvec(4.500 2.405)
\lvec(4.512 2.405) \lvec(4.524 2.406) \lvec(4.536 2.407)
\lvec(4.548 2.408) \lvec(4.560 2.409) \lvec(4.572 2.410)
\lvec(4.584 2.411) \lvec(4.596 2.412) \lvec(4.608 2.413)
\lvec(4.620 2.413) \lvec(4.632 2.414) \lvec(4.644 2.415)
\lvec(4.656 2.416) \lvec(4.668 2.417) \lvec(4.680 2.418)
\lvec(4.692 2.418) \lvec(4.704 2.419) \lvec(4.716 2.420)
\lvec(4.728 2.421) \lvec(4.740 2.422) \lvec(4.752 2.423)
\lvec(4.764 2.423) \lvec(4.776 2.424) \lvec(4.788 2.425)
\lvec(4.800 2.426)
\lpatt (.03 .1)
\move(0.000 0.273) \lvec(0.497 0.273) \lvec(0.504 0.392)
\lvec(0.516 0.469) \lvec(0.528 0.523) \lvec(0.540 0.568)
\lvec(0.552 0.606) \lvec(0.564 0.640) \lvec(0.576 0.672)
\lvec(0.588 0.701) \lvec(0.600 0.728) \lvec(0.612 0.753)
\lvec(0.624 0.777) \lvec(0.636 0.800) \lvec(0.648 0.822)
\lvec(0.660 0.843) \lvec(0.672 0.863) \lvec(0.684 0.883)
\lvec(0.696 0.902) \lvec(0.708 0.920) \lvec(0.720 0.938)
\lvec(0.732 0.955) \lvec(0.744 0.971) \lvec(0.756 0.988)
\lvec(0.768 1.004) \lvec(0.780 1.019) \lvec(0.792 1.034)
\lvec(0.804 1.049) \lvec(0.816 1.063) \lvec(0.828 1.077)
\lvec(0.840 1.091) \lvec(0.852 1.105) \lvec(0.864 1.118)
\lvec(0.876 1.131) \lvec(0.888 1.144) \lvec(0.900 1.156)
\lvec(0.912 1.169) \lvec(0.924 1.181) \lvec(0.936 1.193)
\lvec(0.948 1.204) \lvec(0.960 1.216) \lvec(0.972 1.227)
\lvec(0.984 1.238) \lvec(0.996 1.249) \lvec(1.008 1.260)
\lvec(1.020 1.271) \lvec(1.032 1.281) \lvec(1.044 1.292)
\lvec(1.056 1.302) \lvec(1.068 1.312) \lvec(1.080 1.322)
\lvec(1.092 1.332) \lvec(1.104 1.342) \lvec(1.116 1.351)
\lvec(1.128 1.361) \lvec(1.140 1.370) \lvec(1.152 1.379)
\lvec(1.164 1.388) \lvec(1.176 1.397) \lvec(1.188 1.406)
\lvec(1.200 1.415) \lvec(1.212 1.423) \lvec(1.224 1.432)
\lvec(1.236 1.440) \lvec(1.248 1.449) \lvec(1.260 1.457)
\lvec(1.272 1.465) \lvec(1.284 1.473) \lvec(1.296 1.481)
\lvec(1.308 1.489) \lvec(1.320 1.497) \lvec(1.332 1.504)
\lvec(1.344 1.512) \lvec(1.356 1.520) \lvec(1.368 1.527)
\lvec(1.380 1.534) \lvec(1.392 1.542) \lvec(1.404 1.549)
\lvec(1.416 1.556) \lvec(1.428 1.563) \lvec(1.440 1.570)
\lvec(1.452 1.577) \lvec(1.464 1.584) \lvec(1.476 1.591)
\lvec(1.488 1.598) \lvec(1.500 1.604) \lvec(1.512 1.611)
\lvec(1.524 1.617) \lvec(1.536 1.624) \lvec(1.548 1.630)
\lvec(1.560 1.637) \lvec(1.572 1.643) \lvec(1.584 1.649)
\lvec(1.596 1.655) \lvec(1.608 1.661) \lvec(1.620 1.667)
\lvec(1.632 1.673) \lvec(1.644 1.679) \lvec(1.656 1.685)
\lvec(1.668 1.691) \lvec(1.680 1.697) \lvec(1.692 1.702)
\lvec(1.704 1.708) \lvec(1.716 1.714) \lvec(1.728 1.719)
\lvec(1.740 1.725) \lvec(1.752 1.730) \lvec(1.764 1.735)
\lvec(1.776 1.741) \lvec(1.788 1.746) \lvec(1.800 1.751)
\lvec(1.812 1.757) \lvec(1.824 1.762) \lvec(1.836 1.767)
\lvec(1.848 1.772) \lvec(1.860 1.777) \lvec(1.872 1.782)
\lvec(1.884 1.787) \lvec(1.896 1.792) \lvec(1.908 1.796)
\lvec(1.920 1.801) \lvec(1.932 1.806) \lvec(1.944 1.811)
\lvec(1.956 1.815) \lvec(1.968 1.820) \lvec(1.980 1.825)
\lvec(1.992 1.829) \lvec(2.004 1.834) \lvec(2.016 1.838)
\lvec(2.028 1.843) \lvec(2.040 1.847) \lvec(2.052 1.851)
\lvec(2.064 1.856) \lvec(2.076 1.860) \lvec(2.088 1.864)
\lvec(2.100 1.868) \lvec(2.112 1.873) \lvec(2.124 1.877)
\lvec(2.136 1.881) \lvec(2.148 1.885) \lvec(2.160 1.889)
\lvec(2.172 1.893) \lvec(2.184 1.897) \lvec(2.196 1.901)
\lvec(2.208 1.905) \lvec(2.220 1.909) \lvec(2.232 1.913)
\lvec(2.244 1.916) \lvec(2.256 1.920) \lvec(2.268 1.924)
\lvec(2.280 1.928) \lvec(2.292 1.931) \lvec(2.304 1.935)
\lvec(2.316 1.939) \lvec(2.328 1.942) \lvec(2.340 1.946)
\lvec(2.352 1.949) \lvec(2.364 1.953) \lvec(2.376 1.956)
\lvec(2.388 1.960) \lvec(2.400 1.963) \lvec(2.412 1.967)
\lvec(2.424 1.970) \lvec(2.436 1.974) \lvec(2.448 1.977)
\lvec(2.460 1.980) \lvec(2.472 1.983) \lvec(2.484 1.987)
\lvec(2.496 1.990) \lvec(2.508 1.993) \lvec(2.520 1.996)
\lvec(2.532 1.999) \lvec(2.544 2.003) \lvec(2.556 2.006)
\lvec(2.568 2.009) \lvec(2.580 2.012) \lvec(2.592 2.015)
\lvec(2.604 2.018) \lvec(2.616 2.021) \lvec(2.628 2.024)
\lvec(2.640 2.027) \lvec(2.652 2.030) \lvec(2.664 2.033)
\lvec(2.676 2.036) \lvec(2.688 2.038) \lvec(2.700 2.041)
\lvec(2.712 2.044) \lvec(2.724 2.047) \lvec(2.736 2.050)
\lvec(2.748 2.052) \lvec(2.760 2.055) \lvec(2.772 2.058)
\lvec(2.784 2.060) \lvec(2.796 2.063) \lvec(2.808 2.066)
\lvec(2.820 2.068) \lvec(2.832 2.071) \lvec(2.844 2.074)
\lvec(2.856 2.076) \lvec(2.868 2.079) \lvec(2.880 2.081)
\lvec(2.892 2.084) \lvec(2.904 2.086) \lvec(2.916 2.089)
\lvec(2.928 2.091) \lvec(2.940 2.094) \lvec(2.952 2.096)
\lvec(2.964 2.099) \lvec(2.976 2.101) \lvec(2.988 2.103)
\lvec(3.000 2.106) \lvec(3.012 2.108) \lvec(3.024 2.110)
\lvec(3.036 2.113) \lvec(3.048 2.115) \lvec(3.060 2.117)
\lvec(3.072 2.120) \lvec(3.084 2.122) \lvec(3.096 2.124)
\lvec(3.108 2.126) \lvec(3.120 2.129) \lvec(3.132 2.131)
\lvec(3.144 2.133) \lvec(3.156 2.135) \lvec(3.168 2.137)
\lvec(3.180 2.139) \lvec(3.192 2.142) \lvec(3.204 2.144)
\lvec(3.216 2.146) \lvec(3.228 2.148) \lvec(3.240 2.150)
\lvec(3.252 2.152) \lvec(3.264 2.154) \lvec(3.276 2.156)
\lvec(3.288 2.158) \lvec(3.300 2.160) \lvec(3.312 2.162)
\lvec(3.324 2.164) \lvec(3.336 2.166) \lvec(3.348 2.168)
\lvec(3.360 2.170) \lvec(3.372 2.172) \lvec(3.384 2.174)
\lvec(3.396 2.176) \lvec(3.408 2.177) \lvec(3.420 2.179)
\lvec(3.432 2.181) \lvec(3.444 2.183) \lvec(3.456 2.185)
\lvec(3.468 2.187) \lvec(3.480 2.188) \lvec(3.492 2.190)
\lvec(3.504 2.192) \lvec(3.516 2.194) \lvec(3.528 2.196)
\lvec(3.540 2.197) \lvec(3.552 2.199) \lvec(3.564 2.201)
\lvec(3.576 2.203) \lvec(3.588 2.204) \lvec(3.600 2.206)
\lvec(3.612 2.208) \lvec(3.624 2.209) \lvec(3.636 2.211)
\lvec(3.648 2.213) \lvec(3.660 2.214) \lvec(3.672 2.216)
\lvec(3.684 2.218) \lvec(3.696 2.219) \lvec(3.708 2.221)
\lvec(3.720 2.222) \lvec(3.732 2.224) \lvec(3.744 2.226)
\lvec(3.756 2.227) \lvec(3.768 2.229) \lvec(3.780 2.230)
\lvec(3.792 2.232) \lvec(3.804 2.233) \lvec(3.816 2.235)
\lvec(3.828 2.236) \lvec(3.840 2.238) \lvec(3.852 2.239)
\lvec(3.864 2.241) \lvec(3.876 2.242) \lvec(3.888 2.244)
\lvec(3.900 2.245) \lvec(3.912 2.247) \lvec(3.924 2.248)
\lvec(3.936 2.250) \lvec(3.948 2.251) \lvec(3.960 2.253)
\lvec(3.972 2.254) \lvec(3.984 2.255) \lvec(3.996 2.257)
\lvec(4.008 2.258) \lvec(4.020 2.260) \lvec(4.032 2.261)
\lvec(4.044 2.262) \lvec(4.056 2.264) \lvec(4.068 2.265)
\lvec(4.080 2.266) \lvec(4.092 2.268) \lvec(4.104 2.269)
\lvec(4.116 2.270) \lvec(4.128 2.272) \lvec(4.140 2.273)
\lvec(4.152 2.274) \lvec(4.164 2.276) \lvec(4.176 2.277)
\lvec(4.188 2.278) \lvec(4.200 2.279) \lvec(4.212 2.281)
\lvec(4.224 2.282) \lvec(4.236 2.283) \lvec(4.248 2.284)
\lvec(4.260 2.286) \lvec(4.272 2.287) \lvec(4.284 2.288)
\lvec(4.296 2.289) \lvec(4.308 2.290) \lvec(4.320 2.292)
\lvec(4.332 2.293) \lvec(4.344 2.294) \lvec(4.356 2.295)
\lvec(4.368 2.296) \lvec(4.380 2.298) \lvec(4.392 2.299)
\lvec(4.404 2.300) \lvec(4.416 2.301) \lvec(4.428 2.302)
\lvec(4.440 2.303) \lvec(4.452 2.305) \lvec(4.464 2.306)
\lvec(4.476 2.307) \lvec(4.488 2.308) \lvec(4.500 2.309)
\lvec(4.512 2.310) \lvec(4.524 2.311) \lvec(4.536 2.312)
\lvec(4.548 2.313) \lvec(4.560 2.314) \lvec(4.572 2.316)
\lvec(4.584 2.317) \lvec(4.596 2.318) \lvec(4.608 2.319)
\lvec(4.620 2.320) \lvec(4.632 2.321) \lvec(4.644 2.322)
\lvec(4.656 2.323) \lvec(4.668 2.324) \lvec(4.680 2.325)
\lvec(4.692 2.326) \lvec(4.704 2.327) \lvec(4.716 2.328)
\lvec(4.728 2.329) \lvec(4.740 2.330) \lvec(4.752 2.331)
\lvec(4.764 2.332) \lvec(4.776 2.333) \lvec(4.788 2.334)
\lvec(4.800 2.335) \lpatt()
\end{texdraw}
\end{center}
\caption{$\alpha_0^{\rm opt}$ as a function of $\qc/\qs$ when
$K_2=K_3=\Cp\rho_0^2$; $\Co=\frac{1}{3}\Cp$ (left),
$\Co=\frac{2}{3}\Cp$ (center) or $\Co=\Cp$ (right), and
$b_0/\qs=0$ (bold), $\frac{1}{4},\frac{1}{2},\frac{3}{4},1$
(dotted).}
\label{fig1}
\end{figure}

\section{Bent domains}\label{nonpol}

Let us now consider a general shape, with $\kappa,\tau\neq 0$, in
the absence of spontaneous polarization. In this section, we prove
that the combined effect of intrinsic bending stresses and/or
chirality do not induce shape transitions towards curved domains.

The ground state configuration of the free energy density
$\sigma:=\sigma_{\rm F}+\sigma_{\rm sm}$ is still characterized by
$\rho'\equiv 0$ and $\omega'\equiv{\rm const}$. If we further
introduce the notations
\begin{align*}
A&:=\alpha'+\kappa\cos\varphi &
K_{13}&:=K_1\sin^2\alpha+K_3\cos^2\alpha\\
\Phi&:=(\varphi'-\tau)\sin\alpha-\kappa\cos\alpha\sin\varphi
&K_{23}&:=K_2\sin^2\alpha+K_3\cos^2\alpha\\
q_1&:=\frac{(K_2\qc+K_3b_0\cos\alpha)\sin\alpha}{K_{23}}&
q_2^2&:=\frac{K_2 \qc^2+K_3b_0^2\sin^2\alpha}{K_{23}}\\
\widetilde C&:=\left(\Cp\cos^2\alpha+\Co\sin^2\alpha\right)
\rho_0^2 &\varpi&:=\frac{\Cp}{\widetilde C}\,\rho_0^2\qs\cos\alpha\\
\mathcal{A}&:=\pi r^2\qquad\qquad\qquad\qquad{\rm and}&
f(x)&:=\begin{cases} 2\left(1-\sqrt{1-x^2}\,\right)\Big/x^2&{\rm
if}\ x\in(0,1]\\ 1 &{\rm if}\ x=0\;,
\end{cases}
\end{align*}
the integration of the free energy density over the transverse
coordinates yields:
$$
\mathcal{F} =\mathcal{A}\int_0^\ell ds\left[f(\kappa r)
\left(K_{13}A^2+K_{23}\Phi^2+\widetilde C\omega^{\prime 2}\right)-
2\,(K_{23}q_1\Phi+\widetilde C\varpi\omega')+K_{23}q_2^2
+\Cp\rho_0^2\qs^2\right].
$$
The Euler-Lagrange equation for $\omega$ and the free-boundary
conditions yield \quad $\omega'\equiv\omega'_{\rm opt}
=\dps\frac{\varpi}{f(\kappa r)}\;$. If we insert it in the free
energy, we arrive at
$$
\mathcal{F}=\mathcal{A}\int_0^\ell ds\left( f(\kappa r)\,
\left(K_{13}A^2 +K_{23}\,\Phi^2\right) -2 K_{23} q_1
\Phi+K_{23}q_2^2-\frac{\widetilde C\varpi^2}{f(\kappa r)}
+\Cp\rho_0^2\qs^2\right).
$$
The minimum value of this energy is obtained when $\kappa=0$. To
prove this assertion we notice that $f$ is monotonically
increasing\footnote{The function $f$ is monotonically increasing
since $f'(x)=\left.2\left(2-x^2-2\sqrt{1-x^2}\right)\right/
\left(x^3\sqrt{1-x^2}\right)>0\ \forall x\in(0,1)$.}. In
particular, it is always greater than $f(0)=1$. Furthermore, it is
possible to write the free energy functional as:
\begin{align}
\nonumber\mathcal{F}&=\mathcal{A}\int_0^\ell ds\Biggr( f(\kappa
r)K_{13}A^2 +\big(f(\kappa r)-1\big)K_{23}\Phi^2 +\widetilde
C\varpi^2\,\frac{f(\kappa r)-1}{f(\kappa r)}\\
&+K_{23}\left(\Phi-q_1\right)^2
+\frac{K_2K_3}{K_{23}}\,\big(\qc\cos\alpha-b_0\sin^2\alpha\big)^2
+\frac{\Cp\Co\rho_0^2\qs^2\sin^2\alpha}{
\Cp\cos^2\alpha+\Co\sin^2\alpha} \Biggr)\;.
\label{latter}\end{align}
All the terms depending on the curvature (that is, all terms
appearing in the first row of (\ref{latter})) are minimized if
$\kappa=0$, and thus the ground state shape of $\Omega$ is linear.
When this is the case, the search for the energy minimizer may
proceed as in Section \ref{linear}.

\section{Polarization-induced transitions}

We now focus attention on spontaneously polarized liquid crystals.
First, we insert in the free-energy functional the terms
$\sigma_{\rm pol}$ and $\sigma_{\rm anch}$ introduced in Section
2. Furthermore, the intrinsic bending in the $K_3$-term is to be
replaced by a term $\lambda\bv{P}$, proportional to the
polarization vector. We identify \bv{P} through the angles
$\beta,\phi$ as follows:
$$
\bv{P}=P\,\bv{p}=P\,\big(\sin\beta\,\bv{n}+\cos\beta\cos\phi\,\np
+\cos\beta\sin\phi\,\nt\big)\;.
$$
We will show that these changes induce a spontaneous curvature in
the shape of a smectic-A* capillary, and both a curvature and a
torsion in a smectic-C*.

\subsection{Bent smectic-A*}\label{bent}

In a smectic-A the liquid crystal molecules are orthogonal to the
layers. Let then $\alpha\equiv 0$. We assume that the potentials
$\zeta$ and $\mathcal{G}$ are strong enough to fix the values of
$\rho\equiv{\rm const.}=\rho_0$ and $P\equiv{\rm
const.}=:P_0=:\lambda_0/\lambda$, where $\lambda_0$ has the
dimensions of an inverse length\footnote{The following arguments
can be generalized to the case of non-uniform $P,\rho$, but we
skip those quite longer proofs to shorten our presentation}. In
order to simplify notations, we put
$\zeta(\rho_0)=\mathcal{G}(P_0)=0$. Finally, we define the
quantities $\Gamma:=GP_0^2$ and $\Gamma_1:=G_1P_0^2$, having the
dimensions of nematic elastic constants.

The bulk free energy density $\sigma_{\rm b} =\sigma_{\rm
F}+\sigma_{\rm sm}+\sigma_{\rm pol}$ now reads as:
\begin{align*}
\sigma_{\rm b}&=K_2\qc^2+K_3\left[\left(\lambda_0-\frac{\kappa}{
1-\kappa\xi\cos\vartheta}\right)^2+\frac{2\kappa \lambda_0\, \big(
1-\cos\beta\sin\phi\big)} {1-\kappa\xi\cos\vartheta}
\right]\!+\Cp\rho_0^2
\left(\frac{\omega'}{1-\kappa\xi\cos\vartheta}
-\qs\right)^2\!\\
&+ \Gamma_1\,\frac{(\beta'-\kappa\sin\phi)^2\cos^2\beta}
{(1-\kappa
\xi\cos\vartheta)^2}+\Gamma\,\frac{(\beta'-\kappa\sin\phi)^2
+\big[(\phi'+\tau)\cos\beta+\kappa\sin\beta\cos\phi\big]^2}{(1-\kappa
\xi\cos\vartheta)^2}\;.
\end{align*}
We remark that, in curved domains, the bend elastic term pushes
towards configurations where the spontaneous polarization lies
along the principal normal of the curve. Indeed, the $K_3$-term is
minimized if $\cos\beta\sin\phi=1$ which, together with
$\alpha=0$, implies $\bv{P}=P_0\,\bv{N}$.

The anchoring energy $\sigma_{\rm anch}$ is given by
$$
\sigma_{\rm
anch}=\omega_P\,P_0\,\big(1-\cos\beta\sin(\vartheta+\phi) \big)
\;.
$$
If we integrate the free energy density over the transverse
section of the curve \bv{c}, and then we substitute the
equilibrium value of $\omega'$, we get the free energy density per
unit length:
\begin{align}
&\nonumber\int\sigma_{\rm anch}\,r\,(1-\kappa r\cos\vartheta)\,
d\vartheta+ \int\!\!\!\!\int\sigma_{\rm
b}\,\frac{\xi\,d\xi\,d\vartheta} {1-\kappa\xi\cos\vartheta}
=\mathcal{A}\,\Biggr[ \frac{2\omega_PP_0}{r}+\kappa
\,\omega_PP_0 \cos\beta \sin\phi\\
&\nonumber+K_2\qc^2+K_3\big(\lambda_0^2-2\kappa
\lambda_0\cos\beta\sin\phi+ \kappa^2f(\kappa
r)\big)+\Cp\rho^2_0\qs^2\,
\frac{f(\kappa r)-1}{f(\kappa r)}\\
&+(\Gamma+\Gamma_1\cos^2\beta)\,(\beta'-\kappa\sin\phi)^2f(\kappa
r)+\Gamma\big((\phi'+\tau)\cos\beta+\kappa\sin\beta\cos\phi\big)^2
f(\kappa r)\Biggr].
\label{spsma}
\end{align}
To prove that the spontaneous polarization bends the smectic-A*,
it suffices to find a curved configuration possessing a smaller
free energy than the linear one. We begin by noticing that in the
linear case ($\kappa=\tau=0$) the free energy per unit length
(\ref{spsma}) is minimized when $\beta$ and $\phi$ assume any
constant value. When this is the case, the optimal value for the
free energy is
\begin{equation}
\mathcal{F}_{\rm opt}\Big|_{\kappa,\tau=0}=
\mathcal{A}\ell\,\left( \frac{2\,\omega_P\,P_0}{r}
+K_2\qc^2+K_3\lambda_0^2\right).
\label{spsmamin}
\end{equation}
We are looking for a configuration with a free energy lower than
(\ref{spsmamin}). To this aim, we focus on curves with constant
curvature and torsion. Both the $K_3$-term and the anchoring
energy are minimized if \bv{P} lies in the plane orthogonal to the
layer normal ($\beta=0$). Furthermore, the free energy density is
minimized if the polarization vector lies parallel or
anti-parallel to the principal normal \bv{N}, depending on whether
$K_3$ is greater or smaller than $\omega_P/(2\lambda)$. In the
following, we assume that $K_3 \geq \omega_P/(2\lambda)$. In this
case the minimization process requires $\bv{p}=\bv{N}$
(\emph{i.e.}, $\phi=\frac{\pi}{2}$). Nevertheless, the
considerations below would stand in the case
$K_3<\omega_P/(2\lambda)$, provided we choose
$\phi=-\frac{\pi}{2}$.

\begin{figure}
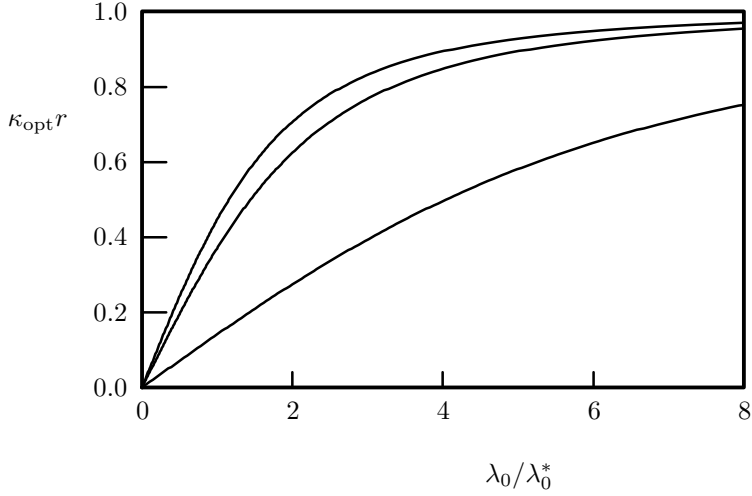

%
%
\begin{center}
\begin{texdraw}
\drawdim truecm \setgray 0
\linewd 0.05
\move (0 0) \lvec (8.0 0)
\lvec (8.0 5.0) \lvec (0 5.0) \lvec (0 0)
\linewd 0.03
\textref h:C v:T \htext (0.00 -.2) {0}
\move (2.000 0) \lvec (2.000 0.200)
\textref h:C v:T \htext (2.00 -.2) {2}
\move (4.000 0) \lvec (4.000 0.200)
\textref h:C v:T \htext (4.00 -.20) {4}
\move (6.000 0) \lvec (6.000 0.200)
\textref h:C v:T \htext (6.00 -.20) {6}
\textref h:C v:T \htext (8.00 -.20) {8}
\textref h:R v:C \htext (-.2 0.00) {0.0}
\move (0 1.000) \lvec (0.320 1.000)
\textref h:R v:C \htext (-.2 1.00) {0.2}
\move (0 2.000) \lvec (0.320 2.000)
\textref h:R v:C \htext (-.2 2.00) {0.4}
\move (0 3.000) \lvec (0.320 3.000)
\textref h:R v:C \htext (-.2 3.00) {0.6}
\move (0 4.000) \lvec (0.320 4.000)
\textref h:R v:C \htext (-.2 4.00) {0.8}
\textref h:R v:C \htext (-.2 5.00) {1.0}
\textref h:C v:T \htext (5 -1) {$\lambda_0/\lambda_0^*$}
\textref h:R v:C \htext (-1 3.50) {$\kappa_{\rm opt} r$}
\move(0.000 0.000) \lvec(0.010 0.025) \lvec(0.020 0.050)
\lvec(0.030 0.075) \lvec(0.040 0.100) \lvec(0.050 0.125)
\lvec(0.060 0.150) \lvec(0.070 0.175) \lvec(0.080 0.200)
\lvec(0.090 0.225) \lvec(0.100 0.250) \lvec(0.110 0.275)
\lvec(0.120 0.300) \lvec(0.130 0.325) \lvec(0.140 0.350)
\lvec(0.150 0.375) \lvec(0.161 0.400) \lvec(0.171 0.425)
\lvec(0.181 0.450) \lvec(0.191 0.475) \lvec(0.201 0.500)
\lvec(0.211 0.525) \lvec(0.221 0.550) \lvec(0.232 0.575)
\lvec(0.242 0.600) \lvec(0.252 0.625) \lvec(0.262 0.650)
\lvec(0.272 0.675) \lvec(0.283 0.700) \lvec(0.293 0.725)
\lvec(0.303 0.750) \lvec(0.314 0.775) \lvec(0.324 0.800)
\lvec(0.335 0.825) \lvec(0.345 0.850) \lvec(0.355 0.875)
\lvec(0.366 0.900) \lvec(0.376 0.925) \lvec(0.387 0.950)
\lvec(0.398 0.975) \lvec(0.408 1.000) \lvec(0.419 1.025)
\lvec(0.430 1.050) \lvec(0.440 1.075) \lvec(0.451 1.100)
\lvec(0.462 1.125) \lvec(0.473 1.150) \lvec(0.484 1.175)
\lvec(0.494 1.200) \lvec(0.505 1.225) \lvec(0.516 1.250)
\lvec(0.527 1.275) \lvec(0.539 1.300) \lvec(0.550 1.325)
\lvec(0.561 1.350) \lvec(0.572 1.375) \lvec(0.583 1.400)
\lvec(0.595 1.425) \lvec(0.606 1.450) \lvec(0.617 1.475)
\lvec(0.629 1.500) \lvec(0.641 1.525) \lvec(0.652 1.550)
\lvec(0.664 1.575) \lvec(0.676 1.600) \lvec(0.687 1.625)
\lvec(0.699 1.650) \lvec(0.711 1.675) \lvec(0.723 1.700)
\lvec(0.735 1.725) \lvec(0.747 1.750) \lvec(0.759 1.775)
\lvec(0.772 1.800) \lvec(0.784 1.825) \lvec(0.797 1.850)
\lvec(0.809 1.875) \lvec(0.822 1.900) \lvec(0.834 1.925)
\lvec(0.847 1.950) \lvec(0.860 1.975) \lvec(0.873 2.000)
\lvec(0.886 2.025) \lvec(0.899 2.050) \lvec(0.912 2.075)
\lvec(0.926 2.100) \lvec(0.939 2.125) \lvec(0.953 2.150)
\lvec(0.966 2.175) \lvec(0.980 2.200) \lvec(0.994 2.225)
\lvec(1.008 2.250) \lvec(1.022 2.275) \lvec(1.036 2.300)
\lvec(1.050 2.325) \lvec(1.065 2.350) \lvec(1.080 2.375)
\lvec(1.094 2.400) \lvec(1.109 2.425) \lvec(1.124 2.450)
\lvec(1.139 2.475) \lvec(1.155 2.500) \lvec(1.170 2.525)
\lvec(1.186 2.550) \lvec(1.202 2.575) \lvec(1.218 2.600)
\lvec(1.234 2.625) \lvec(1.250 2.650) \lvec(1.266 2.675)
\lvec(1.283 2.700) \lvec(1.300 2.725) \lvec(1.317 2.750)
\lvec(1.334 2.775) \lvec(1.352 2.800) \lvec(1.370 2.825)
\lvec(1.387 2.850) \lvec(1.406 2.875) \lvec(1.424 2.900)
\lvec(1.443 2.925) \lvec(1.461 2.950) \lvec(1.481 2.975)
\lvec(1.500 3.000) \lvec(1.520 3.025) \lvec(1.540 3.050)
\lvec(1.560 3.075) \lvec(1.580 3.100) \lvec(1.601 3.125)
\lvec(1.622 3.150) \lvec(1.644 3.175) \lvec(1.666 3.200)
\lvec(1.688 3.225) \lvec(1.711 3.250) \lvec(1.734 3.275)
\lvec(1.757 3.300) \lvec(1.781 3.325) \lvec(1.805 3.350)
\lvec(1.830 3.375) \lvec(1.855 3.400) \lvec(1.880 3.425)
\lvec(1.907 3.450) \lvec(1.933 3.475) \lvec(1.960 3.500)
\lvec(1.988 3.525) \lvec(2.016 3.550) \lvec(2.045 3.575)
\lvec(2.075 3.600) \lvec(2.105 3.625) \lvec(2.136 3.650)
\lvec(2.168 3.675) \lvec(2.200 3.700) \lvec(2.234 3.725)
\lvec(2.268 3.750) \lvec(2.303 3.775) \lvec(2.339 3.800)
\lvec(2.376 3.825) \lvec(2.414 3.850) \lvec(2.453 3.875)
\lvec(2.493 3.900) \lvec(2.534 3.925) \lvec(2.577 3.950)
\lvec(2.621 3.975) \lvec(2.667 4.000) \lvec(2.714 4.025)
\lvec(2.762 4.050) \lvec(2.813 4.075) \lvec(2.865 4.100)
\lvec(2.920 4.125) \lvec(2.976 4.150) \lvec(3.035 4.175)
\lvec(3.096 4.200) \lvec(3.160 4.225) \lvec(3.227 4.250)
\lvec(3.297 4.275) \lvec(3.371 4.300) \lvec(3.448 4.325)
\lvec(3.529 4.350) \lvec(3.615 4.375) \lvec(3.705 4.400)
\lvec(3.802 4.425) \lvec(3.904 4.450) \lvec(4.013 4.475)
\lvec(4.129 4.500) \lvec(4.255 4.525) \lvec(4.390 4.550)
\lvec(4.536 4.575) \lvec(4.695 4.600) \lvec(4.869 4.625)
\lvec(5.060 4.650) \lvec(5.273 4.675) \lvec(5.510 4.700)
\lvec(5.779 4.725) \lvec(6.085 4.750) \lvec(6.440 4.775)
\lvec(6.857 4.800) \lvec(7.359 4.825) \lvec(7.980 4.850)
\lvec(8.000 4.851)
\move(0.000 0.000) \lvec(0.013 0.025) \lvec(0.025 0.050)
\lvec(0.038 0.075) \lvec(0.050 0.100) \lvec(0.063 0.125)
\lvec(0.075 0.150) \lvec(0.088 0.175) \lvec(0.100 0.200)
\lvec(0.113 0.225) \lvec(0.125 0.250) \lvec(0.138 0.275)
\lvec(0.150 0.300) \lvec(0.163 0.325) \lvec(0.175 0.350)
\lvec(0.188 0.375) \lvec(0.201 0.400) \lvec(0.213 0.425)
\lvec(0.226 0.450) \lvec(0.239 0.475) \lvec(0.251 0.500)
\lvec(0.264 0.525) \lvec(0.277 0.550) \lvec(0.289 0.575)
\lvec(0.302 0.600) \lvec(0.315 0.625) \lvec(0.328 0.650)
\lvec(0.341 0.675) \lvec(0.353 0.700) \lvec(0.366 0.725)
\lvec(0.379 0.750) \lvec(0.392 0.775) \lvec(0.405 0.800)
\lvec(0.418 0.825) \lvec(0.431 0.850) \lvec(0.444 0.875)
\lvec(0.457 0.900) \lvec(0.471 0.925) \lvec(0.484 0.950)
\lvec(0.497 0.975) \lvec(0.510 1.000) \lvec(0.524 1.025)
\lvec(0.537 1.050) \lvec(0.550 1.075) \lvec(0.564 1.100)
\lvec(0.577 1.125) \lvec(0.591 1.150) \lvec(0.604 1.175)
\lvec(0.618 1.200) \lvec(0.632 1.225) \lvec(0.645 1.250)
\lvec(0.659 1.275) \lvec(0.673 1.300) \lvec(0.687 1.325)
\lvec(0.701 1.350) \lvec(0.715 1.375) \lvec(0.729 1.400)
\lvec(0.743 1.425) \lvec(0.758 1.450) \lvec(0.772 1.475)
\lvec(0.786 1.500) \lvec(0.801 1.525) \lvec(0.815 1.550)
\lvec(0.830 1.575) \lvec(0.844 1.600) \lvec(0.859 1.625)
\lvec(0.874 1.650) \lvec(0.889 1.675) \lvec(0.904 1.700)
\lvec(0.919 1.725) \lvec(0.934 1.750) \lvec(0.949 1.775)
\lvec(0.965 1.800) \lvec(0.980 1.825) \lvec(0.996 1.850)
\lvec(1.011 1.875) \lvec(1.027 1.900) \lvec(1.043 1.925)
\lvec(1.059 1.950) \lvec(1.075 1.975) \lvec(1.091 2.000)
\lvec(1.107 2.025) \lvec(1.124 2.050) \lvec(1.140 2.075)
\lvec(1.157 2.100) \lvec(1.174 2.125) \lvec(1.191 2.150)
\lvec(1.208 2.175) \lvec(1.225 2.200) \lvec(1.242 2.225)
\lvec(1.260 2.250) \lvec(1.277 2.275) \lvec(1.295 2.300)
\lvec(1.313 2.325) \lvec(1.331 2.350) \lvec(1.349 2.375)
\lvec(1.368 2.400) \lvec(1.386 2.425) \lvec(1.405 2.450)
\lvec(1.424 2.475) \lvec(1.443 2.500) \lvec(1.463 2.525)
\lvec(1.482 2.550) \lvec(1.502 2.575) \lvec(1.522 2.600)
\lvec(1.542 2.625) \lvec(1.563 2.650) \lvec(1.583 2.675)
\lvec(1.604 2.700) \lvec(1.625 2.725) \lvec(1.646 2.750)
\lvec(1.668 2.775) \lvec(1.690 2.800) \lvec(1.712 2.825)
\lvec(1.734 2.850) \lvec(1.757 2.875) \lvec(1.780 2.900)
\lvec(1.803 2.925) \lvec(1.827 2.950) \lvec(1.851 2.975)
\lvec(1.875 3.000) \lvec(1.900 3.025) \lvec(1.925 3.050)
\lvec(1.950 3.075) \lvec(1.976 3.100) \lvec(2.002 3.125)
\lvec(2.028 3.150) \lvec(2.055 3.175) \lvec(2.082 3.200)
\lvec(2.110 3.225) \lvec(2.138 3.250) \lvec(2.167 3.275)
\lvec(2.196 3.300) \lvec(2.226 3.325) \lvec(2.256 3.350)
\lvec(2.287 3.375) \lvec(2.319 3.400) \lvec(2.351 3.425)
\lvec(2.383 3.450) \lvec(2.417 3.475) \lvec(2.450 3.500)
\lvec(2.485 3.525) \lvec(2.521 3.550) \lvec(2.557 3.575)
\lvec(2.594 3.600) \lvec(2.632 3.625) \lvec(2.670 3.650)
\lvec(2.710 3.675) \lvec(2.750 3.700) \lvec(2.792 3.725)
\lvec(2.835 3.750) \lvec(2.878 3.775) \lvec(2.923 3.800)
\lvec(2.970 3.825) \lvec(3.017 3.850) \lvec(3.066 3.875)
\lvec(3.116 3.900) \lvec(3.168 3.925) \lvec(3.221 3.950)
\lvec(3.276 3.975) \lvec(3.333 4.000) \lvec(3.392 4.025)
\lvec(3.453 4.050) \lvec(3.516 4.075) \lvec(3.582 4.100)
\lvec(3.650 4.125) \lvec(3.720 4.150) \lvec(3.794 4.175)
\lvec(3.870 4.200) \lvec(3.950 4.225) \lvec(4.034 4.250)
\lvec(4.121 4.275) \lvec(4.213 4.300) \lvec(4.310 4.325)
\lvec(4.411 4.350) \lvec(4.518 4.375) \lvec(4.632 4.400)
\lvec(4.752 4.425) \lvec(4.880 4.450) \lvec(5.016 4.475)
\lvec(5.162 4.500) \lvec(5.318 4.525) \lvec(5.487 4.550)
\lvec(5.670 4.575) \lvec(5.869 4.600) \lvec(6.086 4.625)
\lvec(6.326 4.650) \lvec(6.591 4.675) \lvec(6.888 4.700)
\lvec(7.223 4.725) \lvec(7.606 4.750) \lvec(8.000 4.772)
\move(0.000 0.000) \lvec(0.035 0.025) \lvec(0.070 0.050)
\lvec(0.105 0.075) \lvec(0.140 0.100) \lvec(0.175 0.125)
\lvec(0.210 0.150) \lvec(0.245 0.175) \lvec(0.280 0.200)
\lvec(0.315 0.225) \lvec(0.350 0.250) \lvec(0.386 0.275)
\lvec(0.421 0.300) \lvec(0.456 0.325) \lvec(0.491 0.350)
\lvec(0.526 0.375) \lvec(0.562 0.400) \lvec(0.597 0.425)
\lvec(0.633 0.450) \lvec(0.668 0.475) \lvec(0.704 0.500)
\lvec(0.739 0.525) \lvec(0.775 0.550) \lvec(0.810 0.575)
\lvec(0.846 0.600) \lvec(0.882 0.625) \lvec(0.918 0.650)
\lvec(0.954 0.675) \lvec(0.990 0.700) \lvec(1.026 0.725)
\lvec(1.062 0.750) \lvec(1.098 0.775) \lvec(1.135 0.800)
\lvec(1.171 0.825) \lvec(1.208 0.850) \lvec(1.244 0.875)
\lvec(1.281 0.900) \lvec(1.318 0.925) \lvec(1.355 0.950)
\lvec(1.392 0.975) \lvec(1.429 1.000) \lvec(1.466 1.025)
\lvec(1.504 1.050) \lvec(1.541 1.075) \lvec(1.579 1.100)
\lvec(1.616 1.125) \lvec(1.654 1.150) \lvec(1.692 1.175)
\lvec(1.731 1.200) \lvec(1.769 1.225) \lvec(1.807 1.250)
\lvec(1.846 1.275) \lvec(1.885 1.300) \lvec(1.924 1.325)
\lvec(1.963 1.350) \lvec(2.002 1.375) \lvec(2.042 1.400)
\lvec(2.081 1.425) \lvec(2.121 1.450) \lvec(2.161 1.475)
\lvec(2.201 1.500) \lvec(2.242 1.525) \lvec(2.282 1.550)
\lvec(2.323 1.575) \lvec(2.364 1.600) \lvec(2.406 1.625)
\lvec(2.447 1.650) \lvec(2.489 1.675) \lvec(2.531 1.700)
\lvec(2.573 1.725) \lvec(2.615 1.750) \lvec(2.658 1.775)
\lvec(2.701 1.800) \lvec(2.744 1.825) \lvec(2.788 1.850)
\lvec(2.832 1.875) \lvec(2.876 1.900) \lvec(2.920 1.925)
\lvec(2.965 1.950) \lvec(3.010 1.975) \lvec(3.055 2.000)
\lvec(3.101 2.025) \lvec(3.147 2.050) \lvec(3.193 2.075)
\lvec(3.240 2.100) \lvec(3.287 2.125) \lvec(3.334 2.150)
\lvec(3.382 2.175) \lvec(3.430 2.200) \lvec(3.478 2.225)
\lvec(3.527 2.250) \lvec(3.577 2.275) \lvec(3.626 2.300)
\lvec(3.677 2.325) \lvec(3.727 2.350) \lvec(3.778 2.375)
\lvec(3.830 2.400) \lvec(3.882 2.425) \lvec(3.935 2.450)
\lvec(3.988 2.475) \lvec(4.041 2.500) \lvec(4.096 2.525)
\lvec(4.150 2.550) \lvec(4.206 2.575) \lvec(4.261 2.600)
\lvec(4.318 2.625) \lvec(4.375 2.650) \lvec(4.433 2.675)
\lvec(4.491 2.700) \lvec(4.550 2.725) \lvec(4.610 2.750)
\lvec(4.670 2.775) \lvec(4.731 2.800) \lvec(4.793 2.825)
\lvec(4.856 2.850) \lvec(4.920 2.875) \lvec(4.984 2.900)
\lvec(5.049 2.925) \lvec(5.115 2.950) \lvec(5.182 2.975)
\lvec(5.250 3.000) \lvec(5.319 3.025) \lvec(5.389 3.050)
\lvec(5.460 3.075) \lvec(5.531 3.100) \lvec(5.604 3.125)
\lvec(5.679 3.150) \lvec(5.754 3.175) \lvec(5.830 3.200)
\lvec(5.908 3.225) \lvec(5.987 3.250) \lvec(6.068 3.275)
\lvec(6.150 3.300) \lvec(6.233 3.325) \lvec(6.318 3.350)
\lvec(6.404 3.375) \lvec(6.492 3.400) \lvec(6.582 3.425)
\lvec(6.673 3.450) \lvec(6.766 3.475) \lvec(6.861 3.500)
\lvec(6.958 3.525) \lvec(7.058 3.550) \lvec(7.159 3.575)
\lvec(7.263 3.600) \lvec(7.368 3.625) \lvec(7.477 3.650)
\lvec(7.588 3.675) \lvec(7.701 3.700) \lvec(7.818 3.725)
\lvec(7.937 3.750) \lvec(8.000 3.763)
\move (-1.85 -1.35) \move (8.1 5.2)
\end{texdraw}
\end{center}
\caption{The preferred curvature of the axis of the smectic-A*
capillary increases with the spontaneous polarization. The inverse
length $\lambda_0$ is proportional to $|\bv{P}|$, and its
reference value $\lambda_0^*$ is defined in (\ref{l0s}). From top
to bottom, the graphs correspond to $\xi=0, 1, 10$.}
\label{kappa}
\end{figure}

With the choices above, the free energy depends only on the
particular values chosen by $\kappa$ and $\tau$, and can be
written as
$$
\frac{\mathcal{F}_{\rm opt}(\kappa,\tau)}{\mathcal{A}\ell}=
A_0-2A_1 \kappa r+A_2\,(\kappa r)^2\,f(\kappa r)-
\frac{A_3}{f(\kappa r)}+B_1\, f(\kappa r)\,(\tau r)^2\;,
$$
where the signs are chosen in a way such that the $A$'s and $B$'s
are non-negative:
\begin{eqnarray*}
A_0&=&K_2\qc^2+K_3\lambda_0^2+\Cp\rho^2_0\qs^2+2\omega_PP_0/r\\
A_1&=&\left(K_3 -\frac{\omega_P}{2\lambda}\right)\frac{\lambda_0}{r}\\
A_2&=&\big(K_3+\Gamma+\Gamma_1\big)\,/\,r^2\\
A_3&=&\Cp\rho^2_0\qs^2\\
B_1&=&\Gamma/r^2\;.
\end{eqnarray*}
The free energy $\mathcal{F}_{\rm opt}$ is clearly minimized when
$\tau=0$ (plane curve). On the contrary, the minimum of
$\mathcal{F}_{\rm opt}$ is attained when the curvature has a
strictly positive value, since
$$
\frac{\mathcal{F}_{\rm opt}(\kappa,0)}{\mathcal{A}\ell}=
(A_0-A_3)-2A_1 \kappa r+\left(A_2+\frac{1}{3}\, A_3\right)(\kappa
r)^2+O(\kappa r)^4\quad{\rm as}\ \kappa r\to 0\;.
$$
We remark that $\mathcal{F}_{\rm opt}$ possesses a unique minimum
as a function of $\kappa$. Indeed, the condition
$\frac{\partial}{\partial \kappa}\mathcal{F}_{\rm opt}=0$ is
equivalent to
\begin{equation}
A_2\,\left(2(\kappa r)\,f(\kappa r)+(\kappa r)^2f'(\kappa r)\right)+
\frac{A_3 f'(\kappa r)}{f^2(\kappa r)}=2A_1
\label{eqk}
\end{equation}
and this equation has one and only one root, since the function at the
left-hand side vanishes when $\kappa r\to 0$, is everywhere strictly
increasing and diverges when $\kappa r\to 1^-$. Let $x:=\kappa
r$. Equation (\ref{eqk}) can be written as
\begin{equation}
2x\,f(x)+x^2f'(x)+
\xi\frac{f'(x)}{f^2(x)}=\frac{\lambda_0}{\lambda_0^*}\;,
\label{xlam}
\end{equation}
with
\begin{equation}
\xi:=\frac{A_3}{A_2}=\frac{\Cp\rho^2_0\qs^2r^2}
{K_3+\Gamma+\Gamma_1} \qquad {\rm and} \qquad
\lambda_0^*:=\frac{A_2\lambda_0}{2A_1}=
\frac{K_3+\Gamma+\Gamma_1}{(2K_3 -\omega_P/\lambda)r}\;.
\label{l0s}
\end{equation}
Figure \ref{kappa} shows how the solutions of (\ref{xlam}) depend
on $\lambda_0$ (which is proportional to the intensity of the
spontaneous polarization) for three different values of the
dimensionless parameter $\xi$. In the absence of spontaneous
polarization the curvature is null. Then, it increases
monotonically with $\lambda_0$. When the spontaneous polarization
makes $\lambda_0$ much greater than its reference value
$\lambda_0^*$, the curvature approaches its maximum allowed value
$r^{-1}$. The curvature increases more rapidly when $\xi$ is
small, that is, in thinner capillaries.

\subsection{Helicoidal smectic-C*}

In this final section we study how the spontaneous polarization
may induce a telephone-cord transition in a smectic-C* capillary.
We focus on a particular, even if quite common, case. We assume
that the smectic part of the free energy is able to fix the
opening angle of the smectic-C* cones to a fixed value:
$\alpha\equiv \alpha_0$. Furthermore, we assume that the
spontaneous polarization of the liquid crystal molecules is always
orthogonal to the director direction: $\beta\equiv 0$, and
determines a constant angle with respect to the principal normal
of the capillary ($\phi\equiv {\rm const.}$). The former of these
assumptions holds, for example, when we are analyzing a system of
banana molecules, in which the polarization is induced by the
curvature of the liquid crystal rods, and it is always orthogonal
to \bv{n}.

Even under the above simplifying assumptions, the bulk free energy
density to be minimized is still quite cumbersome to handle:
\begin{align*}
\sigma_{\rm b}&=\sigma_{\rm
sm}(\alpha_0)+K_1\,\frac{\kappa^2\cos^2\varphi\sin^2\alpha_0}
{(1-\kappa\xi\cos\vartheta)^2}+K_2\left(\qc-\frac{\sin\alpha_0}
{1-\kappa\xi\cos\vartheta}\, \big((\varphi'-\tau)
\sin\alpha_0-\kappa\cos\alpha_0\sin\varphi\big)\right)^2\\
&+K_3\Biggr[ \left(\lambda_0\sin\phi
-\frac{\kappa\cos\varphi\cos\alpha_0
}{1-\kappa\xi\cos\vartheta}\right)^2+\left(\lambda_0\cos\phi-\frac{
(\varphi'-\tau)\sin\alpha_0-\kappa\cos\alpha_0\sin\varphi}
{1-\kappa\xi\cos\vartheta}\,\cos\alpha_0\right)^2\Biggr]\\
&+\frac{\Gamma_1}{(1-\kappa \xi\cos\vartheta)^2}\,\Big[
-\big((\varphi'-\tau)\sin\alpha_0\cos\phi+ \kappa\cos\varphi
\sin\phi-\kappa\cos\alpha_0\sin\varphi\cos\phi\big)
\cos\alpha_0\\
&\qquad\qquad\qquad+\big((\varphi'-\tau)\cos\alpha_0
+\kappa\sin\alpha_0\sin\varphi\big)
\cos\phi\sin\alpha_0\Big]^2\\
&+\frac{\Gamma}{(1-\kappa\xi\cos\vartheta)^2}
\Big[\big((\varphi'-\tau)\sin\alpha_0\cos\phi+ \kappa\cos\varphi
\sin\phi -\kappa\cos\alpha_0\sin\varphi\cos\phi\big)^2\\
&\qquad\qquad\qquad+\big((\varphi'-\tau)\cos\alpha_0
+\kappa\sin\alpha_0\sin\varphi\big)^2\, \Big]\;,
\end{align*}
where $\sigma_{\rm sm}$ represents the smectic part, which fixes
the value of $\alpha_0$. The above expression simplifies if we
introduce the quantities $\bv{x}:=\{x_i,i=1,2,3)$,
$\bv{A}:=\{A_{ij}:i,j=1,2,3\}$, $\bv{b}:=\{b_i,i=1,2,3)$, and
$c\in\reals$, defined as:
\begin{align*}
&x_1:=\frac{\kappa\cos\varphi}{1-\kappa\xi\cos\vartheta}\;,\qquad
x_2:=\frac{(\varphi'-\tau)\sin\alpha_0-\kappa\cos\alpha_0
\sin\varphi}{{1-\kappa\xi\cos\vartheta}}\;,\\
&x_3:=\frac{(\varphi'-\tau)\cos\alpha_0+\kappa\sin\alpha_0
\sin\varphi}{1-\kappa\xi\cos\vartheta}\;,   \\
&A_{11}:=K_1\sin^2\alpha_0+K_3\cos^2\alpha_0+\left(\Gamma
+\Gamma_1\cos^2\alpha_0\right)\sin^2\phi\;,\\
&A_{22}:=K_2\sin^2\alpha_0+K_3\cos^2\alpha_0+\left(\Gamma
+\Gamma_1\cos^2\alpha_0\right)\cos^2\phi\;,\\
&A_{33}:=\Gamma+\Gamma_1\sin^2\alpha_0\cos^2\phi\;, \\
&A_{12}=A_{21}:=\left(\Gamma+\Gamma_1\cos^2\alpha_0\right)
\sin\phi\cos\phi\;,\\
&A_{13}=A_{31}:=-\Gamma_1\sin\alpha_0\cos\alpha_0\sin\phi\cos\phi\;,\\
&A_{23}=A_{32}:=-\Gamma_1\sin\alpha_0\cos\alpha_0\cos^2\phi\;, \\
&b_1:=K_3\lambda_0\sin\phi\cos\alpha_0\;,\qquad
b_2:=K_2\qc\sin\alpha_0+
K_3\lambda_0\cos\phi\cos\alpha_0\;,\qquad b_3:=0\;,\ \\
&c:=\sigma_{\rm sm}(\alpha_0)+K_2\qc^2+K_3\lambda_0^2
\end{align*}
which allow to write
\begin{equation*}
\sigma_{\rm b}=\bv{x}\cdot\bv{Ax}-2\,\bv{b}\cdot\bv{x}+c\;.
\end{equation*}

When $\beta=0$, the scalar product between the polarization
direction and the outside normal to $\partial\Omega$ is given by:
$$
\bv{p}\cdot\bv{\nu}=\cos\alpha_0\sin\phi \cos(\vartheta-\varphi)+
\cos\phi \sin(\vartheta-\varphi)\;.
$$
Thus, the integration of the anchoring energy across the section
orthogonal to the axis of the capillary yields:
\begin{equation*}
\int_0^{2\pi}\!\!\!\sigma_{\rm anch}\,r\,(1-\kappa
r\cos\vartheta)\, d\vartheta=\mathcal{A}\,\Big[2+ \kappa
r\big(\cos\alpha_0\sin\phi \cos\varphi+ \cos\phi
\sin\varphi\big)\Big]\,\frac{\omega_P\,P_0}{r}\;.
\end{equation*}

We now specialize our study to the small curvature (or thin
capillary) regime $\kappa r\ll 1$. In this case we can neglect the
correction to $1$ in the denominators of the $x_i$'s, and the
integration of the bulk free energy density over the transverse
section simply corresponds to a multiplication by $\mathcal{A}$.
This allows us to derive an analytic expression for the free
energy minimizer. In fact, in this case the total free energy can
be written as:
\begin{equation}
\frac{\mathcal{F}}{\mathcal{A}\ell}=\bv{x}\cdot\bv{Ax}-2\,\bv{\tilde
b}\cdot\bv{x}+\tilde c\;,
\label{sigquad}
\end{equation}
provided we define the $\tilde b_i$'s and $\tilde c$ as follows:
\begin{align*}
&\tilde
b_1:=\left(K_3-\frac{\omega_P}{2\lambda}\right)\lambda_0\sin\phi\cos\alpha_0\;,
\qquad \tilde b_2:=K_2\qc\sin\alpha_0+
\left(K_3-\frac{\omega_P}{2\lambda}\right)\lambda_0\cos\phi\cos\alpha_0\;,\\
&\tilde b_3:=\frac{\omega_P\,P_0}{2} \cos\phi\sin\alpha_0\;,\hskip
2.1cm \tilde c:=\sigma_{\rm
sm}(\alpha_0)+K_2\qc^2+K_3\lambda_0^2+\frac{2 \omega_PP_0}{r}\;.
\end{align*}
The functional (\ref{sigquad}) is minimized with respect to the
possible values assumed by the $x_i$'s when
\begin{equation}
\bv{x}=\bv{x}_{\rm opt}:=\bv{A}^{-1}\bv{\tilde b}\;.
\label{xopt}
\end{equation}
(The symmetric matrix \bv{A} is positive definite because of the
positivity of the elastic free energy density). When this is the
case, the free energy takes the value
\begin{equation}
\frac{\mathcal{F}_{\rm opt}}{\mathcal{A}\ell}=\tilde c-\bv{\tilde
b}\cdot\bv{A}^{-1}\bv{\tilde b}\;.
\label{Fopt}
\end{equation}
The $x_i$'s obtained from (\ref{xopt}) fix the constant values of
$\varphi$, $\kappa$, and $\tau$. Indeed, in the thin capillary
limit $\kappa r\ll1$, and setting $\varphi'=0$, the definition of
the $x_i$'s can be written as
\begin{align*}
x_{{\rm opt},1}&=\kappa\cos\varphi&\\
x_{{\rm opt},2}&=-\tau\sin\alpha_0-\kappa\cos\alpha_0\sin\varphi\\
x_{{\rm opt},3}&=
-\tau\cos\alpha_0+\kappa\sin\alpha_0\sin\varphi\;,
\end{align*}
which can be inverted to obtain:
\begin{align*}
\kappa&=\sqrt{x^2_{{\rm opt},1}+\left(x_{{\rm opt},3}\sin\alpha_0-
x_{{\rm opt},2}\cos\alpha_0\right)^2}\\
\tau&=-\left(x_{{\rm opt},2}\sin\alpha_0+x_{{\rm opt},3}\cos\alpha_0\right)
\qquad{\rm and}\\
\varphi&=\arctan\frac{x_{{\rm opt},3}\sin\alpha_0-x_{{\rm opt},2}\cos\alpha_0}
{x_{{\rm opt},1}}\;.
\end{align*}
Thus, at this stage, $\mathcal{F}_{\rm opt}$ in (\ref{Fopt}) still
depends on the constant value attained by $\phi$, the angle that
identifies the polarization direction. Only the minimization of
$\mathcal{F}_{\rm opt}$ with respect to $\phi$ yields the complete
description of the ground state configuration.

In order to illustrate the result of this minimization procedure
we conclude this section by analyzing in detail two particular
cases. In both of them the optimal shape of the smectic capillary
turns out to be a three-dimensional helix, characterized by
non-null values of both the curvature and the torsion of its axis.

\subsubsection{1-constant approximation}

\begin{figure}
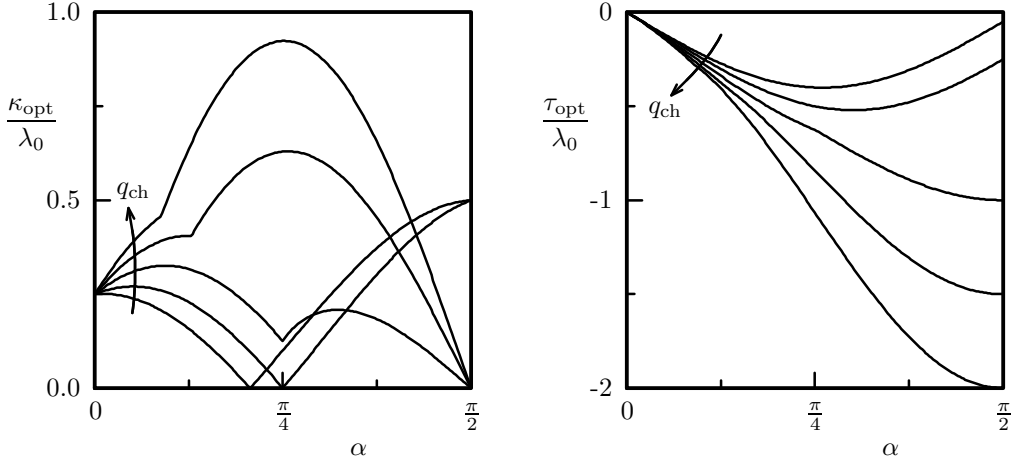

%
%
\null\hfill
\begin{texdraw}
\drawdim truecm \setgray 0
\linewd 0.05
\move (0 0) \lvec (5.0 0)
\lvec (5.0 5.0) \lvec (0 5.0) \lvec (0 0)
\linewd 0.03
\textref h:C v:T \htext (0.00 -.20) {0}
\move (1.250 0) \lvec (1.250 0.100)
\move (2.500 0) \lvec (2.500 0.200)
\textref h:C v:T \htext (2.50 -.20) {$\frac{\pi}{4}$}
\move (3.750 0) \lvec (3.750 0.100)
\textref h:C v:T \htext (5.00 -.20) {$\frac{\pi}{2}$}
\textref h:R v:C \htext (-.20 0.00) {0.0}
\move (0 1.250) \lvec (0.100 1.250)
\move (0 2.500) \lvec (0.200 2.500)
\textref h:R v:C \htext (-.20 2.50) {0.5}
\move (0 3.750) \lvec (0.100 3.750)
\textref h:R v:C \htext (-.20 5.00) {1.0}
\textref h:C v:T \htext (3.5 -.75) {$\alpha$}
\textref h:R v:C \htext (-.5 3.50) {$\dps\frac{\kappa_{\rm opt}}{\lambda_0}$}
\arrowheadtype t:V \arrowheadsize l:0.15 w:.1
\move (0.5 1) \clvec (0.55 1.3)(0.55 1.8)(0.5 2.1)
\ravec (-0.05 0.3)
\textref h:C v:B \htext (0.5 2.5) {$\qc$}
\move(0.000 1.250) \lvec(0.025 1.252) \lvec(0.050 1.253)
\lvec(0.075 1.254) \lvec(0.100 1.254) \lvec(0.125 1.254)
\lvec(0.150 1.253) \lvec(0.175 1.252) \lvec(0.200 1.251)
\lvec(0.225 1.249) \lvec(0.250 1.246) \lvec(0.275 1.244)
\lvec(0.300 1.240) \lvec(0.325 1.236) \lvec(0.350 1.232)
\lvec(0.375 1.227) \lvec(0.400 1.222) \lvec(0.425 1.217)
\lvec(0.450 1.210) \lvec(0.475 1.204) \lvec(0.500 1.197)
\lvec(0.525 1.189) \lvec(0.550 1.181) \lvec(0.575 1.173)
\lvec(0.600 1.164) \lvec(0.625 1.155) \lvec(0.650 1.145)
\lvec(0.675 1.135) \lvec(0.700 1.125) \lvec(0.725 1.114)
\lvec(0.750 1.102) \lvec(0.775 1.091) \lvec(0.800 1.078)
\lvec(0.825 1.066) \lvec(0.850 1.053) \lvec(0.875 1.039)
\lvec(0.900 1.025) \lvec(0.925 1.011) \lvec(0.950 0.996)
\lvec(0.975 0.981) \lvec(1.000 0.965) \lvec(1.025 0.949)
\lvec(1.050 0.933) \lvec(1.075 0.916) \lvec(1.100 0.899)
\lvec(1.125 0.882) \lvec(1.150 0.864) \lvec(1.175 0.846)
\lvec(1.200 0.827) \lvec(1.225 0.808) \lvec(1.250 0.789)
\lvec(1.275 0.770) \lvec(1.300 0.750) \lvec(1.325 0.729)
\lvec(1.350 0.709) \lvec(1.375 0.688) \lvec(1.400 0.666)
\lvec(1.425 0.645) \lvec(1.450 0.623) \lvec(1.475 0.601)
\lvec(1.500 0.578) \lvec(1.525 0.555) \lvec(1.550 0.532)
\lvec(1.575 0.509) \lvec(1.600 0.485) \lvec(1.625 0.461)
\lvec(1.650 0.437) \lvec(1.675 0.413) \lvec(1.700 0.388)
\lvec(1.725 0.363) \lvec(1.750 0.338) \lvec(1.775 0.312)
\lvec(1.800 0.286) \lvec(1.825 0.261) \lvec(1.850 0.234)
\lvec(1.875 0.208) \lvec(1.900 0.181) \lvec(1.925 0.155)
\lvec(1.950 0.128) \lvec(1.975 0.101) \lvec(2.000 0.073)
\lvec(2.025 0.046) \lvec(2.050 0.018) \lvec(2.075 0.010)
\lvec(2.100 0.038) \lvec(2.125 0.066) \lvec(2.150 0.094)
\lvec(2.175 0.122) \lvec(2.200 0.151) \lvec(2.225 0.179)
\lvec(2.250 0.208) \lvec(2.275 0.237) \lvec(2.300 0.266)
\lvec(2.325 0.295) \lvec(2.350 0.324) \lvec(2.375 0.353)
\lvec(2.400 0.383) \lvec(2.425 0.412) \lvec(2.450 0.441)
\lvec(2.475 0.471) \lvec(2.500 0.500) \lvec(2.525 0.529)
\lvec(2.550 0.559) \lvec(2.575 0.588) \lvec(2.600 0.618)
\lvec(2.625 0.647) \lvec(2.650 0.677) \lvec(2.675 0.707)
\lvec(2.700 0.736) \lvec(2.725 0.765) \lvec(2.750 0.795)
\lvec(2.775 0.824) \lvec(2.800 0.854) \lvec(2.825 0.883)
\lvec(2.850 0.912) \lvec(2.875 0.941) \lvec(2.900 0.970)
\lvec(2.925 0.999) \lvec(2.950 1.028) \lvec(2.975 1.057)
\lvec(3.000 1.086) \lvec(3.025 1.114) \lvec(3.050 1.143)
\lvec(3.075 1.171) \lvec(3.100 1.199) \lvec(3.125 1.227)
\lvec(3.150 1.255) \lvec(3.175 1.283) \lvec(3.200 1.310)
\lvec(3.225 1.338) \lvec(3.250 1.365) \lvec(3.275 1.392)
\lvec(3.300 1.419) \lvec(3.325 1.445) \lvec(3.350 1.472)
\lvec(3.375 1.498) \lvec(3.400 1.524) \lvec(3.425 1.550)
\lvec(3.450 1.576) \lvec(3.475 1.601) \lvec(3.500 1.626)
\lvec(3.525 1.651) \lvec(3.550 1.675) \lvec(3.575 1.700)
\lvec(3.600 1.724) \lvec(3.625 1.748) \lvec(3.650 1.771)
\lvec(3.675 1.794) \lvec(3.700 1.817) \lvec(3.725 1.840)
\lvec(3.750 1.862) \lvec(3.775 1.884) \lvec(3.800 1.906)
\lvec(3.825 1.928) \lvec(3.850 1.949) \lvec(3.875 1.970)
\lvec(3.900 1.990) \lvec(3.925 2.010) \lvec(3.950 2.030)
\lvec(3.975 2.049) \lvec(4.000 2.068) \lvec(4.025 2.087)
\lvec(4.050 2.106) \lvec(4.075 2.124) \lvec(4.100 2.141)
\lvec(4.125 2.158) \lvec(4.150 2.175) \lvec(4.175 2.192)
\lvec(4.200 2.208) \lvec(4.225 2.224) \lvec(4.250 2.239)
\lvec(4.275 2.254) \lvec(4.300 2.268) \lvec(4.325 2.282)
\lvec(4.350 2.296) \lvec(4.375 2.309) \lvec(4.400 2.322)
\lvec(4.425 2.335) \lvec(4.450 2.347) \lvec(4.475 2.358)
\lvec(4.500 2.370) \lvec(4.525 2.380) \lvec(4.550 2.391)
\lvec(4.575 2.401) \lvec(4.600 2.410) \lvec(4.625 2.419)
\lvec(4.650 2.428) \lvec(4.675 2.436) \lvec(4.700 2.443)
\lvec(4.725 2.451) \lvec(4.750 2.457) \lvec(4.775 2.464)
\lvec(4.800 2.470) \lvec(4.825 2.475) \lvec(4.850 2.480)
\lvec(4.875 2.484) \lvec(4.900 2.488) \lvec(4.925 2.492)
\lvec(4.950 2.495) \lvec(4.975 2.498) \lvec(5.000 2.500)
\move(0.000 1.250) \lvec(0.025 1.260) \lvec(0.050 1.269)
\lvec(0.075 1.277) \lvec(0.100 1.286) \lvec(0.125 1.293)
\lvec(0.150 1.300) \lvec(0.175 1.307) \lvec(0.200 1.314)
\lvec(0.225 1.319) \lvec(0.250 1.325) \lvec(0.275 1.330)
\lvec(0.300 1.334) \lvec(0.325 1.338) \lvec(0.350 1.341)
\lvec(0.375 1.344) \lvec(0.400 1.347) \lvec(0.425 1.348)
\lvec(0.450 1.350) \lvec(0.475 1.351) \lvec(0.500 1.351)
\lvec(0.525 1.351) \lvec(0.550 1.351) \lvec(0.575 1.350)
\lvec(0.600 1.348) \lvec(0.625 1.346) \lvec(0.650 1.344)
\lvec(0.675 1.341) \lvec(0.700 1.338) \lvec(0.725 1.334)
\lvec(0.750 1.329) \lvec(0.775 1.325) \lvec(0.800 1.319)
\lvec(0.825 1.313) \lvec(0.850 1.307) \lvec(0.875 1.300)
\lvec(0.900 1.293) \lvec(0.925 1.285) \lvec(0.950 1.277)
\lvec(0.975 1.268) \lvec(1.000 1.259) \lvec(1.025 1.250)
\lvec(1.050 1.240) \lvec(1.075 1.229) \lvec(1.100 1.218)
\lvec(1.125 1.207) \lvec(1.150 1.195) \lvec(1.175 1.182)
\lvec(1.200 1.170) \lvec(1.225 1.156) \lvec(1.250 1.143)
\lvec(1.275 1.129) \lvec(1.300 1.114) \lvec(1.325 1.099)
\lvec(1.350 1.084) \lvec(1.375 1.068) \lvec(1.400 1.052)
\lvec(1.425 1.035) \lvec(1.450 1.018) \lvec(1.475 1.001)
\lvec(1.500 0.983) \lvec(1.525 0.964) \lvec(1.550 0.946)
\lvec(1.575 0.927) \lvec(1.600 0.907) \lvec(1.625 0.888)
\lvec(1.650 0.867) \lvec(1.675 0.847) \lvec(1.700 0.826)
\lvec(1.725 0.805) \lvec(1.750 0.783) \lvec(1.775 0.761)
\lvec(1.800 0.739) \lvec(1.825 0.716) \lvec(1.850 0.693)
\lvec(1.875 0.670) \lvec(1.900 0.646) \lvec(1.925 0.622)
\lvec(1.950 0.598) \lvec(1.975 0.574) \lvec(2.000 0.549)
\lvec(2.025 0.524) \lvec(2.050 0.498) \lvec(2.075 0.473)
\lvec(2.100 0.447) \lvec(2.125 0.420) \lvec(2.150 0.394)
\lvec(2.175 0.367) \lvec(2.200 0.340) \lvec(2.225 0.313)
\lvec(2.250 0.286) \lvec(2.275 0.258) \lvec(2.300 0.230)
\lvec(2.325 0.202) \lvec(2.350 0.174) \lvec(2.375 0.145)
\lvec(2.400 0.116) \lvec(2.425 0.088) \lvec(2.450 0.059)
\lvec(2.475 0.029) \lvec(2.500 0.000) \lvec(2.525 0.030)
\lvec(2.550 0.059) \lvec(2.575 0.089) \lvec(2.600 0.119)
\lvec(2.625 0.149) \lvec(2.650 0.179) \lvec(2.675 0.210)
\lvec(2.700 0.240) \lvec(2.725 0.270) \lvec(2.750 0.301)
\lvec(2.775 0.332) \lvec(2.800 0.362) \lvec(2.825 0.393)
\lvec(2.850 0.424) \lvec(2.875 0.455) \lvec(2.900 0.486)
\lvec(2.925 0.517) \lvec(2.950 0.548) \lvec(2.975 0.579)
\lvec(3.000 0.610) \lvec(3.025 0.641) \lvec(3.050 0.672)
\lvec(3.075 0.703) \lvec(3.100 0.734) \lvec(3.125 0.765)
\lvec(3.150 0.796) \lvec(3.175 0.827) \lvec(3.200 0.858)
\lvec(3.225 0.889) \lvec(3.250 0.919) \lvec(3.275 0.950)
\lvec(3.300 0.981) \lvec(3.325 1.011) \lvec(3.350 1.041)
\lvec(3.375 1.072) \lvec(3.400 1.102) \lvec(3.425 1.132)
\lvec(3.450 1.162) \lvec(3.475 1.192) \lvec(3.500 1.221)
\lvec(3.525 1.251) \lvec(3.550 1.280) \lvec(3.575 1.310)
\lvec(3.600 1.339) \lvec(3.625 1.367) \lvec(3.650 1.396)
\lvec(3.675 1.425) \lvec(3.700 1.453) \lvec(3.725 1.481)
\lvec(3.750 1.509) \lvec(3.775 1.537) \lvec(3.800 1.564)
\lvec(3.825 1.591) \lvec(3.850 1.618) \lvec(3.875 1.645)
\lvec(3.900 1.671) \lvec(3.925 1.698) \lvec(3.950 1.723)
\lvec(3.975 1.749) \lvec(4.000 1.775) \lvec(4.025 1.800)
\lvec(4.050 1.824) \lvec(4.075 1.849) \lvec(4.100 1.873)
\lvec(4.125 1.897) \lvec(4.150 1.921) \lvec(4.175 1.944)
\lvec(4.200 1.967) \lvec(4.225 1.990) \lvec(4.250 2.012)
\lvec(4.275 2.034) \lvec(4.300 2.055) \lvec(4.325 2.077)
\lvec(4.350 2.098) \lvec(4.375 2.118) \lvec(4.400 2.138)
\lvec(4.425 2.158) \lvec(4.450 2.177) \lvec(4.475 2.196)
\lvec(4.500 2.215) \lvec(4.525 2.233) \lvec(4.550 2.251)
\lvec(4.575 2.269) \lvec(4.600 2.286) \lvec(4.625 2.302)
\lvec(4.650 2.319) \lvec(4.675 2.334) \lvec(4.700 2.350)
\lvec(4.725 2.365) \lvec(4.750 2.379) \lvec(4.775 2.393)
\lvec(4.800 2.407) \lvec(4.825 2.420) \lvec(4.850 2.433)
\lvec(4.875 2.445) \lvec(4.900 2.457) \lvec(4.925 2.468)
\lvec(4.950 2.479) \lvec(4.975 2.490) \lvec(5.000 2.500)
\move(0.000 1.250) \lvec(0.025 1.269) \lvec(0.050 1.288)
\lvec(0.075 1.307) \lvec(0.100 1.325) \lvec(0.125 1.342)
\lvec(0.150 1.359) \lvec(0.175 1.376) \lvec(0.200 1.392)
\lvec(0.225 1.407) \lvec(0.250 1.422) \lvec(0.275 1.437)
\lvec(0.300 1.451) \lvec(0.325 1.465) \lvec(0.350 1.478)
\lvec(0.375 1.490) \lvec(0.400 1.502) \lvec(0.425 1.513)
\lvec(0.450 1.524) \lvec(0.475 1.535) \lvec(0.500 1.545)
\lvec(0.525 1.554) \lvec(0.550 1.563) \lvec(0.575 1.571)
\lvec(0.600 1.578) \lvec(0.625 1.586) \lvec(0.650 1.592)
\lvec(0.675 1.598) \lvec(0.700 1.604) \lvec(0.725 1.609)
\lvec(0.750 1.613) \lvec(0.775 1.617) \lvec(0.800 1.620)
\lvec(0.825 1.623) \lvec(0.850 1.625) \lvec(0.875 1.627)
\lvec(0.900 1.628) \lvec(0.925 1.628) \lvec(0.950 1.628)
\lvec(0.975 1.628) \lvec(1.000 1.627) \lvec(1.025 1.625)
\lvec(1.050 1.623) \lvec(1.075 1.620) \lvec(1.100 1.616)
\lvec(1.125 1.613) \lvec(1.150 1.608) \lvec(1.175 1.603)
\lvec(1.200 1.597) \lvec(1.225 1.591) \lvec(1.250 1.585)
\lvec(1.275 1.577) \lvec(1.300 1.570) \lvec(1.325 1.561)
\lvec(1.350 1.553) \lvec(1.375 1.543) \lvec(1.400 1.533)
\lvec(1.425 1.523) \lvec(1.450 1.512) \lvec(1.475 1.500)
\lvec(1.500 1.488) \lvec(1.525 1.476) \lvec(1.550 1.463)
\lvec(1.575 1.449) \lvec(1.600 1.435) \lvec(1.625 1.420)
\lvec(1.650 1.405) \lvec(1.675 1.390) \lvec(1.700 1.374)
\lvec(1.725 1.357) \lvec(1.750 1.340) \lvec(1.775 1.322)
\lvec(1.800 1.304) \lvec(1.825 1.286) \lvec(1.850 1.267)
\lvec(1.875 1.247) \lvec(1.900 1.227) \lvec(1.925 1.207)
\lvec(1.950 1.186) \lvec(1.975 1.165) \lvec(2.000 1.143)
\lvec(2.025 1.121) \lvec(2.050 1.098) \lvec(2.075 1.075)
\lvec(2.100 1.052) \lvec(2.125 1.028) \lvec(2.150 1.004)
\lvec(2.175 0.979) \lvec(2.200 0.954) \lvec(2.225 0.929)
\lvec(2.250 0.903) \lvec(2.275 0.877) \lvec(2.300 0.850)
\lvec(2.325 0.823) \lvec(2.350 0.796) \lvec(2.375 0.768)
\lvec(2.400 0.740) \lvec(2.425 0.712) \lvec(2.450 0.683)
\lvec(2.475 0.654) \lvec(2.500 0.625) \lvec(2.525 0.662)
\lvec(2.550 0.697) \lvec(2.575 0.728) \lvec(2.600 0.757)
\lvec(2.625 0.784) \lvec(2.650 0.808) \lvec(2.675 0.831)
\lvec(2.700 0.853) \lvec(2.725 0.872) \lvec(2.750 0.891)
\lvec(2.775 0.908) \lvec(2.800 0.923) \lvec(2.825 0.938)
\lvec(2.850 0.951) \lvec(2.875 0.964) \lvec(2.900 0.975)
\lvec(2.925 0.985) \lvec(2.950 0.994) \lvec(2.975 1.003)
\lvec(3.000 1.010) \lvec(3.025 1.017) \lvec(3.050 1.023)
\lvec(3.075 1.028) \lvec(3.100 1.032) \lvec(3.125 1.035)
\lvec(3.150 1.038) \lvec(3.175 1.040) \lvec(3.200 1.041)
\lvec(3.225 1.042) \lvec(3.250 1.042) \lvec(3.275 1.041)
\lvec(3.300 1.039) \lvec(3.325 1.037) \lvec(3.350 1.035)
\lvec(3.375 1.031) \lvec(3.400 1.027) \lvec(3.425 1.023)
\lvec(3.450 1.018) \lvec(3.475 1.012) \lvec(3.500 1.006)
\lvec(3.525 1.000) \lvec(3.550 0.993) \lvec(3.575 0.985)
\lvec(3.600 0.977) \lvec(3.625 0.968) \lvec(3.650 0.959)
\lvec(3.675 0.949) \lvec(3.700 0.939) \lvec(3.725 0.929)
\lvec(3.750 0.918) \lvec(3.775 0.907) \lvec(3.800 0.895)
\lvec(3.825 0.882) \lvec(3.850 0.870) \lvec(3.875 0.857)
\lvec(3.900 0.843) \lvec(3.925 0.830) \lvec(3.950 0.815)
\lvec(3.975 0.801) \lvec(4.000 0.786) \lvec(4.025 0.771)
\lvec(4.050 0.755) \lvec(4.075 0.739) \lvec(4.100 0.723)
\lvec(4.125 0.707) \lvec(4.150 0.690) \lvec(4.175 0.673)
\lvec(4.200 0.655) \lvec(4.225 0.638) \lvec(4.250 0.620)
\lvec(4.275 0.601) \lvec(4.300 0.583) \lvec(4.325 0.564)
\lvec(4.350 0.545) \lvec(4.375 0.526) \lvec(4.400 0.507)
\lvec(4.425 0.487) \lvec(4.450 0.468) \lvec(4.475 0.448)
\lvec(4.500 0.428) \lvec(4.525 0.407) \lvec(4.550 0.387)
\lvec(4.575 0.366) \lvec(4.600 0.345) \lvec(4.625 0.324)
\lvec(4.650 0.303) \lvec(4.675 0.282) \lvec(4.700 0.261)
\lvec(4.725 0.240) \lvec(4.750 0.218) \lvec(4.775 0.197)
\lvec(4.800 0.175) \lvec(4.825 0.153) \lvec(4.850 0.131)
\lvec(4.875 0.110) \lvec(4.900 0.088) \lvec(4.925 0.066)
\lvec(4.950 0.044) \lvec(4.975 0.022) \lvec(5.000 0.000)
\move(0.000 1.250) \lvec(0.025 1.279) \lvec(0.050 1.308)
\lvec(0.075 1.336) \lvec(0.100 1.364) \lvec(0.125 1.391)
\lvec(0.150 1.418) \lvec(0.175 1.444) \lvec(0.200 1.470)
\lvec(0.225 1.495) \lvec(0.250 1.520) \lvec(0.275 1.544)
\lvec(0.300 1.568) \lvec(0.325 1.591) \lvec(0.350 1.614)
\lvec(0.375 1.636) \lvec(0.400 1.657) \lvec(0.425 1.678)
\lvec(0.450 1.699) \lvec(0.475 1.718) \lvec(0.500 1.738)
\lvec(0.525 1.756) \lvec(0.550 1.774) \lvec(0.575 1.792)
\lvec(0.600 1.809) \lvec(0.625 1.825) \lvec(0.650 1.840)
\lvec(0.675 1.855) \lvec(0.700 1.870) \lvec(0.725 1.884)
\lvec(0.750 1.897) \lvec(0.775 1.909) \lvec(0.800 1.921)
\lvec(0.825 1.933) \lvec(0.850 1.943) \lvec(0.875 1.953)
\lvec(0.900 1.963) \lvec(0.925 1.972) \lvec(0.950 1.980)
\lvec(0.975 1.987) \lvec(1.000 1.994) \lvec(1.025 2.000)
\lvec(1.050 2.006) \lvec(1.075 2.011) \lvec(1.100 2.015)
\lvec(1.125 2.018) \lvec(1.150 2.021) \lvec(1.175 2.024)
\lvec(1.200 2.025) \lvec(1.225 2.026) \lvec(1.250 2.027)
\lvec(1.275 2.026) \lvec(1.300 2.046) \lvec(1.325 2.089)
\lvec(1.350 2.132) \lvec(1.375 2.174) \lvec(1.400 2.214)
\lvec(1.425 2.254) \lvec(1.450 2.293) \lvec(1.475 2.331)
\lvec(1.500 2.368) \lvec(1.525 2.404) \lvec(1.550 2.440)
\lvec(1.575 2.474) \lvec(1.600 2.508) \lvec(1.625 2.541)
\lvec(1.650 2.572) \lvec(1.675 2.603) \lvec(1.700 2.633)
\lvec(1.725 2.663) \lvec(1.750 2.691) \lvec(1.775 2.719)
\lvec(1.800 2.745) \lvec(1.825 2.771) \lvec(1.850 2.796)
\lvec(1.875 2.820) \lvec(1.900 2.844) \lvec(1.925 2.866)
\lvec(1.950 2.888) \lvec(1.975 2.909) \lvec(2.000 2.928)
\lvec(2.025 2.947) \lvec(2.050 2.966) \lvec(2.075 2.983)
\lvec(2.100 2.999) \lvec(2.125 3.015) \lvec(2.150 3.030)
\lvec(2.175 3.044) \lvec(2.200 3.057) \lvec(2.225 3.069)
\lvec(2.250 3.080) \lvec(2.275 3.091) \lvec(2.300 3.100)
\lvec(2.325 3.109) \lvec(2.350 3.117) \lvec(2.375 3.124)
\lvec(2.400 3.130) \lvec(2.425 3.135) \lvec(2.450 3.140)
\lvec(2.475 3.143) \lvec(2.500 3.146) \lvec(2.525 3.148)
\lvec(2.550 3.149) \lvec(2.575 3.149) \lvec(2.600 3.149)
\lvec(2.625 3.147) \lvec(2.650 3.145) \lvec(2.675 3.142)
\lvec(2.700 3.138) \lvec(2.725 3.133) \lvec(2.750 3.127)
\lvec(2.775 3.121) \lvec(2.800 3.113) \lvec(2.825 3.105)
\lvec(2.850 3.096) \lvec(2.875 3.086) \lvec(2.900 3.076)
\lvec(2.925 3.064) \lvec(2.950 3.052) \lvec(2.975 3.039)
\lvec(3.000 3.025) \lvec(3.025 3.011) \lvec(3.050 2.995)
\lvec(3.075 2.979) \lvec(3.100 2.962) \lvec(3.125 2.944)
\lvec(3.150 2.926) \lvec(3.175 2.906) \lvec(3.200 2.886)
\lvec(3.225 2.866) \lvec(3.250 2.844) \lvec(3.275 2.822)
\lvec(3.300 2.799) \lvec(3.325 2.775) \lvec(3.350 2.751)
\lvec(3.375 2.725) \lvec(3.400 2.700) \lvec(3.425 2.673)
\lvec(3.450 2.646) \lvec(3.475 2.618) \lvec(3.500 2.589)
\lvec(3.525 2.560) \lvec(3.550 2.530) \lvec(3.575 2.499)
\lvec(3.600 2.468) \lvec(3.625 2.436) \lvec(3.650 2.404)
\lvec(3.675 2.371) \lvec(3.700 2.337) \lvec(3.725 2.302)
\lvec(3.750 2.268) \lvec(3.775 2.232) \lvec(3.800 2.196)
\lvec(3.825 2.159) \lvec(3.850 2.122) \lvec(3.875 2.084)
\lvec(3.900 2.046) \lvec(3.925 2.007) \lvec(3.950 1.968)
\lvec(3.975 1.928) \lvec(4.000 1.888) \lvec(4.025 1.847)
\lvec(4.050 1.806) \lvec(4.075 1.764) \lvec(4.100 1.722)
\lvec(4.125 1.679) \lvec(4.150 1.636) \lvec(4.175 1.592)
\lvec(4.200 1.549) \lvec(4.225 1.504) \lvec(4.250 1.460)
\lvec(4.275 1.415) \lvec(4.300 1.369) \lvec(4.325 1.323)
\lvec(4.350 1.277) \lvec(4.375 1.231) \lvec(4.400 1.184)
\lvec(4.425 1.137) \lvec(4.450 1.090) \lvec(4.475 1.042)
\lvec(4.500 0.994) \lvec(4.525 0.946) \lvec(4.550 0.898)
\lvec(4.575 0.849) \lvec(4.600 0.800) \lvec(4.625 0.751)
\lvec(4.650 0.702) \lvec(4.675 0.653) \lvec(4.700 0.603)
\lvec(4.725 0.553) \lvec(4.750 0.504) \lvec(4.775 0.454)
\lvec(4.800 0.403) \lvec(4.825 0.353) \lvec(4.850 0.303)
\lvec(4.875 0.253) \lvec(4.900 0.202) \lvec(4.925 0.152)
\lvec(4.950 0.101) \lvec(4.975 0.051) \lvec(5.000 0.000)
\move(0.000 1.250) \lvec(0.025 1.289) \lvec(0.050 1.328)
\lvec(0.075 1.366) \lvec(0.100 1.403) \lvec(0.125 1.440)
\lvec(0.150 1.477) \lvec(0.175 1.513) \lvec(0.200 1.549)
\lvec(0.225 1.584) \lvec(0.250 1.618) \lvec(0.275 1.652)
\lvec(0.300 1.685) \lvec(0.325 1.718) \lvec(0.350 1.750)
\lvec(0.375 1.782) \lvec(0.400 1.813) \lvec(0.425 1.843)
\lvec(0.450 1.873) \lvec(0.475 1.902) \lvec(0.500 1.931)
\lvec(0.525 1.959) \lvec(0.550 1.986) \lvec(0.575 2.013)
\lvec(0.600 2.039) \lvec(0.625 2.064) \lvec(0.650 2.089)
\lvec(0.675 2.113) \lvec(0.700 2.136) \lvec(0.725 2.159)
\lvec(0.750 2.181) \lvec(0.775 2.202) \lvec(0.800 2.222)
\lvec(0.825 2.242) \lvec(0.850 2.261) \lvec(0.875 2.280)
\lvec(0.900 2.326) \lvec(0.925 2.393) \lvec(0.950 2.459)
\lvec(0.975 2.524) \lvec(1.000 2.588) \lvec(1.025 2.651)
\lvec(1.050 2.714) \lvec(1.075 2.775) \lvec(1.100 2.836)
\lvec(1.125 2.896) \lvec(1.150 2.954) \lvec(1.175 3.012)
\lvec(1.200 3.069) \lvec(1.225 3.125) \lvec(1.250 3.180)
\lvec(1.275 3.234) \lvec(1.300 3.287) \lvec(1.325 3.340)
\lvec(1.350 3.391) \lvec(1.375 3.442) \lvec(1.400 3.491)
\lvec(1.425 3.539) \lvec(1.450 3.587) \lvec(1.475 3.633)
\lvec(1.500 3.679) \lvec(1.525 3.724) \lvec(1.550 3.767)
\lvec(1.575 3.810) \lvec(1.600 3.851) \lvec(1.625 3.892)
\lvec(1.650 3.931) \lvec(1.675 3.970) \lvec(1.700 4.007)
\lvec(1.725 4.043) \lvec(1.750 4.079) \lvec(1.775 4.113)
\lvec(1.800 4.146) \lvec(1.825 4.178) \lvec(1.850 4.209)
\lvec(1.875 4.239) \lvec(1.900 4.268) \lvec(1.925 4.295)
\lvec(1.950 4.322) \lvec(1.975 4.347) \lvec(2.000 4.372)
\lvec(2.025 4.395) \lvec(2.050 4.417) \lvec(2.075 4.438)
\lvec(2.100 4.458) \lvec(2.125 4.477) \lvec(2.150 4.494)
\lvec(2.175 4.511) \lvec(2.200 4.526) \lvec(2.225 4.540)
\lvec(2.250 4.553) \lvec(2.275 4.565) \lvec(2.300 4.575)
\lvec(2.325 4.585) \lvec(2.350 4.593) \lvec(2.375 4.600)
\lvec(2.400 4.606) \lvec(2.425 4.611) \lvec(2.450 4.615)
\lvec(2.475 4.618) \lvec(2.500 4.619) \lvec(2.525 4.619)
\lvec(2.550 4.618) \lvec(2.575 4.616) \lvec(2.600 4.613)
\lvec(2.625 4.608) \lvec(2.650 4.603) \lvec(2.675 4.596)
\lvec(2.700 4.588) \lvec(2.725 4.579) \lvec(2.750 4.569)
\lvec(2.775 4.557) \lvec(2.800 4.545) \lvec(2.825 4.531)
\lvec(2.850 4.516) \lvec(2.875 4.500) \lvec(2.900 4.483)
\lvec(2.925 4.465) \lvec(2.950 4.446) \lvec(2.975 4.426)
\lvec(3.000 4.404) \lvec(3.025 4.381) \lvec(3.050 4.358)
\lvec(3.075 4.333) \lvec(3.100 4.307) \lvec(3.125 4.280)
\lvec(3.150 4.252) \lvec(3.175 4.223) \lvec(3.200 4.193)
\lvec(3.225 4.162) \lvec(3.250 4.129) \lvec(3.275 4.096)
\lvec(3.300 4.062) \lvec(3.325 4.026) \lvec(3.350 3.990)
\lvec(3.375 3.953) \lvec(3.400 3.915) \lvec(3.425 3.875)
\lvec(3.450 3.835) \lvec(3.475 3.794) \lvec(3.500 3.752)
\lvec(3.525 3.709) \lvec(3.550 3.665) \lvec(3.575 3.620)
\lvec(3.600 3.574) \lvec(3.625 3.527) \lvec(3.650 3.480)
\lvec(3.675 3.431) \lvec(3.700 3.382) \lvec(3.725 3.332)
\lvec(3.750 3.281) \lvec(3.775 3.229) \lvec(3.800 3.176)
\lvec(3.825 3.123) \lvec(3.850 3.069) \lvec(3.875 3.014)
\lvec(3.900 2.958) \lvec(3.925 2.902) \lvec(3.950 2.844)
\lvec(3.975 2.787) \lvec(4.000 2.728) \lvec(4.025 2.669)
\lvec(4.050 2.609) \lvec(4.075 2.548) \lvec(4.100 2.487)
\lvec(4.125 2.425) \lvec(4.150 2.363) \lvec(4.175 2.300)
\lvec(4.200 2.236) \lvec(4.225 2.172) \lvec(4.250 2.108)
\lvec(4.275 2.042) \lvec(4.300 1.977) \lvec(4.325 1.910)
\lvec(4.350 1.844) \lvec(4.375 1.777) \lvec(4.400 1.709)
\lvec(4.425 1.641) \lvec(4.450 1.573) \lvec(4.475 1.504)
\lvec(4.500 1.435) \lvec(4.525 1.365) \lvec(4.550 1.295)
\lvec(4.575 1.225) \lvec(4.600 1.155) \lvec(4.625 1.084)
\lvec(4.650 1.013) \lvec(4.675 0.942) \lvec(4.700 0.870)
\lvec(4.725 0.798) \lvec(4.750 0.726) \lvec(4.775 0.654)
\lvec(4.800 0.582) \lvec(4.825 0.510) \lvec(4.850 0.437)
\lvec(4.875 0.364) \lvec(4.900 0.292) \lvec(4.925 0.219)
\lvec(4.950 0.146) \lvec(4.975 0.073) \lvec(5.000 0.000)
\move (-1 -1) \move (5.2 5.2)
\end{texdraw}
\hfill
\begin{texdraw}
\drawdim truecm \setgray 0
\linewd 0.05
\move (0 0) \lvec (5.0 0)
\lvec (5.0 5.0) \lvec (0 5.0) \lvec (0 0)
\linewd 0.03
\textref h:C v:T \htext (0.00 -.20) {0}
\move (1.250 0) \lvec (1.250 0.100)
\move (2.500 0) \lvec (2.500 0.200)
\textref h:C v:T \htext (2.50 -.20) {$\frac{\pi}{4}$}
\move (3.750 0) \lvec (3.750 0.100)
\textref h:C v:T \htext (5.00 -.20) {$\frac{\pi}{2}$}
\textref h:R v:C \htext (-.20 0.00) {-2}
\move (0 1.250) \lvec (0.100 1.250)
\move (0 2.500) \lvec (0.200 2.500)
\textref h:R v:C \htext (-.20 2.50) {-1}
\move (0 3.750) \lvec (0.100 3.750)
\textref h:R v:C \htext (-.20 5.00) {0}
\textref h:C v:T \htext (3.5 -.75) {$\alpha$}
\textref h:R v:C \htext (-.5 3.50) {$\dps\frac{\tau_{\rm opt}}{\lambda_0}$}
\arrowheadtype t:V \arrowheadsize l:0.15 w:.1
\move (1.25 4.7) \clvec (1.15 4.5)(1 4.3)(0.8 4.1)
\ravec (-0.2 -0.2)
\textref h:C v:T \htext (0.5 3.8) {$\qc$}
\move(0.000 5.000) \lvec(0.025 4.985) \lvec(0.050 4.971)
\lvec(0.075 4.956) \lvec(0.100 4.941) \lvec(0.125 4.926)
\lvec(0.150 4.911) \lvec(0.175 4.897) \lvec(0.200 4.882)
\lvec(0.225 4.867) \lvec(0.250 4.853) \lvec(0.275 4.838)
\lvec(0.300 4.823) \lvec(0.325 4.809) \lvec(0.350 4.794)
\lvec(0.375 4.779) \lvec(0.400 4.765) \lvec(0.425 4.750)
\lvec(0.450 4.736) \lvec(0.475 4.722) \lvec(0.500 4.707)
\lvec(0.525 4.693) \lvec(0.550 4.679) \lvec(0.575 4.665)
\lvec(0.600 4.650) \lvec(0.625 4.636) \lvec(0.650 4.623)
\lvec(0.675 4.609) \lvec(0.700 4.595) \lvec(0.725 4.581)
\lvec(0.750 4.568) \lvec(0.775 4.554) \lvec(0.800 4.541)
\lvec(0.825 4.527) \lvec(0.850 4.514) \lvec(0.875 4.501)
\lvec(0.900 4.488) \lvec(0.925 4.475) \lvec(0.950 4.462)
\lvec(0.975 4.450) \lvec(1.000 4.437) \lvec(1.025 4.425)
\lvec(1.050 4.412) \lvec(1.075 4.400) \lvec(1.100 4.388)
\lvec(1.125 4.376) \lvec(1.150 4.364) \lvec(1.175 4.353)
\lvec(1.200 4.341) \lvec(1.225 4.330) \lvec(1.250 4.319)
\lvec(1.275 4.308) \lvec(1.300 4.297) \lvec(1.325 4.286)
\lvec(1.350 4.276) \lvec(1.375 4.265) \lvec(1.400 4.255)
\lvec(1.425 4.245) \lvec(1.450 4.235) \lvec(1.475 4.225)
\lvec(1.500 4.216) \lvec(1.525 4.206) \lvec(1.550 4.197)
\lvec(1.575 4.188) \lvec(1.600 4.179) \lvec(1.625 4.171)
\lvec(1.650 4.162) \lvec(1.675 4.154) \lvec(1.700 4.146)
\lvec(1.725 4.138) \lvec(1.750 4.131) \lvec(1.775 4.123)
\lvec(1.800 4.116) \lvec(1.825 4.109) \lvec(1.850 4.102)
\lvec(1.875 4.095) \lvec(1.900 4.089) \lvec(1.925 4.083)
\lvec(1.950 4.077) \lvec(1.975 4.071) \lvec(2.000 4.065)
\lvec(2.025 4.060) \lvec(2.050 4.055) \lvec(2.075 4.050)
\lvec(2.100 4.045) \lvec(2.125 4.040) \lvec(2.150 4.036)
\lvec(2.175 4.032) \lvec(2.200 4.028) \lvec(2.225 4.025)
\lvec(2.250 4.021) \lvec(2.275 4.018) \lvec(2.300 4.015)
\lvec(2.325 4.013) \lvec(2.350 4.010) \lvec(2.375 4.008)
\lvec(2.400 4.006) \lvec(2.425 4.004) \lvec(2.450 4.002)
\lvec(2.475 4.001) \lvec(2.500 4.000) \lvec(2.525 3.999)
\lvec(2.550 3.998) \lvec(2.575 3.998) \lvec(2.600 3.998)
\lvec(2.625 3.998) \lvec(2.650 3.998) \lvec(2.675 3.999)
\lvec(2.700 4.000) \lvec(2.725 4.001) \lvec(2.750 4.002)
\lvec(2.775 4.003) \lvec(2.800 4.005) \lvec(2.825 4.007)
\lvec(2.850 4.009) \lvec(2.875 4.011) \lvec(2.900 4.014)
\lvec(2.925 4.017) \lvec(2.950 4.020) \lvec(2.975 4.023)
\lvec(3.000 4.027) \lvec(3.025 4.030) \lvec(3.050 4.034)
\lvec(3.075 4.038) \lvec(3.100 4.043) \lvec(3.125 4.047)
\lvec(3.150 4.052) \lvec(3.175 4.057) \lvec(3.200 4.063)
\lvec(3.225 4.068) \lvec(3.250 4.074) \lvec(3.275 4.080)
\lvec(3.300 4.086) \lvec(3.325 4.092) \lvec(3.350 4.099)
\lvec(3.375 4.105) \lvec(3.400 4.112) \lvec(3.425 4.120)
\lvec(3.450 4.127) \lvec(3.475 4.135) \lvec(3.500 4.142)
\lvec(3.525 4.150) \lvec(3.550 4.158) \lvec(3.575 4.167)
\lvec(3.600 4.175) \lvec(3.625 4.184) \lvec(3.650 4.193)
\lvec(3.675 4.202) \lvec(3.700 4.211) \lvec(3.725 4.221)
\lvec(3.750 4.230) \lvec(3.775 4.240) \lvec(3.800 4.250)
\lvec(3.825 4.260) \lvec(3.850 4.271) \lvec(3.875 4.281)
\lvec(3.900 4.292) \lvec(3.925 4.303) \lvec(3.950 4.314)
\lvec(3.975 4.325) \lvec(4.000 4.336) \lvec(4.025 4.347)
\lvec(4.050 4.359) \lvec(4.075 4.371) \lvec(4.100 4.382)
\lvec(4.125 4.394) \lvec(4.150 4.406) \lvec(4.175 4.419)
\lvec(4.200 4.431) \lvec(4.225 4.444) \lvec(4.250 4.456)
\lvec(4.275 4.469) \lvec(4.300 4.482) \lvec(4.325 4.495)
\lvec(4.350 4.508) \lvec(4.375 4.521) \lvec(4.400 4.534)
\lvec(4.425 4.548) \lvec(4.450 4.561) \lvec(4.475 4.575)
\lvec(4.500 4.588) \lvec(4.525 4.602) \lvec(4.550 4.616)
\lvec(4.575 4.630) \lvec(4.600 4.644) \lvec(4.625 4.658)
\lvec(4.650 4.672) \lvec(4.675 4.686) \lvec(4.700 4.700)
\lvec(4.725 4.715) \lvec(4.750 4.729) \lvec(4.775 4.744)
\lvec(4.800 4.758) \lvec(4.825 4.773) \lvec(4.850 4.787)
\lvec(4.875 4.802) \lvec(4.900 4.816) \lvec(4.925 4.831)
\lvec(4.950 4.846) \lvec(4.975 4.860) \lvec(5.000 4.875)
\move(0.000 5.000) \lvec(0.025 4.985) \lvec(0.050 4.970)
\lvec(0.075 4.955) \lvec(0.100 4.941) \lvec(0.125 4.925)
\lvec(0.150 4.910) \lvec(0.175 4.895) \lvec(0.200 4.880)
\lvec(0.225 4.865) \lvec(0.250 4.849) \lvec(0.275 4.834)
\lvec(0.300 4.819) \lvec(0.325 4.803) \lvec(0.350 4.788)
\lvec(0.375 4.773) \lvec(0.400 4.757) \lvec(0.425 4.742)
\lvec(0.450 4.726) \lvec(0.475 4.711) \lvec(0.500 4.695)
\lvec(0.525 4.679) \lvec(0.550 4.664) \lvec(0.575 4.648)
\lvec(0.600 4.633) \lvec(0.625 4.617) \lvec(0.650 4.602)
\lvec(0.675 4.587) \lvec(0.700 4.571) \lvec(0.725 4.556)
\lvec(0.750 4.540) \lvec(0.775 4.525) \lvec(0.800 4.510)
\lvec(0.825 4.494) \lvec(0.850 4.479) \lvec(0.875 4.464)
\lvec(0.900 4.449) \lvec(0.925 4.434) \lvec(0.950 4.419)
\lvec(0.975 4.404) \lvec(1.000 4.389) \lvec(1.025 4.375)
\lvec(1.050 4.360) \lvec(1.075 4.345) \lvec(1.100 4.331)
\lvec(1.125 4.316) \lvec(1.150 4.302) \lvec(1.175 4.288)
\lvec(1.200 4.274) \lvec(1.225 4.259) \lvec(1.250 4.246)
\lvec(1.275 4.232) \lvec(1.300 4.218) \lvec(1.325 4.204)
\lvec(1.350 4.191) \lvec(1.375 4.178) \lvec(1.400 4.164)
\lvec(1.425 4.151) \lvec(1.450 4.138) \lvec(1.475 4.125)
\lvec(1.500 4.113) \lvec(1.525 4.100) \lvec(1.550 4.088)
\lvec(1.575 4.076) \lvec(1.600 4.063) \lvec(1.625 4.051)
\lvec(1.650 4.040) \lvec(1.675 4.028) \lvec(1.700 4.017)
\lvec(1.725 4.005) \lvec(1.750 3.994) \lvec(1.775 3.983)
\lvec(1.800 3.972) \lvec(1.825 3.962) \lvec(1.850 3.951)
\lvec(1.875 3.941) \lvec(1.900 3.931) \lvec(1.925 3.921)
\lvec(1.950 3.911) \lvec(1.975 3.902) \lvec(2.000 3.892)
\lvec(2.025 3.883) \lvec(2.050 3.874) \lvec(2.075 3.866)
\lvec(2.100 3.857) \lvec(2.125 3.849) \lvec(2.150 3.841)
\lvec(2.175 3.833) \lvec(2.200 3.825) \lvec(2.225 3.818)
\lvec(2.250 3.810) \lvec(2.275 3.803) \lvec(2.300 3.797)
\lvec(2.325 3.790) \lvec(2.350 3.784) \lvec(2.375 3.777)
\lvec(2.400 3.771) \lvec(2.425 3.766) \lvec(2.450 3.760)
\lvec(2.475 3.755) \lvec(2.500 3.750) \lvec(2.525 3.745)
\lvec(2.550 3.741) \lvec(2.575 3.736) \lvec(2.600 3.732)
\lvec(2.625 3.728) \lvec(2.650 3.725) \lvec(2.675 3.721)
\lvec(2.700 3.718) \lvec(2.725 3.715) \lvec(2.750 3.713)
\lvec(2.775 3.710) \lvec(2.800 3.708) \lvec(2.825 3.706)
\lvec(2.850 3.704) \lvec(2.875 3.703) \lvec(2.900 3.702)
\lvec(2.925 3.701) \lvec(2.950 3.700) \lvec(2.975 3.700)
\lvec(3.000 3.699) \lvec(3.025 3.699) \lvec(3.050 3.700)
\lvec(3.075 3.700) \lvec(3.100 3.701) \lvec(3.125 3.702)
\lvec(3.150 3.703) \lvec(3.175 3.704) \lvec(3.200 3.706)
\lvec(3.225 3.708) \lvec(3.250 3.710) \lvec(3.275 3.713)
\lvec(3.300 3.715) \lvec(3.325 3.718) \lvec(3.350 3.721)
\lvec(3.375 3.725) \lvec(3.400 3.728) \lvec(3.425 3.732)
\lvec(3.450 3.736) \lvec(3.475 3.741) \lvec(3.500 3.745)
\lvec(3.525 3.750) \lvec(3.550 3.755) \lvec(3.575 3.760)
\lvec(3.600 3.766) \lvec(3.625 3.772) \lvec(3.650 3.778)
\lvec(3.675 3.784) \lvec(3.700 3.790) \lvec(3.725 3.797)
\lvec(3.750 3.804) \lvec(3.775 3.811) \lvec(3.800 3.818)
\lvec(3.825 3.825) \lvec(3.850 3.833) \lvec(3.875 3.841)
\lvec(3.900 3.849) \lvec(3.925 3.857) \lvec(3.950 3.866)
\lvec(3.975 3.875) \lvec(4.000 3.884) \lvec(4.025 3.893)
\lvec(4.050 3.902) \lvec(4.075 3.912) \lvec(4.100 3.921)
\lvec(4.125 3.931) \lvec(4.150 3.941) \lvec(4.175 3.952)
\lvec(4.200 3.962) \lvec(4.225 3.973) \lvec(4.250 3.983)
\lvec(4.275 3.994) \lvec(4.300 4.006) \lvec(4.325 4.017)
\lvec(4.350 4.028) \lvec(4.375 4.040) \lvec(4.400 4.052)
\lvec(4.425 4.064) \lvec(4.450 4.076) \lvec(4.475 4.088)
\lvec(4.500 4.101) \lvec(4.525 4.113) \lvec(4.550 4.126)
\lvec(4.575 4.139) \lvec(4.600 4.152) \lvec(4.625 4.165)
\lvec(4.650 4.178) \lvec(4.675 4.191) \lvec(4.700 4.205)
\lvec(4.725 4.218) \lvec(4.750 4.232) \lvec(4.775 4.246)
\lvec(4.800 4.260) \lvec(4.825 4.274) \lvec(4.850 4.288)
\lvec(4.875 4.302) \lvec(4.900 4.317) \lvec(4.925 4.331)
\lvec(4.950 4.346) \lvec(4.975 4.360) \lvec(5.000 4.375)
\move(0.000 5.000) \lvec(0.025 4.985) \lvec(0.050 4.970)
\lvec(0.075 4.955) \lvec(0.100 4.940) \lvec(0.125 4.925)
\lvec(0.150 4.909) \lvec(0.175 4.893) \lvec(0.200 4.878)
\lvec(0.225 4.862) \lvec(0.250 4.846) \lvec(0.275 4.830)
\lvec(0.300 4.813) \lvec(0.325 4.797) \lvec(0.350 4.780)
\lvec(0.375 4.764) \lvec(0.400 4.747) \lvec(0.425 4.730)
\lvec(0.450 4.714) \lvec(0.475 4.697) \lvec(0.500 4.680)
\lvec(0.525 4.663) \lvec(0.550 4.645) \lvec(0.575 4.628)
\lvec(0.600 4.611) \lvec(0.625 4.594) \lvec(0.650 4.576)
\lvec(0.675 4.559) \lvec(0.700 4.541) \lvec(0.725 4.524)
\lvec(0.750 4.506) \lvec(0.775 4.489) \lvec(0.800 4.471)
\lvec(0.825 4.453) \lvec(0.850 4.436) \lvec(0.875 4.418)
\lvec(0.900 4.400) \lvec(0.925 4.383) \lvec(0.950 4.365)
\lvec(0.975 4.347) \lvec(1.000 4.330) \lvec(1.025 4.312)
\lvec(1.050 4.294) \lvec(1.075 4.277) \lvec(1.100 4.259)
\lvec(1.125 4.241) \lvec(1.150 4.224) \lvec(1.175 4.206)
\lvec(1.200 4.189) \lvec(1.225 4.171) \lvec(1.250 4.154)
\lvec(1.275 4.137) \lvec(1.300 4.119) \lvec(1.325 4.102)
\lvec(1.350 4.085) \lvec(1.375 4.068) \lvec(1.400 4.051)
\lvec(1.425 4.034) \lvec(1.450 4.017) \lvec(1.475 4.001)
\lvec(1.500 3.984) \lvec(1.525 3.967) \lvec(1.550 3.951)
\lvec(1.575 3.935) \lvec(1.600 3.918) \lvec(1.625 3.902)
\lvec(1.650 3.886) \lvec(1.675 3.870) \lvec(1.700 3.855)
\lvec(1.725 3.839) \lvec(1.750 3.823) \lvec(1.775 3.808)
\lvec(1.800 3.793) \lvec(1.825 3.778) \lvec(1.850 3.763)
\lvec(1.875 3.748) \lvec(1.900 3.733) \lvec(1.925 3.719)
\lvec(1.950 3.705) \lvec(1.975 3.690) \lvec(2.000 3.677)
\lvec(2.025 3.663) \lvec(2.050 3.649) \lvec(2.075 3.636)
\lvec(2.100 3.622) \lvec(2.125 3.609) \lvec(2.150 3.596)
\lvec(2.175 3.584) \lvec(2.200 3.571) \lvec(2.225 3.559)
\lvec(2.250 3.547) \lvec(2.275 3.535) \lvec(2.300 3.523)
\lvec(2.325 3.512) \lvec(2.350 3.500) \lvec(2.375 3.489)
\lvec(2.400 3.479) \lvec(2.425 3.468) \lvec(2.450 3.458)
\lvec(2.475 3.447) \lvec(2.500 3.438) \lvec(2.525 3.423)
\lvec(2.550 3.408) \lvec(2.575 3.393) \lvec(2.600 3.379)
\lvec(2.625 3.364) \lvec(2.650 3.349) \lvec(2.675 3.335)
\lvec(2.700 3.320) \lvec(2.725 3.305) \lvec(2.750 3.291)
\lvec(2.775 3.276) \lvec(2.800 3.262) \lvec(2.825 3.247)
\lvec(2.850 3.233) \lvec(2.875 3.219) \lvec(2.900 3.204)
\lvec(2.925 3.190) \lvec(2.950 3.176) \lvec(2.975 3.162)
\lvec(3.000 3.148) \lvec(3.025 3.134) \lvec(3.050 3.120)
\lvec(3.075 3.106) \lvec(3.100 3.092) \lvec(3.125 3.079)
\lvec(3.150 3.065) \lvec(3.175 3.052) \lvec(3.200 3.038)
\lvec(3.225 3.025) \lvec(3.250 3.012) \lvec(3.275 2.999)
\lvec(3.300 2.986) \lvec(3.325 2.973) \lvec(3.350 2.960)
\lvec(3.375 2.948) \lvec(3.400 2.935) \lvec(3.425 2.923)
\lvec(3.450 2.911) \lvec(3.475 2.898) \lvec(3.500 2.886)
\lvec(3.525 2.875) \lvec(3.550 2.863) \lvec(3.575 2.851)
\lvec(3.600 2.840) \lvec(3.625 2.829) \lvec(3.650 2.818)
\lvec(3.675 2.807) \lvec(3.700 2.796) \lvec(3.725 2.785)
\lvec(3.750 2.775) \lvec(3.775 2.764) \lvec(3.800 2.754)
\lvec(3.825 2.744) \lvec(3.850 2.734) \lvec(3.875 2.725)
\lvec(3.900 2.715) \lvec(3.925 2.706) \lvec(3.950 2.697)
\lvec(3.975 2.688) \lvec(4.000 2.679) \lvec(4.025 2.670)
\lvec(4.050 2.662) \lvec(4.075 2.654) \lvec(4.100 2.646)
\lvec(4.125 2.638) \lvec(4.150 2.631) \lvec(4.175 2.623)
\lvec(4.200 2.616) \lvec(4.225 2.609) \lvec(4.250 2.602)
\lvec(4.275 2.596) \lvec(4.300 2.589) \lvec(4.325 2.583)
\lvec(4.350 2.577) \lvec(4.375 2.571) \lvec(4.400 2.566)
\lvec(4.425 2.561) \lvec(4.450 2.555) \lvec(4.475 2.551)
\lvec(4.500 2.546) \lvec(4.525 2.541) \lvec(4.550 2.537)
\lvec(4.575 2.533) \lvec(4.600 2.529) \lvec(4.625 2.526)
\lvec(4.650 2.523) \lvec(4.675 2.519) \lvec(4.700 2.517)
\lvec(4.725 2.514) \lvec(4.750 2.512) \lvec(4.775 2.509)
\lvec(4.800 2.507) \lvec(4.825 2.506) \lvec(4.850 2.504)
\lvec(4.875 2.503) \lvec(4.900 2.502) \lvec(4.925 2.501)
\lvec(4.950 2.500) \lvec(4.975 2.500) \lvec(5.000 2.500)
\move(0.000 5.000) \lvec(0.025 4.985) \lvec(0.050 4.970)
\lvec(0.075 4.955) \lvec(0.100 4.939) \lvec(0.125 4.924)
\lvec(0.150 4.908) \lvec(0.175 4.891) \lvec(0.200 4.875)
\lvec(0.225 4.859) \lvec(0.250 4.842) \lvec(0.275 4.825)
\lvec(0.300 4.808) \lvec(0.325 4.790) \lvec(0.350 4.773)
\lvec(0.375 4.755) \lvec(0.400 4.737) \lvec(0.425 4.719)
\lvec(0.450 4.701) \lvec(0.475 4.683) \lvec(0.500 4.664)
\lvec(0.525 4.646) \lvec(0.550 4.627) \lvec(0.575 4.608)
\lvec(0.600 4.589) \lvec(0.625 4.570) \lvec(0.650 4.551)
\lvec(0.675 4.531) \lvec(0.700 4.512) \lvec(0.725 4.492)
\lvec(0.750 4.472) \lvec(0.775 4.452) \lvec(0.800 4.432)
\lvec(0.825 4.412) \lvec(0.850 4.392) \lvec(0.875 4.372)
\lvec(0.900 4.352) \lvec(0.925 4.331) \lvec(0.950 4.311)
\lvec(0.975 4.290) \lvec(1.000 4.270) \lvec(1.025 4.249)
\lvec(1.050 4.229) \lvec(1.075 4.208) \lvec(1.100 4.187)
\lvec(1.125 4.167) \lvec(1.150 4.146) \lvec(1.175 4.125)
\lvec(1.200 4.104) \lvec(1.225 4.083) \lvec(1.250 4.063)
\lvec(1.275 4.042) \lvec(1.300 4.022) \lvec(1.325 4.003)
\lvec(1.350 3.984) \lvec(1.375 3.964) \lvec(1.400 3.944)
\lvec(1.425 3.924) \lvec(1.450 3.904) \lvec(1.475 3.883)
\lvec(1.500 3.863) \lvec(1.525 3.842) \lvec(1.550 3.820)
\lvec(1.575 3.799) \lvec(1.600 3.777) \lvec(1.625 3.755)
\lvec(1.650 3.733) \lvec(1.675 3.711) \lvec(1.700 3.688)
\lvec(1.725 3.665) \lvec(1.750 3.642) \lvec(1.775 3.619)
\lvec(1.800 3.596) \lvec(1.825 3.573) \lvec(1.850 3.549)
\lvec(1.875 3.525) \lvec(1.900 3.501) \lvec(1.925 3.477)
\lvec(1.950 3.453) \lvec(1.975 3.428) \lvec(2.000 3.404)
\lvec(2.025 3.379) \lvec(2.050 3.354) \lvec(2.075 3.330)
\lvec(2.100 3.305) \lvec(2.125 3.280) \lvec(2.150 3.254)
\lvec(2.175 3.229) \lvec(2.200 3.204) \lvec(2.225 3.178)
\lvec(2.250 3.153) \lvec(2.275 3.127) \lvec(2.300 3.102)
\lvec(2.325 3.076) \lvec(2.350 3.050) \lvec(2.375 3.025)
\lvec(2.400 2.999) \lvec(2.425 2.973) \lvec(2.450 2.947)
\lvec(2.475 2.921) \lvec(2.500 2.895) \lvec(2.525 2.870)
\lvec(2.550 2.844) \lvec(2.575 2.818) \lvec(2.600 2.792)
\lvec(2.625 2.766) \lvec(2.650 2.741) \lvec(2.675 2.715)
\lvec(2.700 2.689) \lvec(2.725 2.664) \lvec(2.750 2.638)
\lvec(2.775 2.613) \lvec(2.800 2.587) \lvec(2.825 2.562)
\lvec(2.850 2.536) \lvec(2.875 2.511) \lvec(2.900 2.486)
\lvec(2.925 2.461) \lvec(2.950 2.436) \lvec(2.975 2.412)
\lvec(3.000 2.387) \lvec(3.025 2.362) \lvec(3.050 2.338)
\lvec(3.075 2.314) \lvec(3.100 2.290) \lvec(3.125 2.266)
\lvec(3.150 2.242) \lvec(3.175 2.218) \lvec(3.200 2.195)
\lvec(3.225 2.172) \lvec(3.250 2.148) \lvec(3.275 2.125)
\lvec(3.300 2.103) \lvec(3.325 2.080) \lvec(3.350 2.058)
\lvec(3.375 2.036) \lvec(3.400 2.014) \lvec(3.425 1.992)
\lvec(3.450 1.971) \lvec(3.475 1.949) \lvec(3.500 1.928)
\lvec(3.525 1.907) \lvec(3.550 1.887) \lvec(3.575 1.867)
\lvec(3.600 1.847) \lvec(3.625 1.827) \lvec(3.650 1.807)
\lvec(3.675 1.788) \lvec(3.700 1.769) \lvec(3.725 1.750)
\lvec(3.750 1.732) \lvec(3.775 1.714) \lvec(3.800 1.696)
\lvec(3.825 1.678) \lvec(3.850 1.661) \lvec(3.875 1.644)
\lvec(3.900 1.628) \lvec(3.925 1.611) \lvec(3.950 1.595)
\lvec(3.975 1.580) \lvec(4.000 1.564) \lvec(4.025 1.549)
\lvec(4.050 1.535) \lvec(4.075 1.520) \lvec(4.100 1.506)
\lvec(4.125 1.492) \lvec(4.150 1.479) \lvec(4.175 1.466)
\lvec(4.200 1.454) \lvec(4.225 1.441) \lvec(4.250 1.429)
\lvec(4.275 1.418) \lvec(4.300 1.407) \lvec(4.325 1.396)
\lvec(4.350 1.385) \lvec(4.375 1.375) \lvec(4.400 1.366)
\lvec(4.425 1.356) \lvec(4.450 1.347) \lvec(4.475 1.339)
\lvec(4.500 1.331) \lvec(4.525 1.323) \lvec(4.550 1.315)
\lvec(4.575 1.308) \lvec(4.600 1.302) \lvec(4.625 1.295)
\lvec(4.650 1.290) \lvec(4.675 1.284) \lvec(4.700 1.279)
\lvec(4.725 1.275) \lvec(4.750 1.270) \lvec(4.775 1.266)
\lvec(4.800 1.263) \lvec(4.825 1.260) \lvec(4.850 1.257)
\lvec(4.875 1.255) \lvec(4.900 1.253) \lvec(4.925 1.252)
\lvec(4.950 1.251) \lvec(4.975 1.250) \lvec(5.000 1.250)
\move(0.000 5.000) \lvec(0.025 4.985) \lvec(0.050 4.970)
\lvec(0.075 4.954) \lvec(0.100 4.939) \lvec(0.125 4.923)
\lvec(0.150 4.906) \lvec(0.175 4.890) \lvec(0.200 4.873)
\lvec(0.225 4.855) \lvec(0.250 4.838) \lvec(0.275 4.820)
\lvec(0.300 4.802) \lvec(0.325 4.784) \lvec(0.350 4.765)
\lvec(0.375 4.747) \lvec(0.400 4.728) \lvec(0.425 4.708)
\lvec(0.450 4.689) \lvec(0.475 4.669) \lvec(0.500 4.649)
\lvec(0.525 4.629) \lvec(0.550 4.609) \lvec(0.575 4.588)
\lvec(0.600 4.567) \lvec(0.625 4.546) \lvec(0.650 4.525)
\lvec(0.675 4.503) \lvec(0.700 4.482) \lvec(0.725 4.460)
\lvec(0.750 4.438) \lvec(0.775 4.416) \lvec(0.800 4.394)
\lvec(0.825 4.371) \lvec(0.850 4.349) \lvec(0.875 4.326)
\lvec(0.900 4.305) \lvec(0.925 4.285) \lvec(0.950 4.265)
\lvec(0.975 4.244) \lvec(1.000 4.223) \lvec(1.025 4.201)
\lvec(1.050 4.179) \lvec(1.075 4.156) \lvec(1.100 4.133)
\lvec(1.125 4.109) \lvec(1.150 4.085) \lvec(1.175 4.061)
\lvec(1.200 4.036) \lvec(1.225 4.010) \lvec(1.250 3.985)
\lvec(1.275 3.959) \lvec(1.300 3.932) \lvec(1.325 3.905)
\lvec(1.350 3.878) \lvec(1.375 3.850) \lvec(1.400 3.822)
\lvec(1.425 3.794) \lvec(1.450 3.765) \lvec(1.475 3.736)
\lvec(1.500 3.706) \lvec(1.525 3.676) \lvec(1.550 3.646)
\lvec(1.575 3.616) \lvec(1.600 3.585) \lvec(1.625 3.554)
\lvec(1.650 3.522) \lvec(1.675 3.491) \lvec(1.700 3.459)
\lvec(1.725 3.426) \lvec(1.750 3.394) \lvec(1.775 3.361)
\lvec(1.800 3.328) \lvec(1.825 3.295) \lvec(1.850 3.261)
\lvec(1.875 3.227) \lvec(1.900 3.193) \lvec(1.925 3.159)
\lvec(1.950 3.125) \lvec(1.975 3.090) \lvec(2.000 3.055)
\lvec(2.025 3.021) \lvec(2.050 2.985) \lvec(2.075 2.950)
\lvec(2.100 2.915) \lvec(2.125 2.879) \lvec(2.150 2.843)
\lvec(2.175 2.808) \lvec(2.200 2.772) \lvec(2.225 2.735)
\lvec(2.250 2.699) \lvec(2.275 2.663) \lvec(2.300 2.627)
\lvec(2.325 2.590) \lvec(2.350 2.554) \lvec(2.375 2.517)
\lvec(2.400 2.481) \lvec(2.425 2.444) \lvec(2.450 2.408)
\lvec(2.475 2.371) \lvec(2.500 2.334) \lvec(2.525 2.298)
\lvec(2.550 2.261) \lvec(2.575 2.224) \lvec(2.600 2.188)
\lvec(2.625 2.151) \lvec(2.650 2.115) \lvec(2.675 2.078)
\lvec(2.700 2.042) \lvec(2.725 2.005) \lvec(2.750 1.969)
\lvec(2.775 1.933) \lvec(2.800 1.897) \lvec(2.825 1.861)
\lvec(2.850 1.825) \lvec(2.875 1.789) \lvec(2.900 1.754)
\lvec(2.925 1.718) \lvec(2.950 1.683) \lvec(2.975 1.648)
\lvec(3.000 1.613) \lvec(3.025 1.578) \lvec(3.050 1.544)
\lvec(3.075 1.509) \lvec(3.100 1.475) \lvec(3.125 1.441)
\lvec(3.150 1.407) \lvec(3.175 1.374) \lvec(3.200 1.340)
\lvec(3.225 1.307) \lvec(3.250 1.274) \lvec(3.275 1.242)
\lvec(3.300 1.210) \lvec(3.325 1.178) \lvec(3.350 1.146)
\lvec(3.375 1.115) \lvec(3.400 1.083) \lvec(3.425 1.053)
\lvec(3.450 1.022) \lvec(3.475 0.992) \lvec(3.500 0.962)
\lvec(3.525 0.933) \lvec(3.550 0.904) \lvec(3.575 0.875)
\lvec(3.600 0.846) \lvec(3.625 0.818) \lvec(3.650 0.791)
\lvec(3.675 0.763) \lvec(3.700 0.736) \lvec(3.725 0.710)
\lvec(3.750 0.684) \lvec(3.775 0.658) \lvec(3.800 0.633)
\lvec(3.825 0.608) \lvec(3.850 0.583) \lvec(3.875 0.559)
\lvec(3.900 0.536) \lvec(3.925 0.513) \lvec(3.950 0.490)
\lvec(3.975 0.468) \lvec(4.000 0.446) \lvec(4.025 0.424)
\lvec(4.050 0.404) \lvec(4.075 0.383) \lvec(4.100 0.363)
\lvec(4.125 0.344) \lvec(4.150 0.325) \lvec(4.175 0.307)
\lvec(4.200 0.289) \lvec(4.225 0.271) \lvec(4.250 0.254)
\lvec(4.275 0.238) \lvec(4.300 0.222) \lvec(4.325 0.207)
\lvec(4.350 0.192) \lvec(4.375 0.178) \lvec(4.400 0.164)
\lvec(4.425 0.151) \lvec(4.450 0.138) \lvec(4.475 0.126)
\lvec(4.500 0.114) \lvec(4.525 0.103) \lvec(4.550 0.093)
\lvec(4.575 0.083) \lvec(4.600 0.073) \lvec(4.625 0.064)
\lvec(4.650 0.056) \lvec(4.675 0.048) \lvec(4.700 0.041)
\lvec(4.725 0.035) \lvec(4.750 0.029) \lvec(4.775 0.023)
\lvec(4.800 0.018) \lvec(4.825 0.014) \lvec(4.850 0.010)
\lvec(4.875 0.007) \lvec(4.900 0.005) \lvec(4.925 0.003)
\lvec(4.950 0.001) \lvec(4.975 0.000) \lvec(5.000 0.000)
\move (-1 -1) \move (5.2 5.2)
\end{texdraw}
\hfill\null
\caption{Curvature and torsion of an optimal-shaped
smectic-C* capillary. The plots correspond to the values
$\qc/\lambda_0=0.1,0.5,1.0,1.5,2$.}
\label{kata}
\end{figure}

Let us consider the particular case in which
\begin{equation}
K_1=K_2=K_3=\Gamma=\omega_P/\lambda=:K\qquad {\rm and}\qquad
\Gamma_1 = 0\;,
\label{hyp}
\end{equation}
while keeping the thin capillary regime $\lambda_0 r\ll 1$. The
optimal shape of the capillary axis depends on the tilt angle
$\alpha_0$ of the smectic-C* molecules and on the cholesteric
pitch $\qc$. Figure \ref{kata} illustrates the results. The right
panel (displaying the torsion) proves the three-dimensional
character of the capillary axis. The torsion is enhanced by the
presence of a cholesteric pitch. However, a non-zero $\qc$ is not
a necessary ingredient to obtain three-dimensional shapes. In
fact, if we add $\qc=0$ to (\ref{hyp}), we can derive an
analytical expression for the optimal shape for all values of
$\alpha_0$:
$$
\kappa_{\rm opt}
\big|_{\qc=0}=\frac{|3\cos2\alpha_0-1|}{8}\,\lambda_0 \qquad {\rm
and}\qquad \tau_{\rm
opt}\big|_{\qc=0}=-\frac{3}{8}\,\sin2\alpha_0\;\lambda_0\;.
$$
On the contrary, the role played by $\lambda_0$ (\emph{i.e.\/},
the spontaneous polarization) is crucial. The ratio between either
$\kappa,\tau$ and $\lambda_0$ is finite. Thus, both $\kappa$ and
$\tau$ vanish when $\lambda_0$ does so. This observation is
consistent with the results presented in Section \ref{nonpol},
where we have proved that the optimal capillary shape is linear if
the spontaneous polarization is null. Figure 3 is also coherent
with the result derived in Section \ref{bent} for a smectic-A*: in
the limit $\alpha\to 0$, the torsion vanishes while the curvature
does not.

\subsubsection{Small tilt angle}\label{smalltilt}

Let us now analyse in more detail the small-$\alpha_0$ limit. If
the bend elastic constant prevails again over the effective
anchoring \big($K_3 \geq\omega_P/(2\lambda)$\big), we obtain:
$$
\kappa_{\rm opt}= \frac{K_3-\omega_P/(2\lambda)}
{K_3+\Gamma+\Gamma_1}\,\lambda_0+ O(\alpha_0)\,,\quad \tau_{\rm
opt}=-\frac{K_3} {\Gamma}\left(1- \frac{K_3-\omega_P/(2\lambda)}
{K_3+\Gamma+\Gamma_1}\right)\lambda_0\alpha_0+O(\alpha_0^2)
\quad{\rm as}\ \alpha_0\to 0\;.
$$
These results display the same qualitative features of the
1-constant solution analysed above. The preferred curvature
becomes different from zero as soon as the spontaneous
polarization appears, even when $\alpha_0$ vanishes. On the
contrary, the torsion vanishes when either $\lambda_0$ or
$\alpha_0$ do so. However, a new and interesting result stems from
the computation of the optimal free energy up to $O(\alpha^2)$. We
obtain:
\begin{equation}
\frac{\mathcal{F}_{\rm opt}}{\mathcal{A}\ell}=\sigma_{\rm
sm}(\alpha_0)+ \left[K_3 \lambda_0^2-\frac{
\big(K_3-\frac{\omega_P}{2\lambda}\big)^2\lambda_0^2}{K_3+\Gamma+
\Gamma_1}+K_2\qc^2+\frac{ 2\omega_P\lambda_0}{\lambda
r}\right]-\frac{2K_2\big(K_3-\frac{\omega_P}{2\lambda}\big)\lambda_0
\,\qc\,\alpha_0} {K_3+\Gamma+\Gamma_1}+O(\alpha_0^2)\;.
\label{last}
\end{equation}
The minus sign in front of the first-order term in
$\mathcal{F}_{\rm opt}$ is crucial. It implies that it is possible
to \emph{gain\/} free energy by tilting the director with respect
to the layer normal. This result holds even if $\sigma_{\rm
sm}(\alpha_0)$ pushes towards the smectic-A state, because in that
case $\sigma_{\rm sm}$ is minimum when $\alpha_0=0$, so that it
does not contribute to the $O(\alpha_0)$-term we are discussing.
The structure of the first-order term in $\alpha_0$ shows that
this instabilization of the smectic-A* phase is a combined effect
of both the spontaneous polarization and the cholesteric pitch.
Once $\alpha_0$ becomes non-null, a non-zero value of the torsion
becomes preferred and the ground-state configuration of the
smectic-C* becomes helicoidal.

\section{Discussion}

The present theoretical study proves that telephone-cord
instabilities are to be expected in smectic-C* liquid crystals. We
have derived the ground state configurations and the preferred
shapes of a thin smectic capillary, possibly endowed with
spontaneous polarization. Having in mind the experimental
conditions in which these instabilities have been already
observed, we have imposed free boundary conditions at the external
surface of the capillary for both the nematic and the smectic
variables. Nevertheless, a boundary energy has been inserted in
the free energy functional to take into account polarization
effects on the surrounding liquid.

As long as the spontaneous polarization is absent, the preferred
capillary shape remains linear, as we prove in Section 4. In this
case, our analysis (Section 3) proves that a non-null cholesteric
pitch may induce a SmA-SmC transition, even if the elastic
constant $\Co$ is positive. Figure \ref{fig1} shows how the
optimal value of the tilt angle $\alpha$ depends on the
cholesteric pitch for several different values of he elastic
constants and the intrinsic bending stress.

In Section 5 we have focused on spontaneously polarized smectic
liquid crystals. We have found evidence for possible circular
smectic-A* and helicoidal smectic-C* capillaries. Figure
\ref{kappa} shows how the curvature of a smectic-A* capillary is
expected to increase with the spontaneous polarization. Figure
\ref{kata} displays both the curvature and the torsion as a
function of the tilt angle, for several different values of the
cholesteric pitch (that, however, turns out to be not a key
ingredient in the telephone-cord transition).

In our opinion, the result derived in the final subsection
\S\ref{smalltilt} is particularly challenging. Equation
(\ref{last}) shows that even when the smectic part of the free
energy functional pushes towards the smectic-A phase, it is
possible to save free energy by slightly tilting the nematic
molecules with respect to the layers. Once the molecules are
tilted ($\alpha_0>0$), the preferred torsion becomes non-null, and
a telephone-cord instability originates. This effect arises from a
combined action of the spontaneous polarization and the
cholesteric pitch.

\section*{Acknowledgments}
This work has been supported by the NSF Contract DMS-0128832
\emph{Mathematical Modeling of Advanced Liquid Crystal Materials:
Ferroelectricity and Chirality  \/}. P.B.\ acknowledges the
hospitality of the University of Minnesota, where part of this
work was carried out.

\renewcommand{\theequation}{A1.\arabic{equation}}
\setcounter{equation}{0}  

\section*{Appendix A1: Cylindrical-curvilinear coordinates}

Let $\Omega$ be the domain defined in (\ref{defom}),
$\{\bv{T},\bv{N},\bv{B}\}$ the intrinsic frame associated to
\bv{c}, and $(s,\xi,\vartheta)$ the coordinates introduced in
(\ref{defom}). Let further $\bv{e}_\xi,\bv{e}_\vartheta$ be the
unit vectors defined as:
$$
\bv{e}_\xi:=\cos\vartheta\,\bv{N}+ \sin\vartheta\,\bv{B} \qquad
{\rm and}\qquad \bv{ e}_\vartheta :=-\sin\vartheta\,\bv{N}+
\cos\vartheta\,\bv{B}\;.
$$
When we follow the intrinsic unit-vectors' variation along a curve
$\big(s(t), \xi(t),\vartheta(t)\big)$ in $\Omega$, the
Frenet-Serret formul\ae\ imply: $\bv{\dot T}=\kappa\dot
s\,\bv{N}\,$,\quad $\bv{\dot N}=-\kappa\dot s\,\bv{T}-\tau\dot
s\,\bv{B}\,$,\quad $\bv{\dot B}=\tau\dot s\,\bv{N}\,$, so that
$$
\bv{\dot e}_\xi=\big(\dot\vartheta-\tau\dot s \big)\,
\bv{e}_\vartheta-\kappa\dot s \cos\vartheta\,\bv{T} \qquad{\rm
and}\qquad \bv{\dot e}_\vartheta=-\big(\dot\vartheta-\tau\dot
s\big)\,\bv{e}_\xi+ \kappa\dot s\sin\vartheta\,\bv{T}\;.
$$
We thus obtain
$$
\dot P=\frac{d}{dt}\,\big(\bv{c}(s)+\xi\,\bv{e}_\xi\big)=
\big(1-\kappa\xi \cos\vartheta\big)\,
\dot s\,\bv{T}+\dot\xi\,\bv{e}_\xi+\xi\,\big(\dot\vartheta-\tau\dot s
\big)\, \bv{e}_\vartheta
\;.
$$
For any differentiable real function
$\Psi:\reals^3\to\reals$ we have:
$$
\bv{\nabla}\Psi=\frac{\Psi_{,s}+\tau\Psi_{,\vartheta}}
{1-\kappa\xi \cos\vartheta}\;\bv{T}+
\Psi_{,\xi}\,\bv{e}_\xi+\frac{\Psi_{,\vartheta}}{\xi}
\,\bv{e}_\vartheta\;,
$$
where a comma denotes differentiation with respect to the
indicated variable. In particular, if $\Psi$ depends only on the
arc-length $s$,\quad $\dps\bv{\nabla}\Psi(s)=\frac{1}
{1-\kappa\xi\cos\vartheta}\,\frac{d\Psi}{ds}\,\bv{T}\;$.\\
Furthermore,
\begin{align*}
&\bv{\nabla T}=\phantom{-}\frac{\kappa\cos\vartheta}{1-\kappa\xi\cos\vartheta}\,
\bv{e}_\xi\otimes\bv{T}-\frac{\kappa\sin\vartheta}{1-\kappa\xi\cos\vartheta}\,
\bv{e}_\vartheta\otimes\bv{T}\;,\\
&\bv{\nabla e}_\xi=-\frac{\kappa\cos\vartheta}{1-\kappa\xi\cos\vartheta}
\,\bv{T}\otimes\bv{T}+\frac{1}{\xi}\,\bv{e}_\vartheta\otimes\bv{e}_\vartheta
\;,\\
&\bv{\nabla e}_\vartheta=\phantom{-}\frac{\kappa\sin\vartheta}
{1-\kappa\xi\cos\vartheta}
\,\bv{T}\otimes\bv{T}-\frac{1}{\xi}\,\bv{e}_\xi\otimes\bv{e}_\vartheta
\;,\\
&\bv{\nabla B}=\frac{\tau}{1-\kappa\xi\cos\vartheta}\,\bv{N}\otimes\bv{T}\;\qquad
{\rm and}\\
&\bv{\nabla N}=-\frac{\kappa}{1-\kappa\xi\cos\vartheta}\,\bv{T}\otimes\bv{T}
-\frac{\tau}{1-\kappa\xi\cos\vartheta}\,\bv{B}\otimes\bv{T}\;.
\end{align*}
The volume element in $\Omega$ is given by\quad
$dv=\xi\,|1-\kappa\xi\cos\vartheta|\,ds\,d\xi\,d\vartheta\,$,\quad
so that the curvilinear coordinate system $(s,\xi,\vartheta)$ is
well defined as long as $|1-\kappa\xi|> 0$, which implies
$\kappa(s)\, r<1$ for all $s\in[0,\ell]$, since $\kappa$ is
non-negative by construction. The volume of $\Omega$ is:
$$
{\rm Vol}\,(\Omega)=\int_0^\ell ds \int_0^r d\xi\int_0^{2\pi}
d\vartheta\, \xi \big(1-\kappa\xi\cos\vartheta\big)=\pi r^2 \ell\;.
$$

\end{document}